\documentclass[preprints,review,accept]{mdpi} 
\usepackage{lscape}
\firstpage{1} 
\pubvolume{1}
\issuenum{1}
\articlenumber{0}
\pubyear{2023}
\copyrightyear{2023}
\datereceived{19 March 2023} 
\daterevised{13 May 2023}
\dateaccepted{16 May 2023} 
\datepublished{} 
\hreflink{https://doi.org/} % If needed use \linebreak
\Title{Eighteen years of kilonova discoveries with Swift}

\TitleCitation{Eighteen Years of Kilonova Discoveries with Swift}

\Author{Eleonora Troja $^{1,2}$\orcidA{}}

\AuthorNames{Eleonora Troja}

\AuthorCitation{Troja, E.}
\address{%
$^{1}$ \quad {Department of Physics}, University of Rome Tor Vergata, 
Via Cracovia 50, 00100 Rome, Italy; eleonora.troja@uniroma2.it \\
$^{2}$ \quad Italian National Institute of Astrophysics (INAF), Via Fosso del Cavaliere 100, 00133 Rome, Italy 
}

\corres{Correspondence: eleonora.troja@uniroma2.it}

\abstract{
\textit{Swift} has now completed 18 years of  mission, during which it discovered thousands of gamma-ray bursts as well as new classes of high-energy transient phenomena. 
Its first breakthrough result was the localization of short duration GRBs, which enabled for redshift measurements and kilonova searches. 
\textit{Swift}, in synergy with the \textit{Hubble Space Telescope} and a wide array of ground-based telescopes, provided the first tantalizing evidence of a kilonova in the aftermath of a short GRB. In 2017, \textit{Swift} observations of the gravitational wave event GW170817 captured the
early UV photons from the kilonova AT2017gfo, opening a new window into the physics of kilonovae.
Since then,  \textit{Swift} has continued to expand the sample of known kilonovae, leading to the surprising discovery of a kilonova in a long duration GRB. This article will discuss recent advances in the study of kilonovae driven by the  fundamental contribution of \textit{Swift}. }


\begin{document}

%%%%%%%%%%%%%%%%%%%%%%%%%%%%%%%%%%%%%%%%%%
\section{Introduction}

Kilonovae are a new class of luminous ($L_{pk}\approx$10$^{40-41}$ erg s$^{-1}$) astrophysical transients 
powered by the radioactive decay of heavy elements \citep{Li98,Metzger19}.
These heavy nuclei are synthesized during the rapid decompression of dense and neutron-rich material ejected from compact binary mergers, composed either by two neutron stars (NSs) \citep{Baiotti17,Bauswein13,Korobkin12,Rosswog99,Fre99,Eichler89}
or by a NS and a black hole (BH) \citep{Kyutoku15,Foucart13,Shibata11,Etienne08,Lee07,Rosswog05,Lattimer76,Lattimer74}. 
These mergers are loud sources of gravitational wave (GW) radiation \citep{GWTC3} and progenitors of short-lived flashes of high-energy radiation, known as gamma-ray bursts (GRBs) \citep{Meszaros02, Piran99, Narayan92,Bohdan91}. 
Therefore, kilonovae are expected to accompany GW sources detected by ground-based interferometers as well as GRBs, especially those of short ($<$2 s) duration.
The key observational features of a kilonova are its red color \citep{Tanaka13,Barnes13} and fast evolving timescales ($\sim$day to $\sim$week), which distinguish it from the plethora of astrophysical transients \citep{Cow15,Andreoni21}.

The \textit{Neil Gehrels Swift} mission (hereafter \textit{Swift}) \citep{Gehrels04} played a fundamental role in the discovery and characterization of the first kilonovae.
\textit{Swift}, launched on November 20, 2004, was primarily designed to chase GRBs and localize their rapidly fading afterglow \citep{vanpar97,Costa97,Mes97}. 
The satellite is equipped with a wide-field hard X-ray (15-150 keV) monitor, the Burst Alert Telescope (BAT) \citep{BAT05}, that continuously scans the sky searching for new bursts. 
Once a GRB is discovered, the satellite rapidly slews to its position and, within a couple of minutes, begins observations in the X-ray, ultraviolet (UV) and optical bands with its narrow field instruments, the X-ray Telescope (XRT) \citep{XRT05} and the UltraViolet/Optical Telescope (UVOT) \citep{UVOT05}, respectively. 
This strategy has led to the accurate localization of thousands of GRBs and, in particular, was key to localize short duration GRBs, measure their distance scale, and enable searches for kilonovae with sensitive optical and near-infrared facilities \citep{Gehrels05,Berger14}.  

The identification of kilonovae in a sample of nearby short GRBs \citep{Rossi20,Jin20,Ascenzi19,Lamb19, Troja19b,Troja18b,Tanvir13} provides direct evidence that these bursts are produced by the merger of two compact objects. 
Unexpectedly, signatures of kilonova emission were also found in a small number of nearby long GRBs \citep{Troja22,Rastinejad22,Yang22,Jin16,Yang15,Jin15,Ofek07}, upending the standard paradigm that links the GRB duration to its progenitor system. 
In fact, the canonical classification of GRBs in two phenomenological classes, long duration/soft spectrum bursts and short duration/hard spectrum bursts \citep{K93}, is often translated into a dichotomy of their progenitor systems, collapsing massive stars and NS mergers, respectively.  Whereas this tidy scheme seems to hold for the majority of events,  the large sample of well-localized \textit{Swift} bursts demonstrated that each phenomenological class is a heterogeneous mix of different astrophysical phenomena involving both young and evolved stellar populations \citep{Zhang07}.

In addition to its premiere role in GRB studies,  \textit{Swift} naturally became a workhorse facility for the follow-up of GW sources thanks to its multi-wavelength coverage, rapid response and flexible schedule. 
Thus far, \textit{Swift} observations of GW sources were performed either as rapid large-scale tiling of GW error regions or as single pointings of candidate counterparts identified by other facilities
\citep{Oates21,Klingler21,Page20,Evans17,Evans16}. 
This led to the detection of the early UV
emission from the kilonova AT2017gfo \citep{Evans17}, associated with the binary NS merger GW170817 \citep{Abbott17}. 

As the sensitivity of the GW detectors improves and the rate of detectable events (as well as their distance from us) increases, the detection and identification of the associated kilonova emission becomes a daunting task \citep[e.g.,][]{Chase22}. 
The ideal scenario would be for BAT to trigger on a gamma-ray transient associated with the GW source.   
This could be either the standard prompt GRB emission, 
if the GRB outflow is pointed toward Earth (on-axis), 
or a weaker high-energy signal, if the outflow is misaligned (off-axis) as in the case of GW170817. Shock-breakout flares \citep{Lazzati17,Bromberg18}, precursors \citep{Troja10}, and temporally extended emission \citep{NB06} could be visible from a wider range of viewing angles, and potentially aid gamma-ray monitors in the discovery of off-axis explosions. 
A BAT trigger would deliver an arcminute localization of the GW transient and initiate sensitive observations at X-ray and UV/optical wavelengths with \textit{Swift}'s narrow field instruments.  
These early-time observations have great diagnostic power. X-rays probe the non-thermal radiation from the fastest ejecta and can place initial constraints on the relativistic outflow and its orientation \citep{Ryan15,Ryan20,Evans17,Troja17b}. 
As demonstrated by the case of GW170817, they are also essential to estimate the afterglow contribution at lower energies, thus aiding in the kilonova identification.  
Moreover, during the first few hours after the merger, the kilonova emission from the hot, sub-relativistic ejecta peaks at UV and optical wavelengths.
Its luminosity depends on the ejecta composition and morphology, and is sensitive to the nature of the merger remnant and the properties of the relativistic outflow \citep{Nativi21,Banerjee22,Even20,Wollager19,Waxman18,Piro18,Arcavi18}. Joint GW/BAT triggers, although expected to be rare, would provide us with a treasure trove of information on infant kilonovae that is unlikely to be collected through standard GW follow-up. 

In this paper, I will review recent advances in the study of kilonovae driven by the contribution of \textit{Swift}. Section 2 focuses on \textit{Swift} observations of GW sources with particular regard to the UVOT detection of the kilonova AT2017gfo. 
Section 3 discusses the discovery and status of kilonovae associated with GRBs.

\section{Kilonovae associated with gravitational wave counterparts}

In its early years, \textit{Swift} operated during the observing runs of the Laser Interferometer Gravitational-Wave Observatory (LIGO) in its initial configuration (November 2005 - September 2007) and enhanced configuration (July 2009 - October 2010). 
During the latter run, \textit{Swift} performed the first follow-up observations of candidate GW sources \citep{Evans12}, but none of the candidates turned out to be a real astrophysical event and no electromagnetic counterpart was found. 

Thanks to an increased sensitivity, Advanced LIGO, followed a couple of years later by Advanced Virgo, ushered in a new era of gravitational wave astronomy which began with the discovery of the first binary BH merger (GW150914; \citep{150914}) and culminated with the first multi-messenger observations of a binary NS merger (GW170817; \citep{170817MMA}). 
Unfortunately, at the time of the merger, the location of GW170817 was occulted by the Earth and its associated GRB170817A could not be seen by the BAT. 
Nonetheless, \textit{Swift} responded to the alert  and began observations of the field within 1 hr since the GW trigger, first by tiling the probability peak of the GRB localization, then by targeting known nearby galaxies within the refined GW localization \citep{Gehrels16,Evans17}. 
However, none of these early observations covered NGC4993, the host galaxy of the GW source. 

As the first candidate counterparts from ground-based imaging were publicly released, \textit{Swift} began pointed observations of the most promising sources, and ultimately settled for the target Swope Supernova Survey 17a \citep{Coulter17}, later confirmed to be the counterpart of GW170817 and dubbed AT2017gfo. 

\begin{figure}[H]
\centering
\includegraphics[width=12.5 cm]{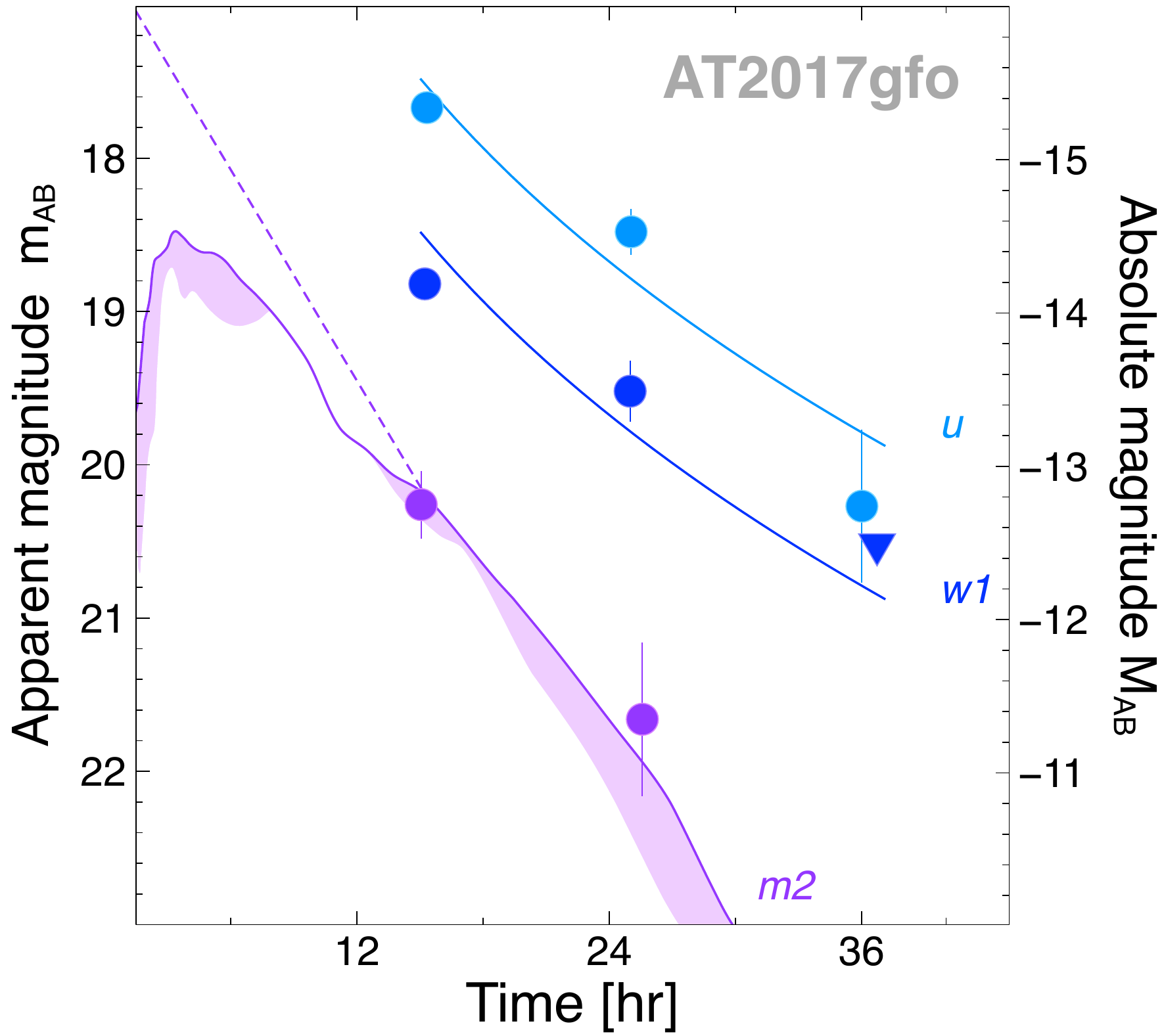}
\caption{\textit{Swift}/UVOT light curves of the kilonova AT2017gfo associated with the GW transient GW170817 \citep{Evans17}. Data are corrected for Galactic extinction along the line of sight, and compared with a range of temporal models. An empirical power-law function is used to illustrate the  decay of the $u$ and $w1$ measurements. Three possible kilonova models were compared to the $m2$ data: one with light r-process elements (solid line), one with light and heavy r-process elements (shaded area), and one including the effects of jet/ejecta interactions (dashed line).
\label{uvotkn}}
\end{figure}

XRT and UVOT observations of AT2017gfo began 0.6 d after the GW trigger, providing key spectral information for the classification of the transient.
\textit{Swift}/UVOT detected a bright UV and optical counterpart (Figure~\ref{uvotkn}) characterized by a 
rapid temporal decay, $F_{\nu} \propto t^{-1.1}$, and a steep spectral index, $F_{\nu} \propto \nu^{-4}$. These properties are different from those observed at optical and nIR wavelengths. 
The steep spectral index and chromatic behavior are not typical of GRB afterglows at this stage of their evolution. An afterglow origin was further disfavored by the lack of an X-ray counterpart down to a 3 $\sigma$ limit of $\lesssim10^{-13}$\,erg\,cm$^{-2}$\,s$^{-1}$ at 1 d post-merger \citep{Evans17}. 

The early behavior of AT2017gfo is instead well described by a thermal spectrum with temperature $T\approx$7,300 K at 0.6 d and emitting radius $R\,\approx\,6\times10^{14}$\,cm at 0.6~d \citep{Evans17}.
According to this model, the bolometric luminosity of the blackbody component is $L_p\,\approx\,6\times10^{41}$\,erg\,s$^{-1}$ and its peak is $t_{p} \lesssim$0.6 d. These two quantities can be used for a rough estimate of the ejecta physical properties, such as its total mass $M_{ej}$ and expansion velocity $v_{ej}$, using a simple analytical model \citep{Grossman14,Arnett82}:
\begin{equation}
M_{ej} \approx 0.015 M_{\odot} \left(\frac{L_p}{6\times10^{41} {\rm ~erg\,s^{-1}}}\right) \left(\frac{t_p}{0.6 {\rm ~d}}\right)^{\alpha}
\end{equation}
\begin{equation}
v_{ej} \approx 0.3\,c \left(\frac{\kappa}{0.3 {\rm ~cm^{2}\,g^{-1}}}\right) \left(\frac{M_{ej}}{0.015 M_{\odot}}\right) 
\left(\frac{t_p}{0.6 {\rm ~d}}\right)^{-2}
\end{equation}
where $\alpha\approx$1.3 is the nuclear energy generation rate \citep{Korobkin12,Goriely11} and $\kappa$ is the opacity of the ejecta, which must be low in order to explain the bright UV emission.

If entirely powered by radioactive decay energy, 
the luminosity and timescale of the UV emission imply a relatively large mass ($\approx 0.015M_{\odot}$) of fast-moving ($\approx$0.3$c$) low-opacity ejecta, hard to explain through the standard channels of mass ejection in compact binary mergers \cite{Rosswog99,Bauswein13,Radice18,Shibata19}.
Although a widespread consensus interprets the UV counterpart of GW170817 as kilonova emission, these results appear in tension with our common understanding of NS mergers. 
Viewing angle effects and uncertainties in the heating rate could alleviate this discrepancy \cite{Korobkin21,Barnes21}. Alternatively, the UV/optical luminosity could be boosted if the merger ejecta gets re-energized either by a long-lived NS remnant \cite{Yu13} or by its interactions with the relativistic outflow \cite{Nativi21,Piro18}.

Early time UV observations of infant kilonovae would be crucial to distinguish between different models and probe the ejecta composition \citep{Arcavi18,Even20}. This is illustrated in Figure~\ref{uvotkn} where the UVOT $m2$ light curve of AT2017gfo is compared with different kilonova models. The effects of jet/ejecta interactions would make the kilonova look bluer and brighter (dashed line) than basic radioactively powered models (solid line). Heavy r-process elements, if present, would instead suppress the UV emission, leaving detectable imprint in the kilonova light curve and spectrum \citep{Even20,Banerjee22}. 
\textit{Swift}/UVOT observations of future GW counterparts can potentially fill this knowledge gap if an accurate localization is promptly available.  
Imaging of the kilonova in its early stages would distinguish between pure radioactive models and 
those requiring an additional energy source. 
However, spectroscopy is ultimately necessary to characterize the ejecta composition
\cite[e. g.][]{Tarumi23,Gillanders21,Watson19,Pian17}.
The predicted range of kilonova luminosities and distance scales are not within the reach of the UVOT grism spectrograph \cite{Kuin05}, and would require ultra-rapid observations with the \textit{Hubble Space Telescope} (HST).  In the immediate future, the ideal scenario to make these observations possible would be a joint GW/BAT trigger, capable of delivering a precise position of the GW source within minutes. 
In the longer term, the launch of a wide-field UV monitor, such as the Ultraviolet Transient Astronomy Satellite (ULTRASAT) \cite{benami2022}, would increase the prospects for an early kilonova detection \cite{dorsman23}.

\section{Kilonovae associated with short GRBs}

One of the most enduring notion in the GRB field is that bursts of short duration are produced by mergers of compact objects. This link was spectacularly demonstrated by the detection of GW170817 and its associated short duration GRB170817A \citep{170817MMA}. 
In its first 18 years of mission (2004-2022), \textit{Swift} has detected over 130 short-duration, hard-spectrum GRBs.  These are defined by the criteria $T_{90}<2 $s and $HR>1$, where  the duration $T_{90}$ is the time interval over which 90\% of the broadband (15-150 keV) fluence is measured, and the hardness ratio $HR$ is the fluence ratio between the hard band (100-150 keV) and the soft (50-100 keV) band \citep{Lien16,Sakamoto08}.
This legacy dataset is the most valuable tool to probe the properties of compact binary mergers and their associated kilonovae across cosmic time. 

Of the $\sim$130 short hard GRBs observed by \textit{Swift}, $\approx$70\% were localized by the XRT, and only $7$\% were also detected in at least one of the \textit{Swift}/UVOT filters, typically the broad \textit{white} filter. These UVOT-detected short bursts have inferred redshifts in the range between 0.111 \citep{Stratta07} and 2.609 \citep{Antonelli09} with a median of $z\approx$0.9, consistent with the general population of short GRBs \citep{Oconnor22}.

\subsection{Kilonovae in the nearby Universe}

\begin{figure}[H]
%\centering
\includegraphics[width=12.5 cm]{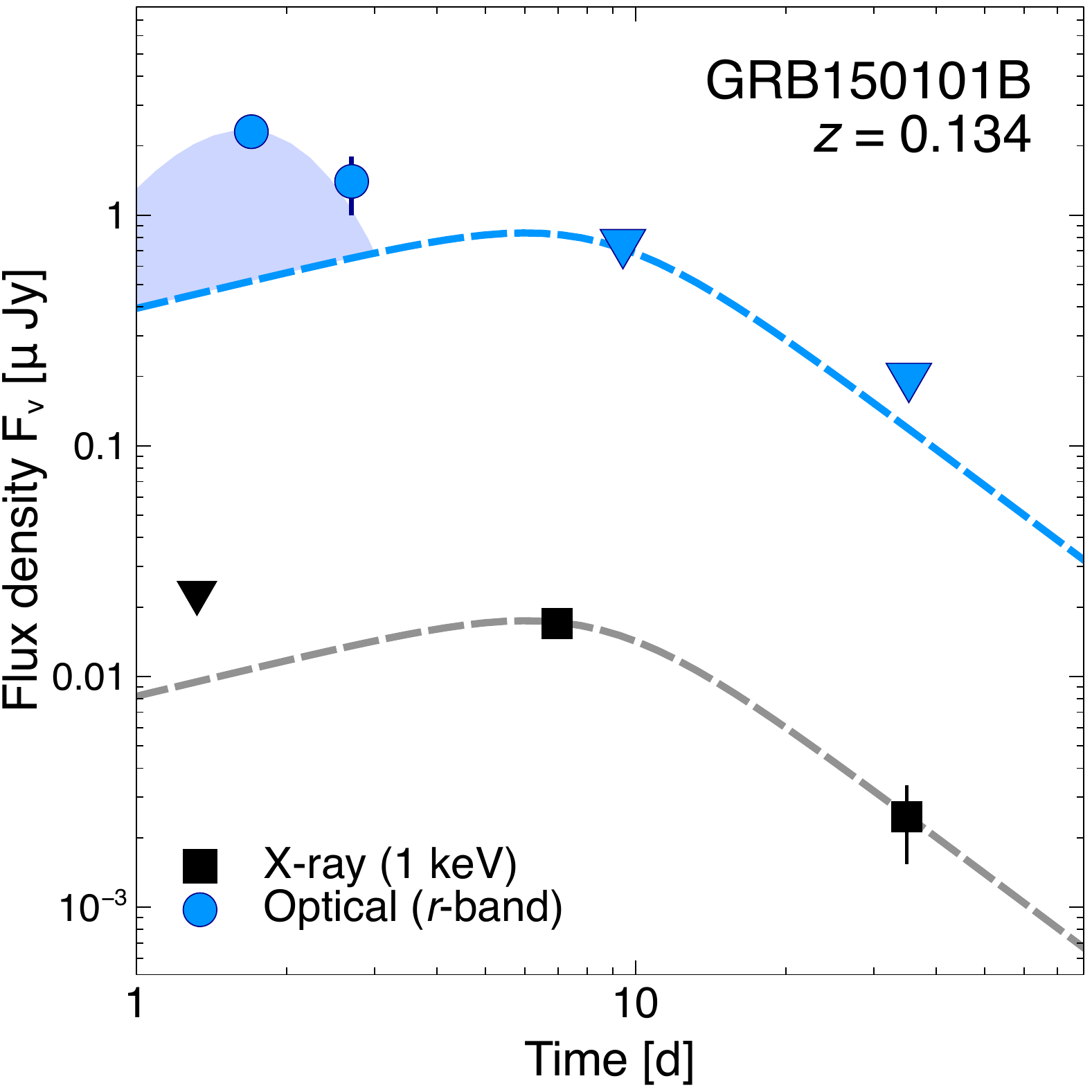}
\caption{Optical (blue circles) and X-ray (black squares) light curves of GRB150101B \cite{Troja18b}. Downward triangles are 3 sigma upper limits. At early times, the optical emission lies above the afterglow model (dashed lines),
and is consistent with a kilonova excess (shaded area).
\label{grb150101b}}
\end{figure}

The discovery of GRB170817A and its delayed X-ray afterglow \cite{Troja17b,Troja19a} showed that the local population of NS mergers would likely appear as low-luminosity short GRBs with no bright X-ray counterpart at early times.
Thus, archival data from \textit{Swift} were re-analyzed in a new light in an attempt to find local bursts similar to GRB170817A. 
This led to the identification of GRB150101B as a possible cosmological analogue \cite{Troja18b}, characterized by a late-peaking afterglow and an early optical emission consistent with a bright blue kilonova (Figure~\ref{grb150101b}). Its prompt gamma-ray properties were also considered similar, in some aspects, to GRB170817A \cite{Burns18}. Its low gamma-ray energy and long lived X-ray emission are characteristic traits of a GRB jet seen far from its axis (off-axis). 

Modeling of the X-ray afterglow points to an energetic jet ($E_{k,iso}\approx$10$^{53}$~erg) with a narrow core ($\theta_{c}\approx$3 deg), 
seen from an angle of $\theta_{v}\approx$13 deg  \cite{Troja18b}. For comparison, the viewing angle inferred for GRB170817A is approximately 20-25 deg \cite{Troja19a,Ryan20,Mooley22}.
This model  shows that the afterglow contribution at optical wavelengths is sub-dominant at $t\lesssim2$ d (dashed lines, Figure~\ref{grb150101b}), suggesting that the observed optical counterpart is mostly powered by kilonova emission (shaded area, Figure~\ref{grb150101b}). The candidate kilonova in GRB150101B has an optical luminosity $L_o \approx 2 \times 10^{41}$~erg~s$^{-1}$\ at 1.5~d, that is roughly 2 times brighter than AT2017gfo at the same epoch. It provided the first evidence, later confirmed by other short GRBs (Section~\ref{sec:cosmological}), that the early blue emission of AT2017gfo is not unusual and could be a rather common component of kilonovae. 
However, as in the case of AT2017gfo, 
 this luminous optical emission implies a massive ($\gtrsim 0.02M_{\odot}$)
 low-opacity ($\kappa\lesssim$1 ~cm$^{2}$~g$^{-1}$)
 ejecta in rapid expansion ($\gtrsim$0.15$c$), whose origin is not fully understood.

Based on the experience of GRB170817A and GRB150101B, subsequent searches for off-axis afterglows mainly focused on short GRBs without an early X-ray detection \cite{Ricci21,Dichiara20,Bartos19}, as this sample could potentially harbor other off-axis short GRBs in the nearby universe.
A third of \textit{Swift} short bursts lack an X-ray counterpart, and are localized by BAT with an accuracy of a few arcminutes. 
These BAT localizations were cross-correlated with galaxy catalogs \cite[e.g.][]{Glade18} and, in a handful of cases, were found to intercept a known nearby galaxy.
In general, the BAT localization error is too large to robustly identify the GRB host galaxy. 
However, the number of matches between short GRBs with no X-ray afterglow and nearby galaxies is slightly higher than the value expected from spurious alignments
\cite{Dichiara20}, the probability to recover the same number of matches 
by chance  being $\approx$2\%.  This lends support to the hypothesis that at least some of these GRBs were hosted in nearby galaxies, constraining the local event rate
to $\lesssim$160~Gpc$^{-3}$~yr$^{-1}$ \cite{Dichiara20}. 
Among the candidate local GRBs, a notable example is the short duration ($T_{90}\approx$0.6~s) GRB190610A \citep{GCN190610}, which occurred during operations of the two LIGO detectors and was localized by BAT close to a nearby ($\approx$165 Mpc) galaxy. Unfortunately, the target's position was too close to the Sun and could not be promptly observed. It was also not optimal for GW detectors, which only reached a distance horizon of 63 Mpc for a NS merger progenitor \citep{LVC21}.

Thus far, these tentative associations between GRBs and local galaxies were not confirmed by any other observational evidence. In particular, no evidence for a bright kilonova was found in archival data, either from \textit{Swift}/UVOT or ground-based imaging \cite{Dichiara20}. The available limits are sparse and mostly obtained in the UV/optical range, thus are not sufficient to rule out the presence of a kilonova fainter or redder than AT2017gfo. Long-term radio monitoring of these bursts can potentially constrain their distance scale by detecting the late-time synchrotron afterglow arising from the interaction of the kilonova ejecta with the surrounding medium \cite{Ghosh22,Ricci21,Bruni21, Bartos19,Fong16,NP11}. 

Explosions seen slightly off-axis, like GW170817, are the most promising targets for studying the early kilonova behavior, which is otherwise outshone by the standard afterglow. 
As \textit{Swift} continues to operate and new facilities, such as \textit{SVOM} (Space-based multi-band astronomical Variable Objects Monitor; \citep{Wei16}) and \textit{Einstein Probe} \citep{Weimin15}, come online, observing campaigns targeting candidate off-axis GRBs could confirm their local origin and identify their kilonovae.

\subsection{Kilonovae at cosmological distances}\label{sec:cosmological}

The majority of \textit{Swift} short bursts lie at cosmological distances, where events like GRB170817A are too faint to be detected. Therefore, the observed population is dominated by on-axis explosions, that is those with a relativistic jet pointing toward Earth. 
Detection of the non-thermal afterglow emission from this jet enables for a rapid and accurate GRB localization \citep{Gehrels05}.
However, the bright afterglow tends to outshine any other emission component at all wavelengths, and represents one of the main challenges in the identification of a kilonova. 
Other factors affecting the search for kilonovae are the presence of dust along the sightline, contamination from the host galaxy light, and the GRB distance scale. Their effect is discussed in more detail below. 

Dust extinction is estimated through the study of the UV/optical and near-infrared (nIR) afterglow spectral energy distribution (SED). 
The afterglow is characterized by a broadband synchrotron spectrum, which, over a narrow energy range,  can be approximated as a power-law function, $F_{\nu} \propto \nu^{-\alpha}$. 
On its way to us, the intrinsic power-law shape is modified 
by dust effects, which mainly suppress the UV and optical flux. 
Therefore, the amount of extinction $A_V$ along the line of sight can be measured by comparing the observed spectral shape with the expected power-law model \citep[e.g.][]{Schady10, Starling07, Kann06}. Due to the faintness of short GRB afterglows, 
the available dataset is sparse and allows for a direct measurement of $A_V$ in just a few cases \citep[e.g.][]{Oconnor21,Troja16,Fong14,Berger09,Soderberg06, Roming06,DePasquale10,Stratta07}. 

An indirect inference of the dust content along the sightline can be derived from X-ray spectroscopy \citep{Galama01,Stratta04}. 
Soft X-rays are absorbed by the intervening gas, generally quantified in terms of the equivalent hydrogen column density $N_H$. 
This measurement can be related to the optical extinction by assuming a gas-to-dust ratio comparable to the Galactic one, $N_H \approx 2.69 \times 10^{21} A_V$ \citep{Nowak12}. 

\begin{figure}[H]
%\centering
\includegraphics[width=12.5 cm]{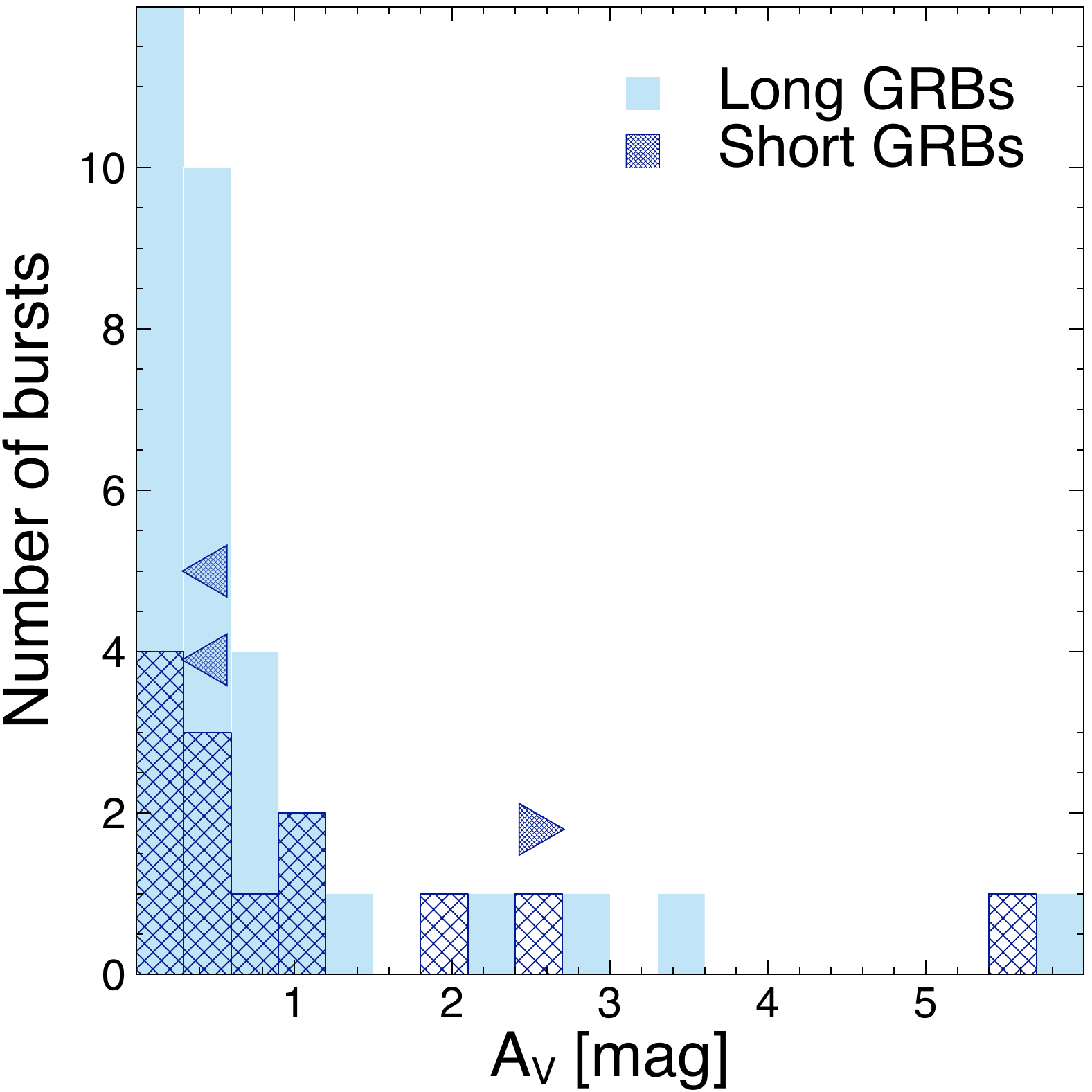}
\caption{Dust extinction in GRB afterglows. Triangles show upper limits (leftward) and lower limits (rightward). Data for long GRBs were taken from \citep{Covino13}.\label{av}}
\end{figure} 

The distribution of extinction values, calculated with both these methods, is reported in Figure~\ref{av}, and shows that a modest amount of dust is present in short GRB environments.
This is sufficient to redden the optical spectrum as kilonovae do. 
Therefore, a red color of the GRB afterglow does not represent unambiguous evidence for a kilonova: 
two or more epochs of multi-color observations are crucial to track the spectral evolution and disentangle the effects of dust from a genuine kilonova.

The GRB location within its host galaxy is another factor affecting the identification of kilonovae. 
This observational bias disfavors the detection of kilonovae in the inner galactic regions in two ways. First, 
contamination from the galaxy's light is higher in these regions and affects our practical capabilities to resolve the transient against the background. 
Second, according to the canonical fireball model \cite{Piran99,Mes97,Kumar15}, the afterglow brightness depends on the density of the surrounding environment as $\propto\,n_0^{1/2}$, which tends to be higher near the galactic core \cite{Oconnor20,BB12}. Indeed, the sample of short GRB afterglows displays a trend of brightness versus galactocentric offset \cite{Oconnor22} and, as mentioned above, bright afterglows hinder the identification of kilonovae. 
Fortunately, short GRBs are widely spread around their galaxies, and only about one third is located within 0.5$"$ from the galaxy's center \cite{Fong13,Berger14,Oconnor22}.

Finally, a major obstacle in the discovery of new kilonovae has been the GRB distance scale. Bright kilonovae can be detected with \textit{HST} up to $z \sim 0.5$, whereas ground-based telescopes can probe them up to $z \lesssim 0.25$. 
Although the median redshift typically quoted for short GRBs is $z \sim 0.5$, this value is derived from measured redshifts and only places a lower limit to the true distance distribution of short GRBs: most bursts with unknown redshift are likely harbored in distant ($z\gtrsim$1) galaxies \cite{Oconnor22,Dichiara21} for which spectroscopic measurements are often unsuccessful. When accounting for this population of events, the median distance scale of short GRBs tends to $z\approx$1.
Therefore, only a small fraction of bursts are sufficiently close for a kilonova detection. For instance, during \textit{Swift}'s lifetime, approximately 100 short GRBs were localized to arcsecond accuracy, but only $\approx$20 events were found at $z$\,$\lesssim$0.5, and less than 10 at $z$\,$<$0.25. A list of these bursts and their properties is reported in Table~\ref{tab1}.

\begin{table}[H] 
\caption{\textit{Swift} short GRBs at $z\lesssim$0.5. Bursts are grouped in two samples based on the robustness of their association with a host galaxy. 
\label{tab1}}
\newcolumntype{C}{>{\centering\arraybackslash}X}
\begin{tabularx}{\textwidth}{CCC}
\toprule
\textbf{GRB}	& \textbf{Redshift}\footnotemark[1]	& \textbf{Comment} \\
\midrule
\multicolumn{3}{c}{Group 1}\\
\midrule
051221A		& 0.5464        & Afterglow dominated \\
070724A		& 0.4571		& Dust obscured \\
080905A		& 0.1218        & Possible kilonova \\
130603B     & 0.3565    	& Candidate kilonova \\
140903A     & 0.3510			& Afterglow dominated \\
150101B     & 0.1341		& Candidate kilonova \\
160821B     & 0.1613		& Candidate kilonova \\ 
170428A     & 0.454         & Afterglow dominated \\
200522A     & 0.5541		& Afterglow dominated \\
%210726A     & 0.35$\pm$0.15   &  No OT \\
\midrule
\multicolumn{3}{c}{Group 2}\\
\midrule
050509B	& 0.225			& No OT \\
060502B		& 0.258			& No OT \\ 
061201		& 0.111/0.084 	& Hostless/Possible kilonova \\ 
070809 		& 0.218/0.473	& Hostless/Possible kilonova \\
090515      & 0.403         & Hostless/Possible kilonova \\
100206A     & 0.4068         & No OT \\
100625A     & 0.452         & No OT \\
101224A     & 0.4536         & No OT \\
120305A     & 0.225         & No OT \\
130822A     & 0.154         & No OT \\
150120A     & 0.4604         & No OT \\
150728A     & 0.461         & No OT \\
160624A     & 0.4833			& No OT \\
210919A     & 0.2415        & No OT \\
\bottomrule
\end{tabularx}
\footnotesize{$^1$ Redshifts were compiled from the literature \cite{Soderberg06,Berger09,Rowlinson10,deugarte14,Troja16,Troja18b,Troja19b,gcn170428A, Oconnor21,Gehrels05,Bloom07,Stratta07,Nugent22,Berger10,Perley12,Fong13,Oconnor22}}

\end{table}

The first group of bursts (Group 1) is characterized by a robust galaxy association, quantified by the chance coincidence probability $P_{cc} \lesssim$2\% \citep{Bloom02}. 
The second group of bursts (Group 2) is characterized by a higher probability of a spurious association, either because of their larger localization error or because the putative host galaxy lies several arcseconds away from the GRB position (hostless bursts). The reported redshift is the value measured for the galaxy with the lowest $P_{cc}$, but it cannot be excluded that the true host is a faint galaxy at higher redshift. 
If two galaxies have comparable probability to be the host, then both their redshifts are listed. 
Although it is unlikely that all the GRB/galaxy associations in Group 2 are spurious, results based on a single burst should be interpreted with caution. 

In Table~\ref{tab1} we consider candidate kilonovae those cases with a robust redshift measurement and a significant optical/nIR excess above the non-thermal afterglow.
Other cases in which the optical luminosity falls within the kilonova range are considered possible kilonovae. However, uncertainties in the distance scale or in the afterglow contribution prevent a clear identification of the kilonova component.

\begin{figure}[H]
%\centering
\includegraphics[width=12.5 cm]{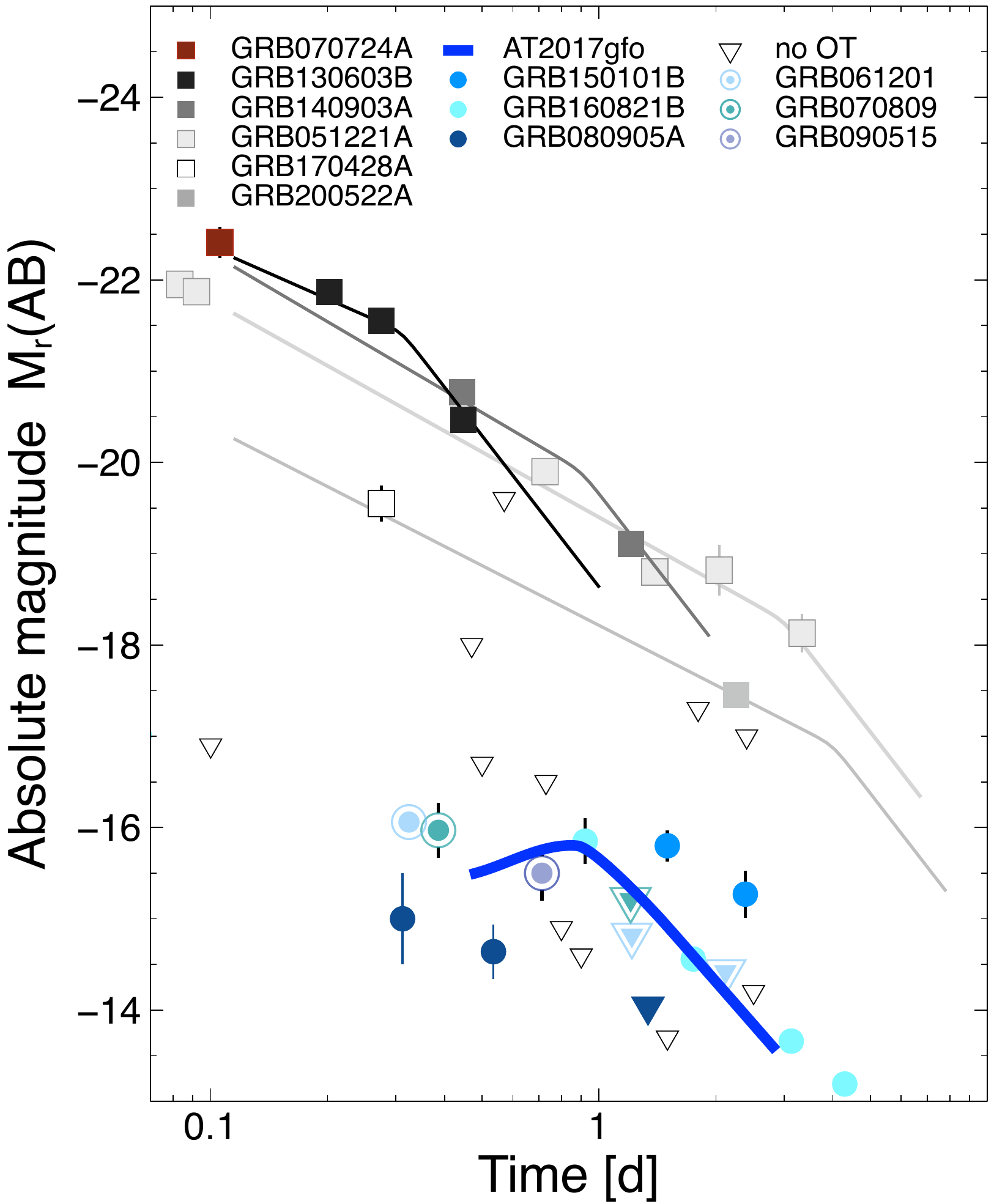}
\caption{Optical light curves of nearby short GRBs. Bursts with bright afterglows are indicated by the squared symbols. 
Solid lines show the best fit afterglow model from multi-wavelength observations. 
Optical counterparts with a possible kilonova contribution are shown by the circles. 
Hostless GRBs are indicated with bulls-eye symbols. 
Downward triangles are 3 sigma upper limits. 
Empty symbols refer to GRBs with no optical counterpart.
\label{otkn}}
\end{figure}

\begin{figure}[H]
%\centering
\includegraphics[width=12.5 cm]{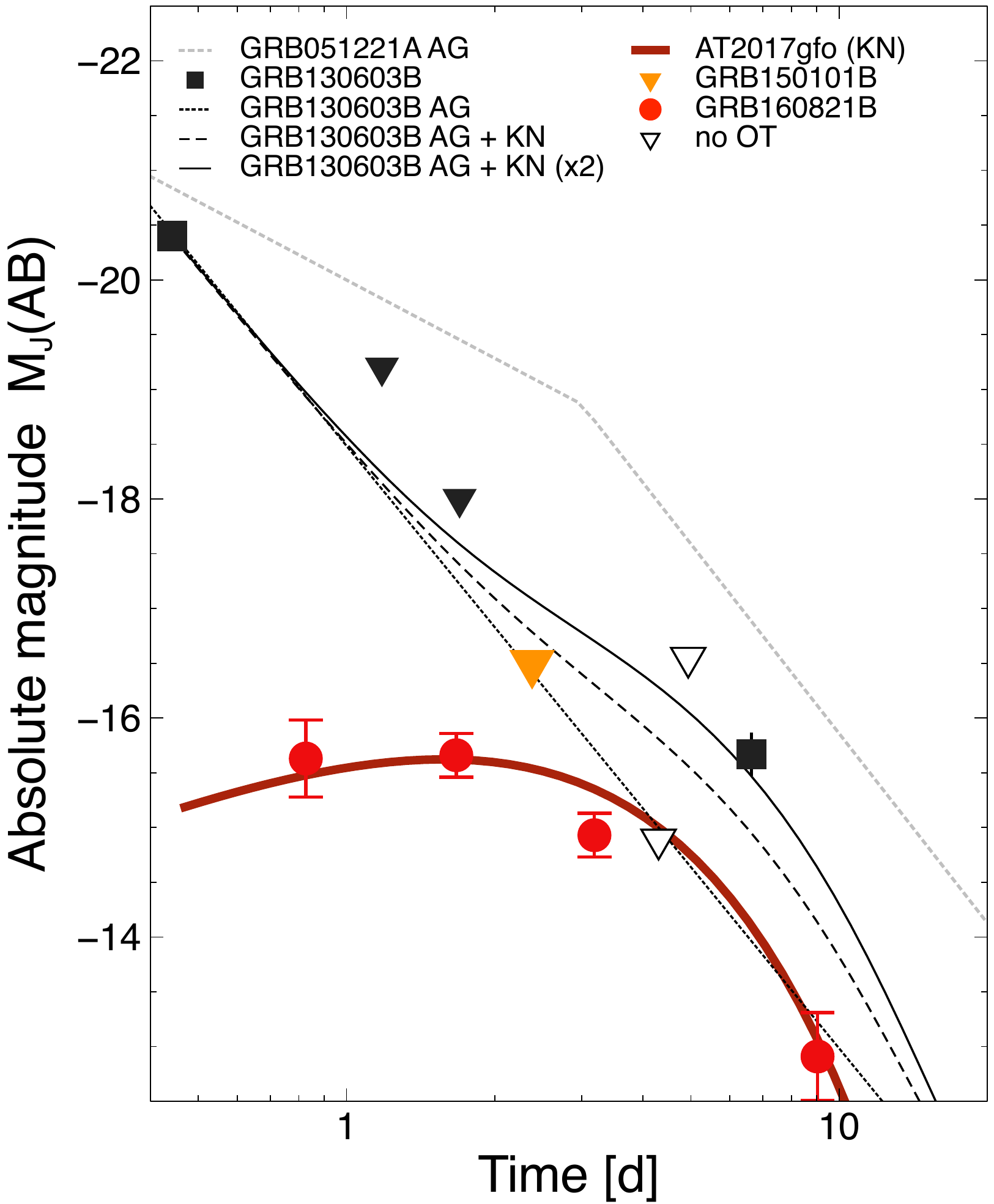}
\caption{Near-infrared light curves of nearby short GRBs. 
The available measurements are few and sparse. Only two short GRBs have late-time detections in the nIR: GRB130603B (squares) and GRB160821B (circles). 
Downward triangles are 3 sigma upper limits. Empty symbols refer to GRBs with no optical counterpart. 
Dotted lines show the best fit afterglow model from multi-wavelength observations. As an illustrative example of a bright afterglow, we report the  model of GRB051221A extrapolated from the X-ray and optical data \cite{Burrows06,Soderberg06}.
\label{irkn}}
\end{figure}

The rest-frame optical and near-infrared light curves of these bursts are shown in Figure~\ref{otkn} and Figure~\ref{irkn}, respectively. 
For comparison, the light curve of the kilonova AT2017gfo is also shown. 

Short GRBs are often observed using the $r$ or $i$ filters, 
therefore the optical light curves were directly derived from the available dataset.  After correcting for Galactic extinction in the burst direction \citep{Schlafly11}, the k-correction to rest-frame $r$-band was calculated using the redshift reported in Table~\ref{tab1} and the observed spectral slope \citep{Hogg02},  then a correction for intrinsic extinction at the GRB site was estimated using a Milky Way extinction law \citep{Pei92}. 
If the optical counterpart was detected only in one filter, a spectral slope of 0.6 was used for the k-correction, and no correction for the unknown intrinsic extinction was applied. 
Coverage at near-infrared wavelengths is sparser and often consists of upper limits. When no detection was available, the nIR light curve was extrapolated from the optical and X-ray data using the measured spectral slope.

At optical wavelengths, AT2017gfo is characterized by an early peak, $M_r\sim-16$ at about 12 hr \citep{Coulter17,lipunov17,valenti17,soares17}, and fast evolving timescales. In some nearby ($z\approx0.2$) GRBs, the optical emission falls in the range of kilonova luminosities and can be used to constrain the kilonova properties.  
A diverse range of behaviors emerges from these observations: a possible kilonova component is visible in  6 short GRBs ($\sim$30\%) although only one event (GRB160821B) closely resembles AT2017gfo \cite{Troja16}. In the case of GRB061201, GRB080905A and GRB070809, the observed light curve, if powered by a kilonova, would require earlier peak times and a faster decay rate than AT2017gfo. 
In a few bursts ($\sim$10\%), including GRB050509B
 and GRB130822A, deep optical limits rule out a kilonova similar to AT2017gfo \citep{Gehrels05,Gompertz18}. 
In all other cases ($\sim$60\%), the limits are unconstraining either due to the bright afterglow or to the shallow depth of the observations. In particular, no useful constraint is derived for $z\gtrsim$0.2.
Exploring kilonovae at farther distances would require a combination of rapid reaction timescales ($\lesssim$1 d) and depth ($\lesssim$27 AB mag), achievable with the next generation of ground-based telescopes.

Searches in the near-infrared appear more promising. The kilonova peaks a few days after the merger when the afterglow has significantly faded. The outflow collimation, which determines the time of the jet-break, plays a significant role in shaping the late-time afterglow evolution. Bursts with bright afterglows and late jet-breaks, like GRB051221A (dotted gray line in Figure~\ref{irkn}), are not suitable for kilonova searches. GRB130603B instead was a good target because of its early jet-break at $\sim$0.5 days. 
Kilonovae span a range of nIR luminosities: from GRB130603B, which is brighter than AT2017gfo by a factor of $\approx$2 (solid black line in Figure~\ref{irkn}),  to GRB160821B, which is fainter than AT2017gfo by a factor of $\approx$2 (note that the data in Figure~\ref{irkn} include both the afterglow and the kilonova contribution). 
The only other useful constraint comes from GRB160624A \citep{Oconnor21}, 
which rules out a kilonova as bright as AT2017gfo.

Despite sustained efforts over the last decade, the available nIR dataset for short GRBs remains limited. 
Even for the closest events at $z\approx$0.1, the range of kilonova luminosities is challenging for any ground-based observatories. However, it is within reach of space-based facilities such as \textit{HST}, \textit{JWST}, and the planned Roman Space Telescope. The detections of GRB130603B and GRB160821B as well as the constraining upper limit from GRB160624A were all obtained by \textit{HST}. The advent of \textit{JWST} can potentially accelerate progress in this area. Thanks to its unprecedented sensitivity, \textit{JWST} can detect a kilonova similar to AT2017gfo up to $z\sim$1, thus allowing us to probe several short GRBs per year rather than one every few years. 
However, these observations would require reaction timescales of 2-3 days and are therefore limited by the small number of disruptive Targets of Opportunity.
% read as "increase the number of ToO on JWST"

\begin{figure}[H]
\centering
\includegraphics[width=9 cm]
{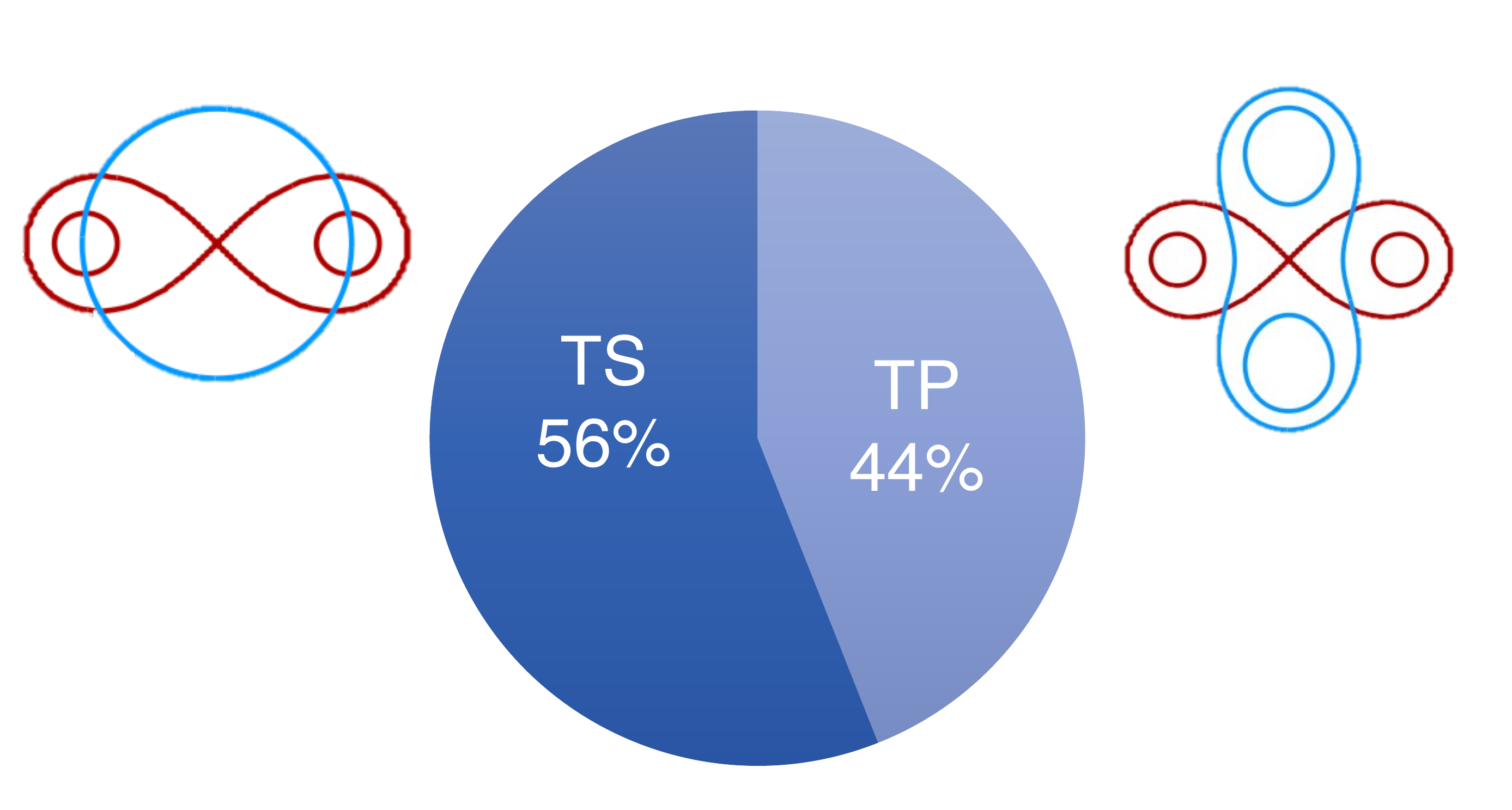}
\vspace{0.4cm}\\
\includegraphics[width=6.5 cm]
{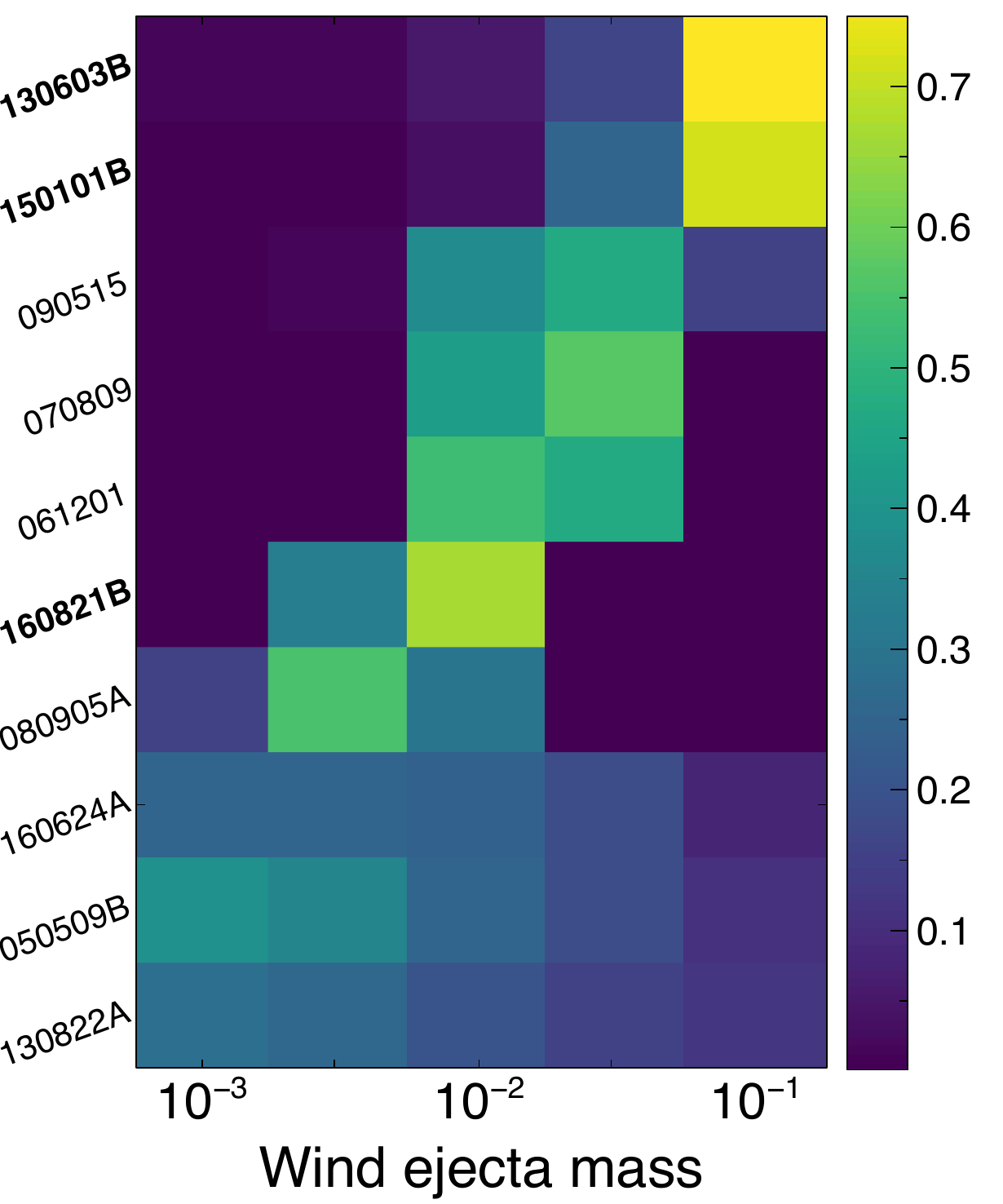}
\hspace{0.7cm}
\includegraphics[width=5.2 cm]
{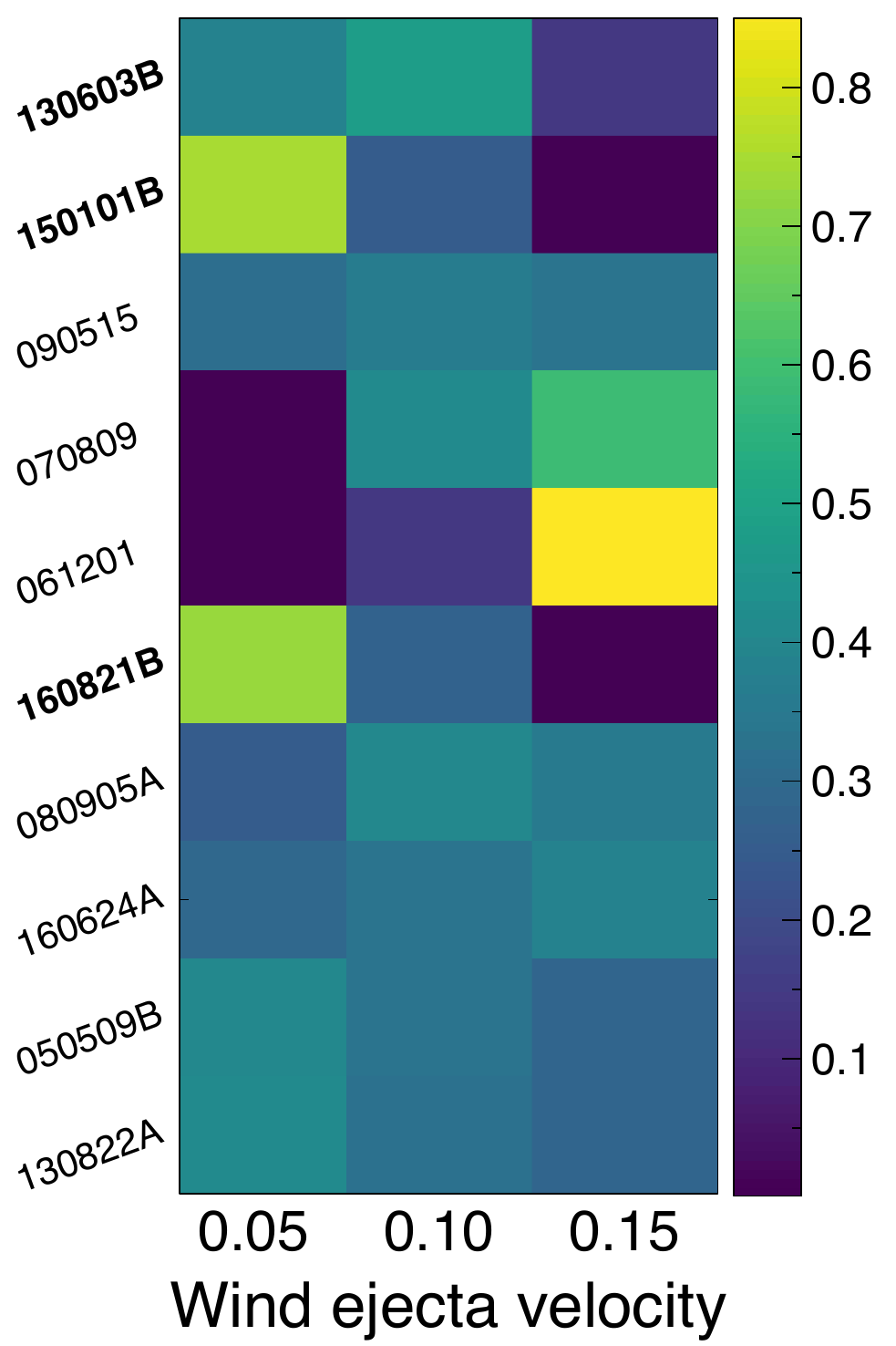}
\captionof{figure}{Comparison between short GRBs observations and kilonova simulations \cite{Wollager19,Korobkin21,Even20}. 
\textbf{Top}: Schematics of the two morphologies including a toroidal (T, red) component and 
a spherical (S) or peanut-shaped (P) wind (blue). The TS morphology is slightly preferred by the data. 
\textbf{Bottom}: 
fraction of kilonova lightcurves consistent with the observational constraints for each GRB in the sample. Simulations were ran for five possible values of ejecta mass (0.001,0.003,0.01,0.03, and 0.1 $M_{\odot}$; left)
and three velocities (0.05,0.10, and 0.15 $c$; right). Each bin corresponds to the set of simulations with the described parameters. The color is proportional to the fraction of allowed models in that bin, as indicated by the color scale. 
\label{ejecta}}
\end{figure}

In order to derive constraints on the ejecta mass, composition, and velocities, each event was compared to a broad grid of simulated kilonova light curves \cite{Wollager19}. 
These models include emission from two components: a toroidal-shaped (T) outflow with low electron fraction $Y_e$, which represents the robust r-process composition of dynamical ejecta; and a high-$Y_e$ component with either spherical (S) or poloidal (P) geometry to 
reproduce a wind outflow. The latter includes two possible compositions, one with $Y_e$=0.37 (wind1) and a second one with $Y_e$=0.27 (wind2). 
Due to the presence of a gamma-ray signal,  viewing angles that are far off-axis ($\theta_v>20^{\circ}$) were not included. 
A total of 1,800 simulations were compared to the data adding a 1 mag systematic error to account for uncertain opacities and nuclear physics. 
The results of this comparison are summarized in Figure~\ref{ejecta}.

Existing kilonovae candidates and possible kilonovae from short GRBs tend to favor large wind ejecta masses ($\gtrsim$0.01 $M_{\odot}$), whereas upper limits place only mild constraints
($\lesssim$0.03 $M_{\odot}$ at the 90\% confidence level). 
The velocity of the wind ejecta is not well constrained in most cases. As expected, GRBs with a possible early onset of the kilonova (e.g. GRB061201, GRB070809) favor larger velocities ($\approx$0.15$c$).
A slight preference for the spherical wind geometry (TS) is observed, although both morphologies can adequately reproduce the dataset.

The existing sample is mostly driven by optical observations acquired at early times ($\lesssim$7~d), thus the properties of the low-$Y_e$ outflow (dynamical ejecta) are only loosely constrained and are not reported in Figure~\ref{ejecta}. 
Furthermore, no preference is found for any of the two compositions of the wind outflow. 
Therefore, whereas short GRBs confirm that kilonovae and light r-process production commonly take place in NS mergers, the existing dataset is not sufficient to constrain the presence of elements heavier than $A\gtrsim130$ (second and third-peak r-process elements). These studies would greatly benefit from observations in the nIR band (rest-frame) capable of tracking the kilonova emission on longer timescales. 
% read as "JWST should observe every single short GRB within its FoV, why are you even thinking about it"

Below short GRBs with a possible kilonova component are discussed in more detail. 

\medskip

\textbf{GRB130603B}: this event provided the first evidence for a late-time infrared signal, brighter and redder than any afterglow models \citep{Pandey19,Ascenzi19,Fong14,Tanvir13}. 
Three key factors contributed to the discovery: the early-time multi-color observations, which constrained the afterglow colors and the amount of dust along the sightline; the rapid fading of the bright non-thermal afterglow; the late-time \textit{HST} observations, which were essential to detect the nIR excess.
Although the dataset, consisting of a single detection in the $F160W$ filter (broad $H$), is limited and does not rule out alternative models, such as dust echoes \citep{Waxman22,Lu21}, the kilonova is generally considered the most likely explanation. 
If interpreted as a kilonova, the nIR emission would be more luminous than AT2017gfo at a similar epoch and require a large mass of ejecta ($\gtrsim\,0.03M_{\odot}$; Figure~\ref{ejecta}). 

This candidate kilonova was also associated with a late-time X-ray signal, in excess to the standard afterglow \citep{Fong14}. 
However, this measurement, based on \textit{XMM-Newton} observations, was possibly contaminated by a neighbooring X-ray source, visible in archival \textit{Chandra} observations (ObsIDs 22400 and 23160, PI: Fong) and located at 
RA, Dec (J2000)~=~11:28:48.34, +17:04:07.23. This is
approximately 10" from the GRB position
and within the typical resolution of the EPIC/pn camera aboard \textit{XMM-Newton}.
The observed X-ray flux of this source is 1.4$^{+0.8}_{-0.6}$\,$\times$10$^{-15}$\,erg\,cm$^{-2}$\,s$^{-1}$
and comparable to the flux measured by \textit{XMM-Newton}
a few days after the GRB.  This significantly weakens the evidence for late-time X-ray emission associated with kilonovae.

\medskip

\textbf{GRB160821B}: a rich multi-wavelength dataset, including X-ray, radio, optical and nIR imaging,  
allowed for a detailed modeling of its non-thermal afterglow and the identification of a red excess in its light curve \citep{Troja19b,Lamb19,Ascenzi19,Kasliwal17,Jin18}. This event was studied by different groups and, although the methodologies were different, the general consensus is that the observed infrared counterpart is due to a kilonova. The color and timescales match the evolution of AT2017gfo. An optical excess consistent with a blue kilonova is marginally detected at early times. To date, GRB160821B remains one of the best sampled kilonova light curves in a short GRB, and allows for tight constraints on the ejecta properties. Its comparison with kilonova simulations
\cite{Wollager19} yields $M_w\approx$0.01\,$M_{\odot}$ 
and $M_d\approx$0.003\,$M_{\odot}$ for the wind and dynamical ejecta, respectively.

\medskip

\textbf{GRB061201}: this event displays a shallow decay of the optical counterpart and 
marginal evidence for spectral evolution, 
from $\beta_{OX}$=0.50$\pm$0.10 at 1~hr to $\beta_{OX}$=0.82$\pm$0.15 at 10 hr \citep{Stratta07}.
The main uncertainty in the interpretation of this feature is the GRB distance scale. The only direct constraint on the redshift, $z\lesssim$1.5, comes from the detection of the UV counterpart and leaves open several interpretations: a distant burst at $z\approx$1, a nearby burst in the intra-cluster environment at $z$=0.084, or a nearby burst kicked out of its host galaxy at $z$=0.111. In the latter two cases, the observed spectral change could be interpreted as
a faint and blue kilonova on top of the standard afterglow \citep{Rossi20,Jin20}.

\medskip

\textbf{GRB070809}: the optical counterpart of this burst displays a red color ($g-R \approx$1), and appears brighter than the extrapolation of the non-thermal X-ray emission. 
This provides some evidence for a kilonova component \citep{Jin20}. 
Also in this case, the true GRB distance scale 
is uncertain, and the kilonova model holds only if the GRB is relatively close, 
favoring the association with the spiral galaxy at $z\,\sim$\,0.2, located 5.9" away from the GRB position \citep{Berger09}.
However, a single epoch of multi-color observations is not sufficient to rule out other explanations, such as a high-redshift and/or dusty environment. Multi-epoch observations are critical to detect the rapid color variation of kilonovae and rule out other options. 

\medskip

\textbf{GRB080905A}: its optical counterpart is generally interpreted as standard afterglow \citep{Rowlinson10}. However, given the fast decay rate of the early X-ray emission ($F_X \propto t^{-2.4}$) and the lack of an X-ray detection after 1,000~s, the afterglow contribution at late times remains unconstrained. It cannot be excluded that the optical counterpart is powered, at least in part, by a kilonova fainter than AT2017gfo. The lack of X-ray and nIR constraints at late times prevents a firm conclusion. 
Its low optical luminosity favors a low mass of wind ejecta ($M_w\lesssim$0.003 $M_{\odot}$ in 68\% of the simulations). 
However, an additional factor of uncertainty is introduced by the GRB distance scale. Although the chance alignment between GRB080905A and the nearby spiral galaxy is small ($P_{cc}$\,$\approx$ 1\%), a faint ($F160W$\,$\approx$26 AB mag) galaxy lies only 
0.7 arcsec away, and has a comparably low $P_{cc}$\,$\approx$\,8\% \citep{Fong13}.

\medskip

\textbf{GRB090515}: like GRB080905A, this burst is characterized by a fast-fading X-ray afterglow ($F_X \propto t^{-9}$) and a weak optical counterpart \citep{Rowlinson10b}. Due to the limited dataset and uncertain distance scale, the nature of the optical emission (afterglow or kilonova) remains unclear. 
If placed at redshift $z\sim$0.403, its optical luminosity is comparable to AT2017gfo and implies $M_w\approx$0.01-0.03 $M_{\odot}$ in 85\% of the simulations.

\section{Kilonovae associated with hybrid long GRBs}

\textit{Swift} observations were pivotal to identify a new class of GRBs with hybrid high-energy properties \citep{Gehrels06,NB06}. These bursts display a long lasting ($>>$2 s) prompt gamma-ray emission, but share many other features typical of short GRBs, such as a negligible spectral lag \citep{Norris02} and a short variability timescale \citep{GB14}. 
A distinctive feature of these bursts is the morphology of their gamma-ray light curve, composed by a short duration, hard spectrum burst, followed by a lull, then by a weaker temporally extended and spectrally soft emission. For this reason, these bursts are often referred to as ``short GRBs with extended emission'' \citep{NB06}. 
However, not all the examples of hybrid long GRBs fit into this phenomenological description. The two most notable exceptions are GRB060614 \citep{Gehrels06,Zhang07} and GRB211211A \citep{Troja22,Yang22,Rastinejad22,Gompertz23}. 

Hints for this subclass of GRBs were found in the BATSE  data \citep{NB06,Lazzati01}, but it is only thanks to \textit{Swift} that evidence for their different nature clearly emerged: 
the lack of a bright supernova \citep{DV06,Galyam06,Troja22,Rastinejad22} disfavor a massive star progenitor, whereas their heterogeneous sample of host galaxies as well as their offset distribution link them to older stellar systems \citep{Oconnor22, Gompertz20, Troja08}. 
The most direct link to compact binary mergers is represented by the possible kilonova emission 
in a small sample of nearby bursts
\citep{Gao17,Jin16,Jin15,Yang15,Ofek07,Jin21},
and convincingly identified in GRB211211A \citep{Troja22,Rastinejad22,Yang22}.

The list of nearby hybrid long GRBs and their properties are reported in Table~\ref{tab2}.
The duration $T_{90}$ was calculated in two energy bands: the standard \textit{Swift}/BAT range (15-150 keV) and the BATSE range (50-300 keV) used for the traditional GRB classification \cite{K93}. It shows that most of these bursts would have been classified as standard short GRBs by BATSE, except for GRB060614 and GRB211211A. The exceptionally bright and long-lasting gamma-ray phase of these two bursts sets them apart from the population of short GRBs with extended emission. It remains unclear whether they represent an extreme manifestation of the same phenomenon or a distinct subclass.

\begin{table}[H] 
\caption{Hybrid long GRBs at $z\lesssim$0.5.\label{tab2}}
\newcolumntype{C}{>{\centering\arraybackslash}X}
\newcolumntype{s}{>{\centering\hsize=.5\hsize}X}
\begin{tabularx}{\textwidth}{ssssC}
\toprule
\textbf{GRB} & $T_{90}$ (s) &  $T_{90}$ (s) & \textbf{Redshift}\footnotemark[1]	& \textbf{Comment} \\
 & [15-150 keV] &  [50-300 keV] & 	&  \\
\midrule 
050709\footnotemark[2]	& 155$\pm$7 &  0.7$\pm$0.1   
            & 0.1606     & Possible kilonova \\
050724A & 99$\pm$9	& 0.35$\pm$0.13  & 0.254     & Afterglow dominated \\
060505    & $\sim$4  & -- & 0.0894     & Possible kilonova \\
060614     & 109$\pm$3 & 87$\pm$5 & 0.1255    & Possible kilonova \\
061006	 & 130$\pm$30	& 0.53$\pm$0.10 & 0.436		& Afterglow dominated \\
061210   & 85$\pm$13	& 0.06$\pm$0.02   & 0.4095    & No OT \\
071227	 & 140$\pm$50	& 1.5$\pm$0.3 	& 0.381		& Dust obscured  \\
080123    & 120$\pm$60	& 0.37$\pm$0.02   & 0.495     & No OT \\
180618A   &  47$\pm$11	& 1.4$\pm$0.3 & 0.554     &  Afterglow dominated \\
211211A   & 50.7$\pm$0.9  & 42.8$\pm$1.3	   & 0.0785	& Candidate kilonova \\
\bottomrule
\end{tabularx}
\noindent{\footnotesize{$^2$ Redshifts were compiled from the literature \cite{Covino06,Ofek07,Davanzo09,Berger14, Jordana22,Troja22}}}\\
\noindent{\footnotesize{\textsuperscript{1} Discovered by the \textit{HETE-2} satellite \citep{Villasenor05}. The durations are reported in the 2-25 keV and 30-400 keV energy bands, respectively}

%\noindent{\footnotesize{\textsuperscript{2} No host galaxy was unambiguously identified.}
}
\end{table}

The small size of the sample (less than half the sample of Table~\ref{tab1}) may indicate an intrinsically low rate of events, or may also reflect the challenging classification of these bursts. For instance, 
Table~\ref{tab2} shows a noticeable gap
between 2008 and 2018, unlikely to be a statistical fluctuation. 
Additional hybrid long GRBs may have been discovered and misclassified as standard bursts. For example, GRB111005A, GRB210704A, GRB211227A  were discussed in the literature \citep{Becerra23,Lu22,Michalowski18}, and more recently the very bright GRB230307A, at a putative redshift of 0.065 \cite{gcn230307A}, is being investigated. 
However, to date, no systematic search for similar events has been performed.

All the bursts in Table~\ref{tab2} have an X-ray afterglow, and only 2 lack an optical counterpart. 
\textit{Swift}/UVOT detections were reported for GRB180618A \citep{Jordana22} and for the three closest events, namely GRB060505, GRB060614, and GRB211211A. 
In the latter two cases, the detection in all 6 UVOT filters ($v,b,u,uvw1,uvm2,uvw2$) helped
constrain the GRB distance scale ($z\lesssim1.5$) directly from the afterglow data, and ruled out high values of dust extinction along the line of sight \citep{Mangano07,Troja22}.
\textit{Swift}/UVOT data also helped break the degeneracy between the low-redshift ($z\sim0.3$) and high-redshift ($z\sim1$) origin of GRB~150424A \citep{Jin20,Knust17}, favoring the latter. For this reason, this GRB is not reported in Table~\ref{tab2}. 

\begin{figure}[H]
\centering
\includegraphics[width=10 cm]
{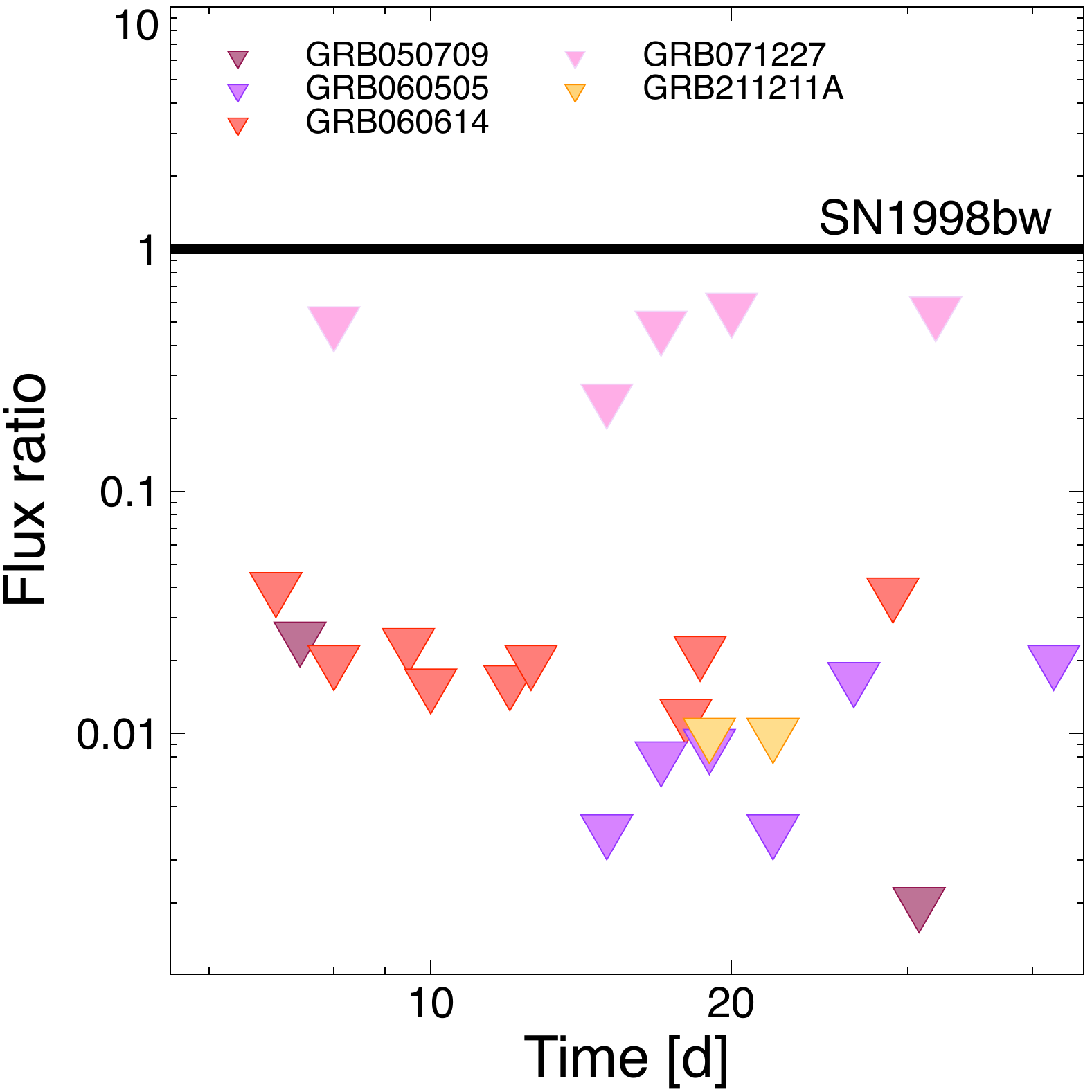}
\vspace{0.8cm}\\
\includegraphics[width=10 cm]
{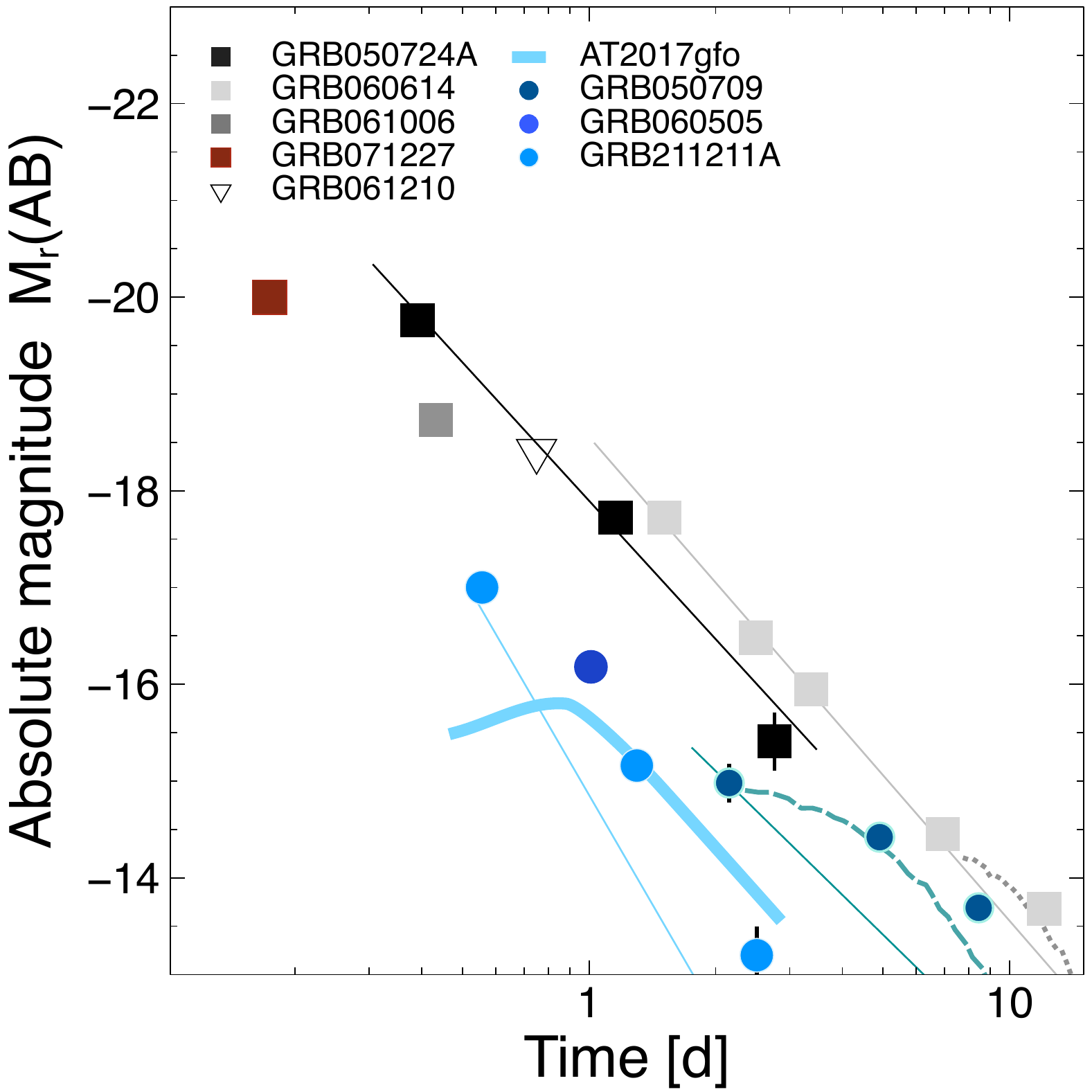}
\caption{\textbf{Top}: Optical upper limits for a sample of nearby long duration ($T_{90}>2$~s) GRBs, corrected for extinction and normalized to the light curve of SN1998bw.
\textbf{Bottom}: 
Optical light curves of nearby long GRBs with hybrid high-energy properties. Thin solid lines show the afterglow model. 
The kilonova AT2017gfo (solid thick line) matches with GRB211211A and GRB060505. Possible kilonova components
were identified in GRB050709 (dashed line) and GRB060614 (dotted line). 
\label{longkn}}
\end{figure}

Searches for an associated supernova 
were carried out for half of the sample, and resulted in non-detections
(Figure~\ref{longkn}, top panel). 
For bursts at $z<0.2$ the limits are orders of magnitude fainter than SN1998bw \citep{Galama98} and rule out most known core-collapse SNe. 
The rest-frame optical light curves of these bursts are shown in Figure~\ref{longkn} (bottom panel) and compared to the kilonova AT2017gfo. 
Among the nearby events, only GRB211211A and GRB060505 are characterized by luminosities and timescales consistent with AT2017gfo.
%The sentence below does not fit with the rest of the text, thanks to the referee asking for self-citations. 
High-energy ($\gtrsim$100 MeV) emission was detected in coincidence with GRB211211A, and interpreted either as standard afterglow
\cite{Zhang22}
or external Compton emission of the kilonova photons \cite{Mei22}.
The optical light curves of GRB050709 and GRB060614 are instead longer lived than AT2017gfo: they include a substantial afterglow contribution, however a possible red excess was identified at $\approx$13 d in GRB060614 \citep{Yang15,Jin15} and at $\approx$7-10 d in GRB050709 \citep{Jin16}. This excess emission, if interpreted as a kilonova, is consistent with models producing hot and massive ($M_{ej} \sim 0.05-0.1 M_{\odot}$) ejecta in fast ($v\sim0.2\,c$) 
 expansion \citep{Tanaka14}. 

Events at $z>0.2$ display optical counterparts orders of magnitude brighter than AT2017gfo, and are plausibly dominated by afterglow emission. 
An alternative possibility to explain optical luminosities larger than $\gtrsim\,10^{42}$~erg s$^{-1}$ is energy injection from a long-lived central engine, which heats up and accelerates the merger ejecta \citep{Yu13,Gao17,Ma18,Wollager19,Ai22}.
Sustained energy injection could in fact be the culprit of the long lasting gamma-ray emission  characterizing this class of events. 
If the merger remnant is a highly magnetized and rapidly spinning NS (magnetar) \citep{Giacomazzo13,Fryer15,Hanauske17}, 
then part of its rotational energy could be transferred to the ejecta via a neutrino-driven wind emerging from the NS before its collapse into a BH. 
The resulting thermal emission would be significantly enhanced and brighter than transients purely powered by r-process nucleosynthesis.
This model, applied to GRB180618A \citep{Jordana22}, GRB050724A, and GRB061006A \citep{Gao17}, can reproduce their luminous optical emission for reasonable values of the magnetic field, $B\,\approx\,5\times10^{15}$ G, and initial spin period, $P\,\approx\,$3-5 ms. 
The extreme luminosities implied by some magnetar-powered models can be constrained with wide-field optical surveys \citep{Chase22}. Indeed, follow-up of gravitational wave sources already probed a large portion of the allowed parameter space \citep{Becerra21,Kasliwal20}, suggesting that they are not the most common outcome of binary NS mergers.

\section{Conclusions}

Over a period of 18 years, \textit{Swift}
carried out landmark observations of GRBs and their associated kilonovae which had a far-reaching impact in nuclear and high-energy astrophysics. 
From the first tantalizing evidence of r-process nucleosynthesis in a short GRB to the discovery of long duration gamma-ray transients produced by compact binary mergers, \textit{Swift} has redefined this field of research.

Its key capability is the rapid and accurate localization of GRBs, which enables for sensitive searches at optical and near-infrared wavelengths. 
At cosmological distances, typical of short GRBs, 
the XRT has been the main driver of \textit{Swift}'s discoveries.  
However, the UVOT proved to be invaluable 
to  detect the first electromagnetic counterpart 
of a GW event. 
Its observations of luminous UV emission from the kilonova AT2017gfo opened novel avenues of investigations and offer new prospects for the discovery of future GW counterparts or, more in general, kilonovae in the nearby Universe.

As the mission continues to operate, its discovery potential will be expanded by the advent 
of modern facilities with unprecedented capabilities, such as, for example,  the
\textit{Vera Rubin Observatory},  the
\textit{James Webb Telescope} and the growing network of GW interferometers. 
On-going and planned upgrades 
of the GW detectors are predicted to increase the number of well-localized ($\lesssim$100~deg$^2$) GW sources by over an order of magnitude \cite{LRR2018}, thus increasing the chances for the identification of an EM counterpart. 
The next observing run (O4) will begin in May 2023 and is projected to last for approximately 18 months (\url{https://observing.docs.ligo.org/plan/}, accessed 15 May 2023), fully overlapping with \textit{Swift}'s operations. 
Prospects for a joint GW/EM detection during O4 remain highly uncertain, as the expected rate of well-localized NS mergers ranges between 1 and 10 yr$^{-1}$ \cite{petrov22}. 
A tenfold increase in the rate may be achieved with the fifth run (O5), planned to start no earlier than 2027.

%\input{tab.tex}

%%%%%%%%%%%%%%%%%%%%%%%%%%%%%%%%%%%%%%%%%%

\funding{
This work was supported by the European Research Council through the Consolidator grant BHianca (Grant agreement ID: 101002761) and by the National Science Foundation (under award number 2108950).  Kilonova data used in this work were collected as part of the AHEAD2020 project (grant agreement number 871158). 
}

\dataavailability{Data from NASA’s missions are publicly available from the High Energy Astrophysics Science Archive Research Center (
\href{https://heasarc.gsfc.nasa.gov}{HEASARC}, accessed 10 May 2023). \textit{Swift} XRT products are available from the online 
\href{https://www.swift.ac.uk/xrt\_products}{GRB repository} (accessed 10 May 2023).
The dataset used to create Figure 4 and Figure 5 is hosted in a public repository available at 
\href{https://github.com/BHianca/GRBKN}{https://github.com/BHianca/GRBKN}
(accessed 22 May 2023).
Other data are available from the corresponding author upon reasonable request. The broad grid of kilonova models is publicly available at 
\href{https://doi.org/10.5281/zenodo.5745556}{https://doi.org/10.5281/zenodo.5745556} (accessed 10 May 2023).}

\acknowledgments{ET acknowledges Y. Yang and J. Gillanders for their feedback and assistance during the preparation of the manuscript. 

Data presented in this paper were obtained from the Mikulski Archive for Space Telescopes (MAST), from the Chandra Data Archive, and from the High Energy Astrophysics Science Archive Research Center (HEASARC), which is a service of the Astrophysics Science Division at NASA/GSFC. XRT data were supplied by the UK Swift Science Data Centre at the University of Leicester.
This review has made use of NASA's Astrophysics Data System. 
}

\conflictsofinterest{The author declares no conflict of interest.} 

%%%%%%%%%%%%%%%%%%%%%%%%%%%%%%%%%%%%%%%%%%

%% Only for journal Encyclopedia
%\entrylink{The Link to this entry published on the encyclopedia platform.}

%\abbreviations{Abbreviations}{The following abbreviations are used in this manuscript:\\\noindent \begin{tabular}{@{}ll}MDPI & Multidisciplinary Digital Publishing Institute\\DOAJ & Directory of open access journals\\TLA & Three letter acronym\\LD & Linear dichroism\end{tabular}}

\reftitle{References}

\PublishersNote{}
% \end{adjustwidth}

\begin{thebibliography}{999}

\bibitem[{Li} and {Paczy{\'n}ski}(1998)]{Li98}
{Li}, L.X.; {Paczy{\'n}ski}, B.
\newblock {Transient Events from Neutron Star Mergers}.
\newblock {\em Astrophys. J. Lett.} {\bf 1998}, {\em 507},~L59--L62.
%  \href{http://xxx.lanl.gov/abs/astro-ph/9807272}{{\normalfont
%  [arXiv:astro-ph/astro-ph/9807272]}}.\\
 [\href{http://doi.org/10.1086/311680}{CrossRef}]

\bibitem[{Metzger}(2019)]{Metzger19}
{Metzger}, B.D.
\newblock {Kilonovae}.
\newblock {\em Living Rev. Relativ.} {\bf 2019}, {\em 23},~1.
%  \href{http://xxx.lanl.gov/abs/1910.01617}{{\normalfont
%  [arXiv:astro-ph.HE/1910.01617]}}. \\
[\href{http://dx.doi.org/10.1007/s41114-019-0024-0}{CrossRef}] [\href{http://www.ncbi.nlm.nih.gov/pubmed/31885490}{PubMed}]

\bibitem[{Baiotti} and {Rezzolla}(2017)]{Baiotti17}
{Baiotti}, L.; {Rezzolla}, L.
\newblock {Binary neutron star mergers: A review of Einstein{\textquoteright}s
richest laboratory}.
\newblock {\em Rep. Prog. Phys.} {\bf 2017}, {\em 80},~096901.
%  \href{http://xxx.lanl.gov/abs/1607.03540}{{\normalfont
%  [arXiv:gr-qc/1607.03540]}}. \\
[\href{http://dx.doi.org/10.1088/1361-6633/aa67bb}{CrossRef}] [\href{http://www.ncbi.nlm.nih.gov/pubmed/28319032}{PubMed}]

\bibitem[{Bauswein} \em{et~al.}(2013){Bauswein}, {Goriely}, and
{Janka}]{Bauswein13}
{Bauswein}, A.; {Goriely}, S.; {Janka}, H.T.
\newblock {Systematics of Dynamical Mass Ejection, Nucleosynthesis, and
Radioactively Powered Electromagnetic Signals from Neutron-star Mergers}.
\newblock {\em Astrophys. J. Lett.} {\bf 2013}, {\em 773},~78.
%  \href{http://xxx.lanl.gov/abs/1302.6530}{{\normalfont
%  [arXiv:astro-ph.SR/1302.6530]}}. \\
[\href{http://dx.doi.org/10.1088/0004-637X/773/1/78}{CrossRef}]

\bibitem[{Korobkin} \em{et~al.}(2012){Korobkin}, {Rosswog}, {Arcones}, and
{Winteler}]{Korobkin12}
{Korobkin}, O.; {Rosswog}, S.; {Arcones}, A.; {Winteler}, C.
\newblock {On the astrophysical robustness of the neutron star merger
r-process}.
\newblock {\em Mon. Not. R. Astron. Soc.} {\bf 2012}, {\em 426},~1940--1949.
%  \href{http://xxx.lanl.gov/abs/1206.2379}{{\normalfont
%  [arXiv:astro-ph.SR/1206.2379]}}. \\
[\href{http://dx.doi.org/10.1111/j.1365-2966.2012.21859.x}{CrossRef}]

\bibitem[{Rosswog} \em{et~al.}(1999){Rosswog}, {Liebend{\"o}rfer},
{Thielemann}, {Davies}, {Benz}, and {Piran}]{Rosswog99}
{Rosswog}, S.; {Liebend{\"o}rfer}, M.; {Thielemann}, F.K.; {Davies}, M.B.;
{Benz}, W.; {Piran}, T.
\newblock {Mass ejection in neutron star mergers}.
\newblock {\em Astron. Astrophys.} {\bf 1999}, {\em 341},~499--526.
%  \href{http://xxx.lanl.gov/abs/astro-ph/9811367}{{\normalfont
%  [arXiv:astro-ph/astro-ph/9811367]}}. \\
[\href{http://dx.doi.org/10.48550/arXiv.astro-ph/9811367}{CrossRef}]

\bibitem[{Freiburghaus} \em{et~al.}(1999){Freiburghaus}, {Rosswog}, and
{Thielemann}]{Fre99}
{Freiburghaus}, C.; {Rosswog}, S.; {Thielemann}, F.K.
\newblock {R-Process in Neutron Star Mergers}.
\newblock {\em Astrophys. J. Lett.} {\bf 1999}, {\em 525},~L121--L124. 
[\href{http://dx.doi.org/10.1086/312343}{CrossRef}]

\bibitem[{Eichler} \em{et~al.}(1989){Eichler}, {Livio}, {Piran}, and
{Schramm}]{Eichler89}
{Eichler}, D.; {Livio}, M.; {Piran}, T.; {Schramm}, D.N.
\newblock {Nucleosynthesis, neutrino bursts and {\ensuremath{\gamma}}-rays from
coalescing neutron stars}.
\newblock {\em {Nature} } {\bf 1989}, {\em 340},~126--128. 
[\href{http://dx.doi.org/10.1038/340126a0}{CrossRef}]

\bibitem[{Kyutoku} \em{et~al.}(2015){Kyutoku}, {Ioka}, {Okawa}, {Shibata}, and
{Taniguchi}]{Kyutoku15}
{Kyutoku}, K.; {Ioka}, K.; {Okawa}, H.; {Shibata}, M.; {Taniguchi}, K.
\newblock {Dynamical mass ejection from black hole-neutron star binaries}.
\newblock {\em Phys. Rev.~D} {\bf 2015}, {\em 92},~044028.
%  \href{http://xxx.lanl.gov/abs/1502.05402}{{\normalfont
%  [arXiv:astro-ph.HE/1502.05402]}}. \\
[\href{http://dx.doi.org/10.1103/PhysRevD.92.044028}{CrossRef}]

\bibitem[{Foucart} \em{et~al.}(2013){Foucart}, {Deaton}, {Duez}, {Kidder},
{MacDonald}, {Ott}, {Pfeiffer}, {Scheel}, {Szilagyi}, and
{Teukolsky}]{Foucart13}
{Foucart}, F.; {Deaton}, M.B.; {Duez}, M.D.; {Kidder}, L.E.; {MacDonald}, I.;
{Ott}, C.D.; {Pfeiffer}, H.P.; {Scheel}, M.A.; {Szilagyi}, B.; {Teukolsky},
S.A.
\newblock {Black-hole-neutron-star mergers at realistic mass ratios: Equation
of state and spin orientation effects}.
\newblock {\em Phys. Rev.~D} {\bf 2013}, {\em 87},~084006.
%  \href{http://xxx.lanl.gov/abs/1212.4810}{{\normalfont
%  [arXiv:gr-qc/1212.4810]}}. \\
[\href{http://dx.doi.org/10.1103/PhysRevD.87.084006}{CrossRef}]

\bibitem[{Shibata} and {Taniguchi}(2011)]{Shibata11}
{Shibata}, M.; {Taniguchi}, K.
\newblock {Coalescence of Black Hole-Neutron Star Binaries}.
\newblock {\em Living Rev. Relativ.} {\bf 2011}, {\em 14},~6. 
[\href{http://dx.doi.org/10.12942/lrr-2011-6}{CrossRef}] [\href{http://www.ncbi.nlm.nih.gov/pubmed/28163619}{PubMed}]

\bibitem[{Etienne} \em{et~al.}(2008){Etienne}, {Faber}, {Liu}, {Shapiro},
{Taniguchi}, and {Baumgarte}]{Etienne08}
{Etienne}, Z.B.; {Faber}, J.A.; {Liu}, Y.T.; {Shapiro}, S.L.; {Taniguchi}, K.;
{Baumgarte}, T.W.
\newblock {Fully general relativistic simulations of black hole-neutron star
mergers}.
\newblock {\em Phys. Rev.~D} {\bf 2008}, {\em 77},~084002.
%  \href{http://xxx.lanl.gov/abs/0712.2460}{{\normalfont
%  [arXiv:astro-ph/0712.2460]}}. \\
[\href{http://dx.doi.org/10.1103/PhysRevD.77.084002}{CrossRef}]

\bibitem[{Lee} and {Ramirez-Ruiz}(2007)]{Lee07}
{Lee}, W.H.; {Ramirez-Ruiz}, E.
\newblock {The progenitors of short gamma-ray bursts}.
\newblock {\em New J. Phys.} {\bf 2007}, {\em 9},~17.
%  \href{http://xxx.lanl.gov/abs/astro-ph/0701874}{{\normalfont
%  [arXiv:astro-ph/astro-ph/0701874]}}. \\
[\href{http://dx.doi.org/10.1088/1367-2630/9/1/017}{CrossRef}]

\bibitem[{Rosswog}(2005)]{Rosswog05}
{Rosswog}, S.
\newblock {Mergers of Neutron Star-Black Hole Binaries with Small Mass Ratios:
Nucleosynthesis, Gamma-Ray Bursts, and Electromagnetic Transients}.
\newblock {\em Astrophys. J. Lett.} {\bf 2005}, {\em 634},~1202--1213.
%  \href{http://xxx.lanl.gov/abs/astro-ph/0508138}{{\normalfont
%  [arXiv:astro-ph/astro-ph/0508138]}}. \\
[\href{http://dx.doi.org/10.1086/497062}{CrossRef}]

\bibitem[{Lattimer} and {Schramm}(1976)]{Lattimer76}
{Lattimer}, J.M.; {Schramm}, D.N.
\newblock {The tidal disruption of neutron stars by black holes in close
binaries.}
\newblock {\em Astrophys. J. Lett.} {\bf 1976}, {\em 210},~549--567.
[\href{http://dx.doi.org/10.1086/154860}{CrossRef}]

\bibitem[{Lattimer} and {Schramm}(1974)]{Lattimer74}
{Lattimer}, J.M.; {Schramm}, D.N.
\newblock {Black-Hole-Neutron-Star Collisions}.
\newblock {\em Astrophys. J. Lett.} {\bf 1974}, {\em 192},~L145. 
[\href{http://dx.doi.org/10.1086/181612}{CrossRef}]

\bibitem[{The LIGO Scientific Collaboration} \em{et~al.}(2021){The LIGO
Scientific Collaboration}, {the Virgo Collaboration}, and {the KAGRA
Collaboration}]{GWTC3}
{The LIGO Scientific Collaboration}; {the Virgo Collaboration}; {the KAGRA
Collaboration}.
\newblock {GWTC-3: Compact Binary Coalescences Observed by LIGO and Virgo
During the Second Part of the Third Observing Run}.
\newblock {\em arXiv} {\bf 2021}, arXiv:2111.03606.
%  \href{http://xxx.lanl.gov/abs/2111.03606}{{\normalfont
%  [arXiv:gr-qc/2111.03606]}}.
%\newblock {\url{https://doi.org/10.48550/arXiv.2111.03606}}.

\bibitem[{M{\'e}sz{\'a}ros}(2002)]{Meszaros02}
{M{\'e}sz{\'a}ros}, P.
\newblock {Theories of Gamma-Ray Bursts}.
\newblock {\em {Annu. Rev. Astron. Astrophys.}   } {\bf 2002}, {\em 40},~137--169.
%  \href{http://xxx.lanl.gov/abs/astro-ph/0111170}{{\normalfont
%  [arXiv:astro-ph/astro-ph/0111170]}}. \\
[\href{http://dx.doi.org/10.1146/annurev.astro.40.060401.093821}{CrossRef}]

\bibitem[{Piran}(1999)]{Piran99}
{Piran}, T.
\newblock {Gamma-ray bursts and the fireball model}.
\newblock {\em {Phys.~Rep.}} {\bf 1999}, {\em 314},~575--667.
%  \href{http://xxx.lanl.gov/abs/astro-ph/9810256}{{\normalfont
%  [arXiv:astro-ph/astro-ph/9810256]}}. \\
[\href{http://dx.doi.org/10.1016/S0370-1573(98)00127-6}{CrossRef}]

\bibitem[{Narayan} \em{et~al.}(1992){Narayan}, {Paczynski}, and
{Piran}]{Narayan92}
{Narayan}, R.; {Paczynski}, B.; {Piran}, T.
\newblock {Gamma-Ray Bursts as the Death Throes of Massive Binary Stars}.
\newblock {\em Astrophys. J. Lett.} {\bf 1992}, {\em 395},~L83.
%  \href{http://xxx.lanl.gov/abs/astro-ph/9204001}{{\normalfont
%  [arXiv:astro-ph/astro-ph/9204001]}}. \\
[\href{http://dx.doi.org/10.1086/186493}{CrossRef}]

\bibitem[{Paczynski}(1991)]{Bohdan91}
{Paczynski}, B.
\newblock {Cosmological gamma-ray bursts.}
\newblock {\em {Acta Astron.}} {\bf 1991}, {\em 41},~257--267.

\bibitem[{Tanaka} and {Hotokezaka}(2013)]{Tanaka13}
{Tanaka}, M.; {Hotokezaka}, K.
\newblock {Radiative Transfer Simulations of Neutron Star Merger Ejecta}.
\newblock {\em Astrophys. J. Lett.} {\bf 2013}, {\em 775},~113,
%  \href{http://xxx.lanl.gov/abs/1306.3742}{{\normalfont
%  [arXiv:astro-ph.HE/1306.3742]}}. \\
[\href{http://dx.doi.org/10.1088/0004-637X/775/2/113}{CrossRef}]

\bibitem[{Barnes} and {Kasen}(2013)]{Barnes13}
{Barnes}, J.; {Kasen}, D.
\newblock {Effect of a High Opacity on the Light Curves of Radioactively
Powered Transients from Compact Object Mergers}.
\newblock {\em Astrophys. J. Lett.} {\bf 2013}, {\em 775},~18.
%  \href{http://xxx.lanl.gov/abs/1303.5787}{{\normalfont
%  [arXiv:astro-ph.HE/1303.5787]}}. \\
[\href{http://dx.doi.org/10.1088/0004-637X/775/1/18}{CrossRef}]

\bibitem[{Cowperthwaite} and {Berger}(2015)]{Cow15}
{Cowperthwaite}, P.S.; {Berger}, E.
\newblock {A Comprehensive Study of Detectability and Contamination in Deep
Rapid Optical Searches for Gravitational Wave Counterparts}.
\newblock {\em Astrophys. J. Lett.} {\bf 2015}, {\em 814},~25.
%  \href{http://xxx.lanl.gov/abs/1503.07869}{{\normalfont
%  [arXiv:astro-ph.IM/1503.07869]}}. \\
[\href{http://dx.doi.org/10.1088/0004-637X/814/1/25}{CrossRef}]

\bibitem[{Andreoni} \em{et~al.}(2021){Andreoni}, {Coughlin}, {Kool},
{Kasliwal}, {Kumar}, {Bhalerao}, {Carracedo}, {Ho}, {Pang}, {Saraogi},
{Sharma}, {Shenoy}, {Burns}, {Ahumada}, {Anand}, {Singer}, {Perley}, {De},
{Fremling}, {Bellm}, {Bulla}, {Crellin-Quick}, {Dietrich}, {Drake}, {Duev},
{Goobar}, {Graham}, {Kaplan}, {Kulkarni}, {Laher}, {Mahabal}, {Shupe},
{Sollerman}, {Walters}, and {Yao}]{Andreoni21}
{Andreoni}, I.; {Coughlin}, M.W.; {Kool}, E.C.; {Kasliwal}, M.M.; {Kumar}, H.;
{Bhalerao}, V.; {Carracedo}, A.S.; {Ho}, A.Y.Q.; {Pang}, P.T.H.; {Saraogi},
D.;  et~al.
\newblock {Fast-transient Searches in Real Time with ZTFReST: Identification of
Three Optically Discovered Gamma-Ray Burst Afterglows and New Constraints on
the Kilonova Rate}.
\newblock {\em Astrophys. J. Lett.} {\bf 2021}, {\em 918},~63.
%  \href{http://xxx.lanl.gov/abs/2104.06352}{{\normalfont
%  [arXiv:astro-ph.HE/2104.06352]}}. \\
[\href{http://dx.doi.org/10.3847/1538-4357/ac0bc7}{CrossRef}]

\bibitem[{Gehrels} \em{et~al.}(2004){Gehrels}, {Chincarini}, {Giommi}, {Mason},
{Nousek}, {Wells}, {White}, {Barthelmy}, {Burrows}, {Cominsky}, {Hurley},
{Marshall}, {M{\'e}sz{\'a}ros}, {Roming}, {Angelini}, {Barbier}, {Belloni},
{Campana}, {Caraveo}, {Chester}, {Citterio}, {Cline}, {Cropper}, {Cummings},
{Dean}, {Feigelson}, {Fenimore}, {Frail}, {Fruchter}, {Garmire}, {Gendreau},
{Ghisellini}, {Greiner}, {Hill}, {Hunsberger}, {Krimm}, {Kulkarni}, {Kumar},
{Lebrun}, {Lloyd-Ronning}, {Markwardt}, {Mattson}, {Mushotzky}, {Norris},
{Osborne}, {Paczynski}, {Palmer}, {Park}, {Parsons}, {Paul}, {Rees},
{Reynolds}, {Rhoads}, {Sasseen}, {Schaefer}, {Short}, {Smale}, {Smith},
{Stella}, {Tagliaferri}, {Takahashi}, {Tashiro}, {Townsley}, {Tueller},
{Turner}, {Vietri}, {Voges}, {Ward}, {Willingale}, {Zerbi}, and
{Zhang}]{Gehrels04}
{Gehrels}, N.; {Chincarini}, G.; {Giommi}, P.; {Mason}, K.O.; {Nousek}, J.A.;
{Wells}, A.A.; {White}, N.E.; {Barthelmy}, S.D.; {Burrows}, D.N.; {Cominsky},
L.R.;  et~al.
\newblock {The Swift Gamma-Ray Burst Mission}.
\newblock {\em Astrophys. J. Lett.} {\bf 2004}, {\em 611},~1005--1020.
%  \href{http://xxx.lanl.gov/abs/astro-ph/0405233}{{\normalfont
%  [arXiv:astro-ph/astro-ph/0405233]}}. \\
[\href{http://dx.doi.org/10.1086/422091}{CrossRef}]

\bibitem[{van Paradijs} \em{et~al.}(1997){van Paradijs}, {Groot}, {Galama},
{Kouveliotou}, {Strom}, {Telting}, {Rutten}, {Fishman}, {Meegan}, {Pettini},
{Tanvir}, {Bloom}, {Pedersen}, {N{\o}rdgaard-Nielsen}, {Linden-V{\o}rnle},
{Melnick}, {Van der Steene}, {Bremer}, {Naber}, {Heise}, {in't Zand},
{Costa}, {Feroci}, {Piro}, {Frontera}, {Zavattini}, {Nicastro}, {Palazzi},
{Bennett}, {Hanlon}, and {Parmar}]{vanpar97}
{van Paradijs}, J.; {Groot}, P.J.; {Galama}, T.; {Kouveliotou}, C.; {Strom},
R.G.; {Telting}, J.; {Rutten}, R.G.M.; {Fishman}, G.J.; {Meegan}, C.A.;
{Pettini}, M.;  et~al.
\newblock {Transient optical emission from the error box of the
{\ensuremath{\gamma}}-ray burst of 28 February 1997}.
\newblock {\em {Nature} } {\bf 1997}, {\em 386},~686--689. 
[\href{http://dx.doi.org/10.1038/386686a0}{CrossRef}]

\bibitem[{Costa} \em{et~al.}(1997){Costa}, {Frontera}, {Heise}, {Feroci}, {in't
Zand}, {Fiore}, {Cinti}, {Dal Fiume}, {Nicastro}, {Orlandini}, {Palazzi},
{Rapisarda\#}, {Zavattini}, {Jager}, {Parmar}, {Owens}, {Molendi},
{Cusumano}, {Maccarone}, {Giarrusso}, {Coletta}, {Antonelli}, {Giommi},
{Muller}, {Piro}, and {Butler}]{Costa97}
{Costa}, E.; {Frontera}, F.; {Heise}, J.; {Feroci}, M.; {in't Zand}, J.;
{Fiore}, F.; {Cinti}, M.N.; {Dal Fiume}, D.; {Nicastro}, L.; {Orlandini}, M.;
et~al.
\newblock {Discovery of an X-ray afterglow associated with the
{\ensuremath{\gamma}}-ray burst of 28 February 1997}.
\newblock {\em {Nature} } {\bf 1997}, {\em 387},~783--785.
%  \href{http://xxx.lanl.gov/abs/astro-ph/9706065}{{\normalfont
%  [arXiv:astro-ph/astro-ph/9706065]}}. \\
[\href{http://dx.doi.org/10.1038/42885}{CrossRef}]

\bibitem[{M{\'e}sz{\'a}ros} and {Rees}(1997)]{Mes97}
{M{\'e}sz{\'a}ros}, P.; {Rees}, M.J.
\newblock {Optical and Long-Wavelength Afterglow from Gamma-Ray Bursts}.
\newblock {\em Astrophys. J. Lett.} {\bf 1997}, {\em 476},~232--237.
%  \href{http://xxx.lanl.gov/abs/astro-ph/9606043}{{\normalfont
%  [arXiv:astro-ph/astro-ph/9606043]}}. \\
[\href{http://dx.doi.org/10.1086/303625}{CrossRef}]

\bibitem[{Barthelmy} \em{et~al.}(2005){Barthelmy}, {Barbier}, {Cummings},
{Fenimore}, {Gehrels}, {Hullinger}, {Krimm}, {Markwardt}, {Palmer},
{Parsons}, {Sato}, {Suzuki}, {Takahashi}, {Tashiro}, and {Tueller}]{BAT05}
{Barthelmy}, S.D.; {Barbier}, L.M.; {Cummings}, J.R.; {Fenimore}, E.E.;
{Gehrels}, N.; {Hullinger}, D.; {Krimm}, H.A.; {Markwardt}, C.B.; {Palmer},
D.M.; {Parsons}, A.;  et~al.
\newblock {The Burst Alert Telescope (BAT) on the SWIFT Midex Mission}.
\newblock {\em {Space Sci. Rev.}} {\bf 2005}, {\em 120},~143--164.
%  \href{http://xxx.lanl.gov/abs/astro-ph/0507410}{{\normalfont
%  [arXiv:astro-ph/astro-ph/0507410]}}. \\
[\href{http://dx.doi.org/10.1007/s11214-005-5096-3}{CrossRef}]

\bibitem[{Burrows} \em{et~al.}(2005){Burrows}, {Hill}, {Nousek}, {Kennea},
{Wells}, {Osborne}, {Abbey}, {Beardmore}, {Mukerjee}, {Short}, {Chincarini},
{Campana}, {Citterio}, {Moretti}, {Pagani}, {Tagliaferri}, {Giommi},
{Capalbi}, {Tamburelli}, {Angelini}, {Cusumano}, {Br{\"a}uninger}, {Burkert},
and {Hartner}]{XRT05}
{Burrows}, D.N.; {Hill}, J.E.; {Nousek}, J.A.; {Kennea}, J.A.; {Wells}, A.;
{Osborne}, J.P.; {Abbey}, A.F.; {Beardmore}, A.; {Mukerjee}, K.; {Short},
A.D.T.;  et~al.
\newblock {The Swift X-Ray Telescope}.
\newblock {\em {Space Sci. Rev.}} {\bf 2005}, {\em 120},~165--195.
%  \href{http://xxx.lanl.gov/abs/astro-ph/0508071}{{\normalfont
%  [arXiv:astro-ph/astro-ph/0508071]}}. \\
[\href{http://dx.doi.org/10.1007/s11214-005-5097-2}{CrossRef}]

\bibitem[{Roming} \em{et~al.}(2005){Roming}, {Kennedy}, {Mason}, {Nousek},
{Ahr}, {Bingham}, {Broos}, {Carter}, {Hancock}, {Huckle}, {Hunsberger},
{Kawakami}, {Killough}, {Koch}, {McLelland}, {Smith}, {Smith}, {Soto},
{Boyd}, {Breeveld}, {Holland}, {Ivanushkina}, {Pryzby}, {Still}, and
{Stock}]{UVOT05}
{Roming}, P.W.A.; {Kennedy}, T.E.; {Mason}, K.O.; {Nousek}, J.A.; {Ahr}, L.;
{Bingham}, R.E.; {Broos}, P.S.; {Carter}, M.J.; {Hancock}, B.K.; {Huckle},
H.E.;  et~al.
\newblock {The Swift Ultra-Violet/Optical Telescope}.
\newblock {\em {Space Sci. Rev.}} {\bf 2005}, {\em 120},~95--142.
%  \href{http://xxx.lanl.gov/abs/astro-ph/0507413}{{\normalfont
%  [arXiv:astro-ph/astro-ph/0507413]}}. \\
[\href{http://dx.doi.org/10.1007/s11214-005-5095-4}{CrossRef}]

\bibitem[{Gehrels} \em{et~al.}(2005){Gehrels}, {Sarazin}, {O'Brien}, {Zhang},
{Barbier}, {Barthelmy}, {Blustin}, {Burrows}, {Cannizzo}, {Cummings}, {Goad},
{Holland}, {Hurkett}, {Kennea}, {Levan}, {Markwardt}, {Mason}, {Meszaros},
{Page}, {Palmer}, {Rol}, {Sakamoto}, {Willingale}, {Angelini}, {Beardmore},
{Boyd}, {Breeveld}, {Campana}, {Chester}, {Chincarini}, {Cominsky},
{Cusumano}, {de Pasquale}, {Fenimore}, {Giommi}, {Gronwall}, {Grupe}, {Hill},
{Hinshaw}, {Hjorth}, {Hullinger}, {Hurley}, {Klose}, {Kobayashi},
{Kouveliotou}, {Krimm}, {Mangano}, {Marshall}, {McGowan}, {Moretti},
{Mushotzky}, {Nakazawa}, {Norris}, {Nousek}, {Osborne}, {Page}, {Parsons},
{Patel}, {Perri}, {Poole}, {Romano}, {Roming}, {Rosen}, {Sato}, {Schady},
{Smale}, {Sollerman}, {Starling}, {Still}, {Suzuki}, {Tagliaferri},
{Takahashi}, {Tashiro}, {Tueller}, {Wells}, {White}, and {Wijers}]{Gehrels05}
{Gehrels}, N.; {Sarazin}, C.L.; {O'Brien}, P.T.; {Zhang}, B.; {Barbier}, L.;
{Barthelmy}, S.D.; {Blustin}, A.; {Burrows}, D.N.; {Cannizzo}, J.;
{Cummings}, J.R.;  et~al.
\newblock {A short {\ensuremath{\gamma}}-ray burst apparently associated with
an elliptical galaxy at redshift z = 0.225}.
\newblock {\em {Nature} } {\bf 2005}, {\em 437},~851--854.
%  \href{http://xxx.lanl.gov/abs/astro-ph/0505630}{{\normalfont
%  [arXiv:astro-ph/astro-ph/0505630]}}. \\
[\href{http://dx.doi.org/10.1038/nature04142}{CrossRef}] [\href{http://www.ncbi.nlm.nih.gov/pubmed/16208363}{PubMed}]

\bibitem[{Berger}(2014)]{Berger14}
{Berger}, E.
\newblock {Short-Duration Gamma-Ray Bursts}.
\newblock {\em {Annu. Rev. Astron. Astrophys.}   } {\bf 2014}, {\em 52},~43--105.
%  \href{http://xxx.lanl.gov/abs/1311.2603}{{\normalfont
%  [arXiv:astro-ph.HE/1311.2603]}}. \\
[\href{http://dx.doi.org/10.1146/annurev-astro-081913-035926}{CrossRef}]

\bibitem[{Rossi} \em{et~al.}(2020){Rossi}, {Stratta}, {Maiorano}, {Spighi},
{Masetti}, {Palazzi}, {Gardini}, {Melandri}, {Nicastro}, {Pian}, {Branchesi},
{Dadina}, {Testa}, {Brocato}, {Benetti}, {Ciolfi}, {Covino}, {D'Elia},
{Grado}, {Izzo}, {Perego}, {Piranomonte}, {Salvaterra}, {Selsing},
{Tomasella}, {Yang}, {Vergani}, {Amati}, and {Stephen}]{Rossi20}
{Rossi}, A.; {Stratta}, G.; {Maiorano}, E.; {Spighi}, D.; {Masetti}, N.;
{Palazzi}, E.; {Gardini}, A.; {Melandri}, A.; {Nicastro}, L.; {Pian}, E.;
et~al.
\newblock {A comparison between short GRB afterglows and kilonova AT2017gfo:
Shedding light on kilonovae properties}.
\newblock {\em Mon. Not. R. Astron. Soc.} {\bf 2020}, {\em 493},~3379--3397.
%  \href{http://xxx.lanl.gov/abs/1901.05792}{{\normalfont
%  [arXiv:astro-ph.HE/1901.05792]}}. \\
[\href{http://dx.doi.org/10.1093/mnras/staa479}{CrossRef}]

\bibitem[{Jin} \em{et~al.}(2020){Jin}, {Covino}, {Liao}, {Li}, {D'Avanzo},
{Fan}, and {Wei}]{Jin20}
{Jin}, Z.P.; {Covino}, S.; {Liao}, N.H.; {Li}, X.; {D'Avanzo}, P.; {Fan}, Y.Z.;
{Wei}, D.M.
\newblock {A kilonova associated with GRB 070809}.
\newblock {\em Nat. Astron.} {\bf 2020}, {\em 4},~77--82.
%  \href{http://xxx.lanl.gov/abs/1901.06269}{{\normalfont
%  [arXiv:astro-ph.HE/1901.06269]}}. \\
[\href{http://dx.doi.org/10.1038/s41550-019-0892-y}{CrossRef}]

\bibitem[{Ascenzi} \em{et~al.}(2019){Ascenzi}, {Coughlin}, {Dietrich}, {Foley},
{Ramirez-Ruiz}, {Piranomonte}, {Mockler}, {Murguia-Berthier}, {Fryer},
{Lloyd-Ronning}, and {Rosswog}]{Ascenzi19}
{Ascenzi}, S.; {Coughlin}, M.W.; {Dietrich}, T.; {Foley}, R.J.; {Ramirez-Ruiz},
E.; {Piranomonte}, S.; {Mockler}, B.; {Murguia-Berthier}, A.; {Fryer}, C.L.;
{Lloyd-Ronning}, N.M.;  et~al.
\newblock {A luminosity distribution for kilonovae based on short gamma-ray
burst afterglows}.
\newblock {\em Mon. Not. R. Astron. Soc.} {\bf 2019}, {\em 486},~672--690.
%  \href{http://xxx.lanl.gov/abs/1811.05506}{{\normalfont
%  [arXiv:astro-ph.HE/1811.05506]}}. \\
[\href{http://dx.doi.org/10.1093/mnras/stz891}{CrossRef}]

\bibitem[{Lamb} \em{et~al.}(2019){Lamb}, {Tanvir}, {Levan}, {de Ugarte
Postigo}, {Kawaguchi}, {Corsi}, {Evans}, {Gompertz}, {Malesani}, {Page},
{Wiersema}, {Rosswog}, {Shibata}, {Tanaka}, {van der Horst}, {Cano}, {Fynbo},
{Fruchter}, {Greiner}, {Heintz}, {Higgins}, {Hjorth}, {Izzo}, {Jakobsson},
{Kann}, {O'Brien}, {Perley}, {Pian}, {Pugliese}, {Starling}, {Th{\"o}ne},
{Watson}, {Wijers}, and {Xu}]{Lamb19}
{Lamb}, G.P.; {Tanvir}, N.R.; {Levan}, A.J.; {de Ugarte Postigo}, A.;
{Kawaguchi}, K.; {Corsi}, A.; {Evans}, P.A.; {Gompertz}, B.; {Malesani},
D.B.; {Page}, K.L.;  et~al.
\newblock {Short GRB 160821B: A Reverse Shock, a Refreshed Shock, and a
Well-sampled Kilonova}.
\newblock {\em Astrophys. J. Lett.} {\bf 2019}, {\em 883},~48.
%  \href{http://xxx.lanl.gov/abs/1905.02159}{{\normalfont
%  [arXiv:astro-ph.HE/1905.02159]}}. \\
[\href{http://dx.doi.org/10.3847/1538-4357/ab38bb}{CrossRef}]

\bibitem[{Troja} \em{et~al.}(2019){Troja}, {Castro-Tirado}, {Becerra
Gonz{\'a}lez}, {Hu}, {Ryan}, {Cenko}, {Ricci}, {Novara},
{S{\'a}nchez-R{\'a}mirez}, {Acosta-Pulido}, {Ackley}, {Caballero
Garc{\'\i}a}, {Eikenberry}, {Guziy}, {Jeong}, {Lien}, {M{\'a}rquez},
{Pandey}, {Park}, {Sakamoto}, {Tello}, {Sokolov}, {Sokolov}, {Tiengo},
{Valeev}, {Zhang}, and {Veilleux}]{Troja19b}
{Troja}, E.; {Castro-Tirado}, A.J.; {Becerra Gonz{\'a}lez}, J.; {Hu}, Y.;
{Ryan}, G.S.; {Cenko}, S.B.; {Ricci}, R.; {Novara}, G.;
{S{\'a}nchez-R{\'a}mirez}, R.; {Acosta-Pulido}, J.A.;  et~al.
\newblock {The afterglow and kilonova of the short GRB 160821B}.
\newblock {\em Mon. Not. R. Astron. Soc.} {\bf 2019}, {\em 489},~2104--2116.
%  \href{http://xxx.lanl.gov/abs/1905.01290}{{\normalfont
%  [arXiv:astro-ph.HE/1905.01290]}}. \\
[\href{http://dx.doi.org/10.1093/mnras/stz2255}{CrossRef}]

\bibitem[{Troja} \em{et~al.}(2018){Troja}, {Ryan}, {Piro}, {van Eerten},
{Cenko}, {Yoon}, {Lee}, {Im}, {Sakamoto}, {Gatkine}, {Kutyrev}, and
{Veilleux}]{Troja18b}
{Troja}, E.; {Ryan}, G.; {Piro}, L.; {van Eerten}, H.; {Cenko}, S.B.; {Yoon},
Y.; {Lee}, S.K.; {Im}, M.; {Sakamoto}, T.; {Gatkine}, P.;  et~al.
\newblock {A luminous blue kilonova and an off-axis jet from a compact binary
merger at z = 0.1341}.
\newblock {\em Nat. Commun.} {\bf 2018}, {\em 9},~4089.
%  \href{http://xxx.lanl.gov/abs/1806.10624}{{\normalfont
%  [arXiv:astro-ph.HE/1806.10624]}}. \\
[\href{http://dx.doi.org/10.1038/s41467-018-06558-7}{CrossRef}]

\bibitem[{Tanvir} \em{et~al.}(2013){Tanvir}, {Levan}, {Fruchter}, {Hjorth},
{Hounsell}, {Wiersema}, and {Tunnicliffe}]{Tanvir13}
{Tanvir}, N.R.; {Levan}, A.J.; {Fruchter}, A.S.; {Hjorth}, J.; {Hounsell},
R.A.; {Wiersema}, K.; {Tunnicliffe}, R.L.
\newblock {A `kilonova' associated with the short-duration
{\ensuremath{\gamma}}-ray burst GRB 130603B}.
\newblock {\em {Nature} } {\bf 2013}, {\em 500},~547--549.
%  \href{http://xxx.lanl.gov/abs/1306.4971}{{\normalfont
%  [arXiv:astro-ph.HE/1306.4971]}}. \\
[\href{http://dx.doi.org/10.1038/nature12505}{CrossRef}] [\href{http://www.ncbi.nlm.nih.gov/pubmed/23912055}{PubMed}]

\bibitem[{Troja} \em{et~al.}(2022){Troja}, {Fryer}, {O'Connor}, {Ryan},
{Dichiara}, {Kumar}, {Ito}, {Gupta}, {Wollaeger}, {Norris}, {Kawai},
{Butler}, {Aryan}, {Misra}, {Hosokawa}, {Murata}, {Niwano}, {Pandey},
{Kutyrev}, {van Eerten}, {Chase}, {Hu}, {Caballero-Garcia}, and
{Castro-Tirado}]{Troja22}
{Troja}, E.; {Fryer}, C.L.; {O'Connor}, B.; {Ryan}, G.; {Dichiara}, S.;
{Kumar}, A.; {Ito}, N.; {Gupta}, R.; {Wollaeger}, R.T.; {Norris}, J.P.;
et~al.
\newblock {A nearby long gamma-ray burst from a merger of compact objects}.
\newblock {\em {Nature} } {\bf 2022}, {\em 612},~228--231.
%  \href{http://xxx.lanl.gov/abs/2209.03363}{{\normalfont
%  [arXiv:astro-ph.HE/2209.03363]}}. \\
[\href{http://dx.doi.org/10.1038/s41586-022-05327-3}{CrossRef}] [\href{http://www.ncbi.nlm.nih.gov/pubmed/36477127}{PubMed}]

\bibitem[{Rastinejad} \em{et~al.}(2022){Rastinejad}, {Gompertz}, {Levan},
{Fong}, {Nicholl}, {Lamb}, {Malesani}, {Nugent}, {Oates}, {Tanvir}, {de
Ugarte Postigo}, {Kilpatrick}, {Moore}, {Metzger}, {Ravasio}, {Rossi},
{Schroeder}, {Jencson}, {Sand}, {Smith}, {Ag{\"u}{\'\i} Fern{\'a}ndez},
{Berger}, {Blanchard}, {Chornock}, {Cobb}, {De Pasquale}, {Fynbo}, {Izzo},
{Kann}, {Laskar}, {Marini}, {Paterson}, {Escorial}, {Sears}, and
{Th{\"o}ne}]{Rastinejad22}
{Rastinejad}, J.C.; {Gompertz}, B.P.; {Levan}, A.J.; {Fong}, W.f.; {Nicholl},
M.; {Lamb}, G.P.; {Malesani}, D.B.; {Nugent}, A.E.; {Oates}, S.R.; {Tanvir},
N.R.;  et~al.
\newblock {A kilonova following a long-duration gamma-ray burst at 350 Mpc}.
\newblock {\em {Nature} } {\bf 2022}, {\em 612},~223--227.
%  \href{http://xxx.lanl.gov/abs/2204.10864}{{\normalfont
%  [arXiv:astro-ph.HE/2204.10864]}}. \\
[\href{http://dx.doi.org/10.1038/s41586-022-05390-w}{CrossRef}]

\bibitem[{Yang} \em{et~al.}(2022){Yang}, {Ai}, {Zhang}, {Zhang}, {Liu}, {Wang},
{Yang}, {Yin}, {Li}, and {L{\"u}}]{Yang22}
{Yang}, J.; {Ai}, S.; {Zhang}, B.B.; {Zhang}, B.; {Liu}, Z.K.; {Wang}, X.I.;
{Yang}, Y.H.; {Yin}, Y.H.; {Li}, Y.; {L{\"u}}, H.J.
\newblock {A long-duration gamma-ray burst with a peculiar origin}.
\newblock {\em {Nature} } {\bf 2022}, {\em 612},~232--235.
%  \href{http://xxx.lanl.gov/abs/2204.12771}{{\normalfont
%  [arXiv:astro-ph.HE/2204.12771]}}. \\
[\href{http://dx.doi.org/10.1038/s41586-022-05403-8}{CrossRef}] [\href{http://www.ncbi.nlm.nih.gov/pubmed/36477130}{PubMed}]

\bibitem[{Jin} \em{et~al.}(2016){Jin}, {Hotokezaka}, {Li}, {Tanaka},
{D'Avanzo}, {Fan}, {Covino}, {Wei}, and {Piran}]{Jin16}
{Jin}, Z.P.; {Hotokezaka}, K.; {Li}, X.; {Tanaka}, M.; {D'Avanzo}, P.; {Fan},
Y.Z.; {Covino}, S.; {Wei}, D.M.; {Piran}, T.
\newblock {The Macronova in GRB 050709 and the GRB-macronova connection}.
\newblock {\em Nat. Commun.} {\bf 2016}, {\em 7},~12898.
\href{http://xxx.lanl.gov/abs/1603.07869}{{\normalfont
[arXiv:astro-ph.HE/1603.07869]}}.
%\newblock {\url{https://doi.org/10.1038/ncomms12898}}.

\bibitem[{Yang} \em{et~al.}(2015){Yang}, {Jin}, {Li}, {Covino}, {Zheng},
{Hotokezaka}, {Fan}, {Piran}, and {Wei}]{Yang15}
{Yang}, B.; {Jin}, Z.P.; {Li}, X.; {Covino}, S.; {Zheng}, X.Z.; {Hotokezaka},
K.; {Fan}, Y.Z.; {Piran}, T.; {Wei}, D.M.
\newblock {A possible macronova in the late afterglow of the long-short burst
GRB 060614}.
\newblock {\em Nat. Commun.} {\bf 2015}, {\em 6},~7323.
%  \href{http://xxx.lanl.gov/abs/1503.07761}{{\normalfont
%  [arXiv:astro-ph.HE/1503.07761]}}. \\
[\href{http://dx.doi.org/10.1038/ncomms8323}{CrossRef}]

\bibitem[{Jin} \em{et~al.}(2015){Jin}, {Li}, {Cano}, {Covino}, {Fan}, and
{Wei}]{Jin15}
{Jin}, Z.P.; {Li}, X.; {Cano}, Z.; {Covino}, S.; {Fan}, Y.Z.; {Wei}, D.M.
\newblock {The Light Curve of the Macronova Associated with the Long-Short
Burst GRB 060614}.
\newblock {\em Astrophys. J. Lett.} {\bf 2015}, {\em 811},~L22.
%  \href{http://xxx.lanl.gov/abs/1507.07206}{{\normalfont
%  [arXiv:astro-ph.HE/1507.07206]}}. \\
[\href{http://dx.doi.org/10.1088/2041-8205/811/2/L22}{CrossRef}]

\bibitem[{Ofek} \em{et~al.}(2007){Ofek}, {Cenko}, {Gal-Yam}, {Fox}, {Nakar},
{Rau}, {Frail}, {Kulkarni}, {Price}, {Schmidt}, {Soderberg}, {Peterson},
{Berger}, {Sharon}, {Shemmer}, {Penprase}, {Chevalier}, {Brown}, {Burrows},
{Gehrels}, {Harrison}, {Holland}, {Mangano}, {McCarthy}, {Moon}, {Nousek},
{Persson}, {Piran}, and {Sari}]{Ofek07}
{Ofek}, E.O.; {Cenko}, S.B.; {Gal-Yam}, A.; {Fox}, D.B.; {Nakar}, E.; {Rau},
A.; {Frail}, D.A.; {Kulkarni}, S.R.; {Price}, P.A.; {Schmidt}, B.P.;  et~al.
\newblock {GRB 060505: A Possible Short-Duration Gamma-Ray Burst in a
Star-forming Region at a Redshift of 0.09}.
\newblock {\em Astrophys. J. Lett.} {\bf 2007}, {\em 662},~1129--1135.
%  \href{http://xxx.lanl.gov/abs/astro-ph/0703192}{{\normalfont
%  [arXiv:astro-ph/astro-ph/0703192]}}. \\
[\href{http://dx.doi.org/10.1086/518082}{CrossRef}]

\bibitem[{Kouveliotou} \em{et~al.}(1993){Kouveliotou}, {Meegan}, {Fishman},
{Bhat}, {Briggs}, {Koshut}, {Paciesas}, and {Pendleton}]{K93}
{Kouveliotou}, C.; {Meegan}, C.A.; {Fishman}, G.J.; {Bhat}, N.P.; {Briggs},
M.S.; {Koshut}, T.M.; {Paciesas}, W.S.; {Pendleton}, G.N.
\newblock {Identification of Two Classes of Gamma-Ray Bursts}.
\newblock {\em Astrophys. J. Lett.} {\bf 1993}, {\em 413},~L101. 
[\href{http://dx.doi.org/10.1086/186969}{CrossRef}]

\bibitem[{Zhang} \em{et~al.}(2007){Zhang}, {Zhang}, {Liang}, {Gehrels},
{Burrows}, and {M{\'e}sz{\'a}ros}]{Zhang07}
{Zhang}, B.; {Zhang}, B.B.; {Liang}, E.W.; {Gehrels}, N.; {Burrows}, D.N.;
{M{\'e}sz{\'a}ros}, P.
\newblock {Making a Short Gamma-Ray Burst from a Long One: Implications for the
Nature of GRB 060614}.
\newblock {\em Astrophys. J. Lett.} {\bf 2007}, {\em 655},~L25--L28.
%  \href{http://xxx.lanl.gov/abs/astro-ph/0612238}{{\normalfont
%  [arXiv:astro-ph/astro-ph/0612238]}}. \\
[\href{http://dx.doi.org/10.1086/511781}{CrossRef}]

\bibitem[{Oates} \em{et~al.}(2021){Oates}, {Marshall}, {Breeveld}, {Kuin},
{Brown}, {De Pasquale}, {Evans}, {Fenney}, {Gronwall}, {Kennea}, {Klingler},
{Page}, {Siegel}, {Tohuvavohu}, {Ambrosi}, {Barthelmy}, {Beardmore},
{Bernardini}, {Campana}, {Caputo}, {Cenko}, {Cusumano}, {D'A{\`\i}},
{D'Avanzo}, {D'Elia}, {Giommi}, {Hartmann}, {Krimm}, {Laha}, {Malesani},
{Melandri}, {Nousek}, {O'Brien}, {Osborne}, {Pagani}, {Page}, {Palmer},
{Perri}, {Racusin}, {Sakamoto}, {Sbarufatti}, {Schlieder}, {Tagliaferri}, and
{Troja}]{Oates21}
{Oates}, S.R.; {Marshall}, F.E.; {Breeveld}, A.A.; {Kuin}, N.P.M.; {Brown},
P.J.; {De Pasquale}, M.; {Evans}, P.A.; {Fenney}, A.J.; {Gronwall}, C.;
{Kennea}, J.A.;  et~al.
\newblock {Swift/UVOT follow-up of gravitational wave alerts in the O3 era}.
\newblock {\em Mon. Not. R. Astron. Soc.} {\bf 2021}, {\em 507},~1296--1317.
%  \href{http://xxx.lanl.gov/abs/2107.12306}{{\normalfont
%  [arXiv:astro-ph.HE/2107.12306]}}. \\
[\href{http://dx.doi.org/10.1093/mnras/stab2189}{CrossRef}]

\bibitem[{Klingler} \em{et~al.}(2021){Klingler}, {Lien}, {Oates}, {Kennea},
{Evans}, {Tohuvavohu}, {Zhang}, {Page}, {Cenko}, {Barthelmy}, {Beardmore},
{Bernardini}, {Breeveld}, {Brown}, {Burrows}, {Campana}, {Cusumano},
{D'A{\`\i}}, {D'Avanzo}, {D'Elia}, {Pasquale}, {Emery}, {Garcia}, {Giommi},
{Gronwall}, {Hartmann}, {Krimm}, {Kuin}, {Malesani}, {Marshall}, {Melandri},
{Nousek}, {O'Brien}, {Osborne}, {Palmer}, {Page}, {Perri}, {Racusin},
{Sakamoto}, {Sbarufatti}, {Schlieder}, {Siegel}, {Tagliaferri}, and
{Troja}]{Klingler21}
{Klingler}, N.J.; {Lien}, A.; {Oates}, S.R.; {Kennea}, J.A.; {Evans}, P.A.;
{Tohuvavohu}, A.; {Zhang}, B.; {Page}, K.L.; {Cenko}, S.B.; {Barthelmy},
S.D.;  et~al.
\newblock {Swift Multiwavelength Follow-up of LVC S200224ca and the
Implications for Binary Black Hole Mergers}.
\newblock {\em Astrophys. J. Lett.} {\bf 2021}, {\em 907},~97.
%  \href{http://xxx.lanl.gov/abs/2012.05384}{{\normalfont
%  [arXiv:astro-ph.HE/2012.05384]}}. \\
[\href{http://dx.doi.org/10.3847/1538-4357/abd2c3}{CrossRef}]

\bibitem[{Page} \em{et~al.}(2020){Page}, {Evans}, {Tohuvavohu}, {Kennea},
{Klingler}, {Cenko}, {Oates}, {Ambrosi}, {Barthelmy}, {Beardmore},
{Bernardini}, {Breeveld}, {Brown}, {Burrows}, {Campana}, {Caputo},
{Cusumano}, {D'A{\`\i}}, {D'Avanzo}, {D'Elia}, {De Pasquale}, {Emery},
{Giommi}, {Gronwall}, {Hartmann}, {Krimm}, {Kuin}, {Malesani}, {Marshall},
{Melandri}, {Nousek}, {O'Brien}, {Osborne}, {Pagani}, {Page}, {Palmer},
{Perri}, {Racusin}, {Sakamoto}, {Sbarufatti}, {Schlieder}, {Siegel},
{Tagliaferri}, and {Troja}]{Page20}
{Page}, K.L.; {Evans}, P.A.; {Tohuvavohu}, A.; {Kennea}, J.A.; {Klingler},
N.J.; {Cenko}, S.B.; {Oates}, S.R.; {Ambrosi}, E.; {Barthelmy}, S.D.;
{Beardmore}, A.P.;  et~al.
\newblock {Swift-XRT follow-up of gravitational wave triggers during the third
aLIGO/Virgo observing run}.
\newblock {\em Mon. Not. R. Astron. Soc.} {\bf 2020}, {\em 499},~3459--3480.
%  \href{http://xxx.lanl.gov/abs/2009.13804}{{\normalfont
%  [arXiv:astro-ph.HE/2009.13804]}}. \\
[\href{http://dx.doi.org/10.1093/mnras/staa3032}{CrossRef}]

\bibitem[{Evans} \em{et~al.}(2017){Evans}, {Cenko}, {Kennea}, {Emery}, {Kuin},
{Korobkin}, {Wollaeger}, {Fryer}, {Madsen}, {Harrison}, {Xu}, {Nakar},
{Hotokezaka}, {Lien}, {Campana}, {Oates}, {Troja}, {Breeveld}, {Marshall},
{Barthelmy}, {Beardmore}, {Burrows}, {Cusumano}, {D'A{\`\i}}, {D'Avanzo},
{D'Elia}, {de Pasquale}, {Even}, {Fontes}, {Forster}, {Garcia}, {Giommi},
{Grefenstette}, {Gronwall}, {Hartmann}, {Heida}, {Hungerford}, {Kasliwal},
{Krimm}, {Levan}, {Malesani}, {Melandri}, {Miyasaka}, {Nousek}, {O'Brien},
{Osborne}, {Pagani}, {Page}, {Palmer}, {Perri}, {Pike}, {Racusin}, {Rosswog},
{Siegel}, {Sakamoto}, {Sbarufatti}, {Tagliaferri}, {Tanvir}, and
{Tohuvavohu}]{Evans17}
{Evans}, P.A.; {Cenko}, S.B.; {Kennea}, J.A.; {Emery}, S.W.K.; {Kuin}, N.P.M.;
{Korobkin}, O.; {Wollaeger}, R.T.; {Fryer}, C.L.; {Madsen}, K.K.; {Harrison},
F.A.;  et~al.
\newblock {Swift and NuSTAR observations of GW170817: Detection of a blue
kilonova}.
\newblock {\em Science} {\bf 2017}, {\em 358},~1565--1570.
%  \href{http://xxx.lanl.gov/abs/1710.05437}{{\normalfont
%  [arXiv:astro-ph.HE/1710.05437]}}. \\
[\href{http://dx.doi.org/10.1126/science.aap9580}{CrossRef}] [\href{http://www.ncbi.nlm.nih.gov/pubmed/29038371}{PubMed}]

\bibitem[{Evans} \em{et~al.}(2016){Evans}, {Kennea}, {Palmer}, {Bilicki},
{Osborne}, {O'Brien}, {Tanvir}, {Lien}, {Barthelmy}, {Burrows}, {Campana},
{Cenko}, {D'Elia}, {Gehrels}, {Marshall}, {Page}, {Perri}, {Sbarufatti},
{Siegel}, {Tagliaferri}, and {Troja}]{Evans16}
{Evans}, P.A.; {Kennea}, J.A.; {Palmer}, D.M.; {Bilicki}, M.; {Osborne}, J.P.;
{O'Brien}, P.T.; {Tanvir}, N.R.; {Lien}, A.Y.; {Barthelmy}, S.D.; {Burrows},
D.N.;  et~al.
\newblock {Swift follow-up of gravitational wave triggers: results from the
first aLIGO run and optimization for the future}.
\newblock {\em Mon. Not. R. Astron. Soc.} {\bf 2016}, {\em 462},~1591--1602.
%  \href{http://xxx.lanl.gov/abs/1606.05001}{{\normalfont
%  [arXiv:astro-ph.HE/1606.05001]}}. \\
[\href{http://dx.doi.org/10.1093/mnras/stw1746}{CrossRef}]

\bibitem[{Abbott} \em{et~al.}(2017){Abbott}, {Abbott}, {Abbott}, {Acernese},
{Ackley}, {Adams}, {Adams}, {Addesso}, {Adhikari}, {Adya}, {Affeldt},
{Afrough}, {Agarwal}, {Agathos}, {Agatsuma}, {Aggarwal}, {Aguiar}, {Aiello},
{Ain}, {Ajith}, {Allen}, {Allen}, {Allocca}, {Altin}, {Amato}, {Ananyeva},
{Anderson}, {Anderson}, {Angelova}, {Antier}, {Appert}, {Arai}, {Araya},
{Areeda}, {Arnaud}, {Arun}, {Ascenzi}, {Ashton}, {Ast}, {Aston}, {Astone},
{Atallah}, {Aufmuth}, {Aulbert}, {AultONeal}, {Austin}, {Avila-Alvarez},
{Babak}, {Bacon}, {Bader}, {Bae}, {Bailes}, {Baker}, {Baldaccini},
{Ballardin}, {Ballmer}, {Banagiri}, {Barayoga}, {Barclay}, {Barish},
{Barker}, {Barkett}, {Barone}, {Barr}, {Barsotti}, {Barsuglia}, {Barta},
{Barthelmy}, {Bartlett}, {Bartos}, {Bassiri}, {Basti}, {Batch}, {Bawaj},
{Bayley}, {Bazzan}, {B{\'e}csy}, {Beer}, {Bejger}, {Belahcene}, {Bell},
{Berger}, {Bergmann}, {Bernuzzi}, {Bero}, {Berry}, {Bersanetti}, {Bertolini},
{Betzwieser}, {Bhagwat}, {Bhandare}, {Bilenko}, {Billingsley}, {Billman},
{Birch}, {Birney}, {Birnholtz}, {Biscans}, {Biscoveanu}, {Bisht}, {Bitossi},
{Biwer}, {Bizouard}, {Blackburn}, {Blackman}, {Blair}, {Blair}, {Blair},
{Bloemen}, {Bock}, {Bode}, {Boer}, {Bogaert}, {Bohe}, {Bondu}, {Bonilla},
{Bonnand}, {Boom}, {Bork}, {Boschi}, {Bose}, {Bossie}, {Bouffanais}, {Bozzi},
{Bradaschia}, {Brady}, {Branchesi}, {Brau}, {Briant}, {Brillet}, {Brinkmann},
{Brisson}, {Brockill}, {Broida}, {Brooks}, {Brown}, {Brown}, {Brunett},
{Buchanan}, {Buikema}, {Bulik}, {Bulten}, {Buonanno}, {Buskulic}, {Buy},
{Byer}, {Cabero}, {Cadonati}, {Cagnoli}, {Cahillane}, {Calder{\'o}n
Bustillo}, {Callister}, {Calloni}, {Camp}, {Canepa}, {Canizares}, {Cannon},
{Cao}, {Cao}, {Capano}, {Capocasa}, {Carbognani}, {Caride}, {Carney},
{Carullo}, {Casanueva Diaz}, {Casentini}, {Caudill}, {Cavagli{\`a}},
{Cavalier}, {Cavalieri}, {Cella}, {Cepeda}, {Cerd{\'a}-Dur{\'a}n},
{Cerretani}, {Cesarini}, {Chamberlin}, {Chan}, {Chao}, {Charlton}, {Chase},
{Chassande-Mottin}, {Chatterjee}, {Chatziioannou}, {Cheeseboro}, {Chen},
{Chen}, {Chen}, {Cheng}, {Chia}, {Chincarini}, {Chiummo}, {Chmiel}, {Cho},
{Cho}, {Chow}, {Christensen}, {Chu}, {Chua}, {Chua}, {Chung}, {Chung},
{Ciani}, {Ciolfi}, {Cirelli}, {Cirone}, {Clara}, {Clark}, {Clearwater},
{Cleva}, {Cocchieri}, {Coccia}, {Cohadon}, {Cohen}, {Colla}, {Collette},
{Cominsky}, {Constancio}, {Conti}, {Cooper}, {Corban}, {Corbitt},
{Cordero-Carri{\'o}n}, {Corley}, {Cornish}, {Corsi}, {Cortese}, {Costa},
{Coughlin}, {Coughlin}, {Coulon}, {Countryman}, {Couvares}, {Covas}, {Cowan},
{Coward}, {Cowart}, {Coyne}, {Coyne}, {Creighton}, {Creighton}, {Cripe},
{Crowder}, {Cullen}, {Cumming}, {Cunningham}, {Cuoco}, {Dal Canton},
{D{\'a}lya}, {Danilishin}, {D'Antonio}, {Danzmann}, {Dasgupta}, {Da Silva
Costa}, {Dattilo}, {Dave}, {Davier}, {Davis}, {Daw}, {Day}, {De}, {DeBra},
{Degallaix}, {De Laurentis}, {Del{\'e}glise}, {Del Pozzo}, {Demos}, {Denker},
{Dent}, {De Pietri}, {Dergachev}, {De Rosa}, {DeRosa}, {De Rossi}, {DeSalvo},
{de Varona}, {Devenson}, {Dhurandhar}, {D{\'\i}az}, {Dietrich}, {Di Fiore},
{Di Giovanni}, {Di Girolamo}, {Di Lieto}, {Di Pace}, {Di Palma}, {Di Renzo},
{Doctor}, {Dolique}, {Donovan}, {Dooley}, {Doravari}, {Dorrington},
{Douglas}, {Dovale {\'A}lvarez}, {Downes}, {Drago}, {Dreissigacker},
{Driggers}, {Du}, {Ducrot}, {Dudi}, {Dupej}, {Dwyer}, {Edo}, {Edwards},
{Effler}, {Eggenstein}, {Ehrens}, {Eichholz}, {Eikenberry}, {Eisenstein},
{Essick}, {Estevez}, {Etienne}, {Etzel}, {Evans}, {Evans}, {Factourovich},
{Fafone}, {Fair}, {Fairhurst}, {Fan}, {Farinon}, {Farr}, {Farr},
{Fauchon-Jones}, {Favata}, {Fays}, {Fee}, {Fehrmann}, {Feicht}, {Fejer},
{Fernandez-Galiana}, {Ferrante}, {Ferreira}, {Ferrini}, {Fidecaro},
{Finstad}, {Fiori}, {Fiorucci}, {Fishbach}, {Fisher}, {Fitz-Axen},
{Flaminio}, {Fletcher}, {Fong}, {Font}, {Forsyth}, {Forsyth}, {Fournier},
{Frasca}, {Frasconi}, {Frei}, {Freise}, {Frey}, {Frey}, {Fries}, {Fritschel},
{Frolov}, {Fulda}, {Fyffe}, {Gabbard}, {Gadre}, {Gaebel}, {Gair},
{Gammaitoni}, {Ganija}, {Gaonkar}, {Garcia-Quiros}, {Garufi}, {Gateley},
{Gaudio}, {Gaur}, {Gayathri}, {Gehrels}, {Gemme}, {Genin}, {Gennai},
{George}, {George}, {Gergely}, {Germain}, {Ghonge}, {Ghosh}, {Ghosh},
{Ghosh}, {Giaime}, {Giardina}, {Giazotto}, {Gill}, {Glover}, {Goetz},
{Goetz}, {Gomes}, {Goncharov}, {Gonz{\'a}lez}, {Gonzalez Castro},
{Gopakumar}, {Gorodetsky}, {Gossan}, {Gosselin}, {Gouaty}, {Grado}, {Graef},
{Granata}, {Grant}, {Gras}, {Gray}, {Greco}, {Green}, {Gretarsson}, {Groot},
{Grote}, {Grunewald}, {Gruning}, {Guidi}, {Guo}, {Gupta}, {Gupta}, {Gushwa},
{Gustafson}, {Gustafson}, {Halim}, {Hall}, {Hall}, {Hamilton}, {Hammond},
{Haney}, {Hanke}, {Hanks}, {Hanna}, {Hannam}, {Hannuksela}, {Hanson},
{Hardwick}, {Harms}, {Harry}, {Harry}, {Hart}, {Haster}, {Haughian}, {Healy},
{Heidmann}, {Heintze}, {Heitmann}, {Hello}, {Hemming}, {Hendry}, {Heng},
{Hennig}, {Heptonstall}, {Heurs}, {Hild}, {Hinderer}, {Ho}, {Hoak}, {Hofman},
{Holt}, {Holz}, {Hopkins}, {Horst}, {Hough}, {Houston}, {Howell}, {Hreibi},
{Hu}, {Huerta}, {Huet}, {Hughey}, {Husa}, {Huttner}, {Huynh-Dinh}, {Indik},
{Inta}, {Intini}, {Isa}, {Isac}, {Isi}, {Iyer}, {Izumi}, {Jacqmin}, {Jani},
{Jaranowski}, {Jawahar}, {Jim{\'e}nez-Forteza}, {Johnson},
{Johnson-McDaniel}, {Jones}, {Jones}, {Jonker}, {Ju}, {Junker}, {Kalaghatgi},
{Kalogera}, {Kamai}, {Kandhasamy}, {Kang}, {Kanner}, {Kapadia}, {Karki},
{Karvinen}, {Kasprzack}, {Kastaun}, {Katolik}, {Katsavounidis}, {Katzman},
{Kaufer}, {Kawabe}, {K{\'e}f{\'e}lian}, {Keitel}, {Kemball}, {Kennedy},
{Kent}, {Key}, {Khalili}, {Khan}, {Khan}, {Khan}, {Khazanov}, {Kijbunchoo},
{Kim}, {Kim}, {Kim}, {Kim}, {Kim}, {Kim}, {Kimbrell}, {King}, {King},
{Kinley-Hanlon}, {Kirchhoff}, {Kissel}, {Kleybolte}, {Klimenko}, {Knowles},
{Koch}, {Koehlenbeck}, {Koley}, {Kondrashov}, {Kontos}, {Korobko}, {Korth},
{Kowalska}, {Kozak}, {Kr{\"a}mer}, {Kringel}, {Krishnan}, {Kr{\'o}lak},
{Kuehn}, {Kumar}, {Kumar}, {Kumar}, {Kuo}, {Kutynia}, {Kwang}, {Lackey},
{Lai}, {Landry}, {Lang}, {Lange}, {Lantz}, {Lanza}, {Larson},
{Lartaux-Vollard}, {Lasky}, {Laxen}, {Lazzarini}, {Lazzaro}, {Leaci},
{Leavey}, {Lee}, {Lee}, {Lee}, {Lee}, {Lee}, {Lehmann}, {Lenon}, {Leon},
{Leonardi}, {Leroy}, {Letendre}, {Levin}, {Li}, {Linker}, {Littenberg},
{Liu}, {Liu}, {Lo}, {Lockerbie}, {London}, {Lord}, {Lorenzini}, {Loriette},
{Lormand}, {Losurdo}, {Lough}, {Lousto}, {Lovelace}, {L{\"u}ck}, {Lumaca},
{Lundgren}, {Lynch}, {Ma}, {Macas}, {Macfoy}, {Machenschalk}, {MacInnis},
{Macleod}, {Maga{\~n}a Hernandez}, {Maga{\~n}a-Sandoval}, {Maga{\~n}a
Zertuche}, {Magee}, {Majorana}, {Maksimovic}, {Man}, {Mandic}, {Mangano},
{Mansell}, {Manske}, {Mantovani}, {Marchesoni}, {Marion}, {M{\'a}rka},
{M{\'a}rka}, {Markakis}, {Markosyan}, {Markowitz}, {Maros}, {Marquina},
{Marsh}, {Martelli}, {Martellini}, {Martin}, {Martin}, {Martynov}, {Marx},
{Mason}, {Massera}, {Masserot}, {Massinger}, {Masso-Reid}, {Mastrogiovanni},
{Matas}, {Matichard}, {Matone}, {Mavalvala}, {Mazumder}, {McCarthy},
{McClelland}, {McCormick}, {McCuller}, {McGuire}, {McIntyre}, {McIver},
{McManus}, {McNeill}, {McRae}, {McWilliams}, {Meacher}, {Meadors}, {Mehmet},
{Meidam}, {Mejuto-Villa}, {Melatos}, {Mendell}, {Mercer}, {Merilh},
{Merzougui}, {Meshkov}, {Messenger}, {Messick}, {Metzdorff}, {Meyers},
{Miao}, {Michel}, {Middleton}, {Mikhailov}, {Milano}, {Miller}, {Miller},
{Miller}, {Millhouse}, {Milovich-Goff}, {Minazzoli}, {Minenkov}, {Ming},
{Mishra}, {Mitra}, {Mitrofanov}, {Mitselmakher}, {Mittleman}, {Moffa},
{Moggi}, {Mogushi}, {Mohan}, {Mohapatra}, {Molina}, {Montani}, {Moore},
{Moraru}, {Moreno}, {Morisaki}, {Morriss}, {Mours}, {Mow-Lowry}, {Mueller},
{Muir}, {Mukherjee}, {Mukherjee}, {Mukherjee}, {Mukund}, {Mullavey}, {Munch},
{Mu{\~n}iz}, {Muratore}, {Murray}, {Nagar}, {Napier}, {Nardecchia},
{Naticchioni}, {Nayak}, {Neilson}, {Nelemans}, {Nelson}, {Nery}, {Neunzert},
{Nevin}, {Newport}, {Newton}, {Ng}, {Nguyen}, {Nguyen}, {Nichols}, {Nielsen},
{Nissanke}, {Nitz}, {Noack}, {Nocera}, {Nolting}, {North}, {Nuttall},
{Oberling}, {O'Dea}, {Ogin}, {Oh}, {Oh}, {Ohme}, {Okada}, {Oliver},
{Oppermann}, {Oram}, {O'Reilly}, {Ormiston}, {Ortega}, {O'Shaughnessy},
{Ossokine}, {Ottaway}, {Overmier}, {Owen}, {Pace}, {Page}, {Page}, {Pai},
{Pai}, {Palamos}, {Palashov}, {Palomba}, {Pal-Singh}, {Pan}, {Pan}, {Pang},
{Pang}, {Pankow}, {Pannarale}, {Pant}, {Paoletti}, {Paoli}, {Papa}, {Parida},
{Parker}, {Pascucci}, {Pasqualetti}, {Passaquieti}, {Passuello}, {Patil},
{Patricelli}, {Pearlstone}, {Pedraza}, {Pedurand}, {Pekowsky}, {Pele},
{Penn}, {Perez}, {Perreca}, {Perri}, {Pfeiffer}, {Phelps}, {Piccinni},
{Pichot}, {Piergiovanni}, {Pierro}, {Pillant}, {Pinard}, {Pinto}, {Pirello},
{Pitkin}, {Poe}, {Poggiani}, {Popolizio}, {Porter}, {Post}, {Powell},
{Prasad}, {Pratt}, {Pratten}, {Predoi}, {Prestegard}, {Prijatelj},
{Principe}, {Privitera}, {Prix}, {Prodi}, {Prokhorov}, {Puncken}, {Punturo},
{Puppo}, {P{\"u}rrer}, {Qi}, {Quetschke}, {Quintero}, {Quitzow-James},
{Raab}, {Rabeling}, {Radkins}, {Raffai}, {Raja}, {Rajan}, {Rajbhandari},
{Rakhmanov}, {Ramirez}, {Ramos-Buades}, {Rapagnani}, {Raymond}, {Razzano},
{Read}, {Regimbau}, {Rei}, {Reid}, {Reitze}, {Ren}, {Reyes}, {Ricci},
{Ricker}, {Rieger}, {Riles}, {Rizzo}, {Robertson}, {Robie}, {Robinet},
{Rocchi}, {Rolland}, {Rollins}, {Roma}, {Romano}, {Romano}, {Romel}, {Romie},
{Rosi{\'n}ska}, {Ross}, {Rowan}, {R{\"u}diger}, {Ruggi}, {Rutins}, {Ryan},
{Sachdev}, {Sadecki}, {Sadeghian}, {Sakellariadou}, {Salconi}, {Saleem},
{Salemi}, {Samajdar}, {Sammut}, {Sampson}, {Sanchez}, {Sanchez},
{Sanchis-Gual}, {Sandberg}, {Sanders}, {Sassolas}, {Sathyaprakash},
{Saulson}, {Sauter}, {Savage}, {Sawadsky}, {Schale}, {Scheel}, {Scheuer},
{Schmidt}, {Schmidt}, {Schnabel}, {Schofield}, {Sch{\"o}nbeck}, {Schreiber},
{Schuette}, {Schulte}, {Schutz}, {Schwalbe}, {Scott}, {Scott}, {Seidel},
{Sellers}, {Sengupta}, {Sentenac}, {Sequino}, {Sergeev}, {Shaddock},
{Shaffer}, {Shah}, {Shahriar}, {Shaner}, {Shao}, {Shapiro}, {Shawhan},
{Sheperd}, {Shoemaker}, {Shoemaker}, {Siellez}, {Siemens}, {Sieniawska},
{Sigg}, {Silva}, {Singer}, {Singh}, {Singhal}, {Sintes}, {Slagmolen},
{Smith}, {Smith}, {Smith}, {Somala}, {Son}, {Sonnenberg}, {Sorazu},
{Sorrentino}, {Souradeep}, {Spencer}, {Srivastava}, {Staats}, {Staley},
{Steinke}, {Steinlechner}, {Steinlechner}, {Steinmeyer}, {Stevenson},
{Stone}, {Stops}, {Strain}, {Stratta}, {Strigin}, {Strunk}, {Sturani},
{Stuver}, {Summerscales}, {Sun}, {Sunil}, {Suresh}, {Sutton}, {Swinkels},
{Szczepa{\'n}czyk}, {Tacca}, {Tait}, {Talbot}, {Talukder}, {Tanner},
{T{\'a}pai}, {Taracchini}, {Tasson}, {Taylor}, {Taylor}, {Tewari}, {Theeg},
{Thies}, {Thomas}, {Thomas}, {Thomas}, {Thorne}, {Thorne}, {Thrane},
{Tiwari}, {Tiwari}, {Tokmakov}, {Toland}, {Tonelli}, {Tornasi},
{Torres-Forn{\'e}}, {Torrie}, {T{\"o}yr{\"a}}, {Travasso}, {Traylor},
{Trinastic}, {Tringali}, {Trozzo}, {Tsang}, {Tse}, {Tso}, {Tsukada}, {Tsuna},
{Tuyenbayev}, {Ueno}, {Ugolini}, {Unnikrishnan}, {Urban}, {Usman},
{Vahlbruch}, {Vajente}, {Valdes}, {Vallisneri}, {van Bakel}, {van Beuzekom},
{van den Brand}, {Van Den Broeck}, {Vander-Hyde}, {van der Schaaf}, {van
Heijningen}, {van Veggel}, {Vardaro}, {Varma}, {Vass}, {Vas{\'u}th},
{Vecchio}, {Vedovato}, {Veitch}, {Veitch}, {Venkateswara}, {Venugopalan},
{Verkindt}, {Vetrano}, {Vicer{\'e}}, {Viets}, {Vinciguerra}, {Vine}, {Vinet},
{Vitale}, {Vo}, {Vocca}, {Vorvick}, {Vyatchanin}, {Wade}, {Wade}, {Wade},
{Walet}, {Walker}, {Wallace}, {Walsh}, {Wang}, {Wang}, {Wang}, {Wang},
{Wang}, {Ward}, {Warner}, {Was}, {Watchi}, {Weaver}, {Wei}, {Weinert},
{Weinstein}, {Weiss}, {Wen}, {Wessel}, {We{\ss}els}, {Westerweck},
{Westphal}, {Wette}, {Whelan}, {Whitcomb}, {Whiting}, {Whittle}, {Wilken},
{Williams}, {Williams}, {Williamson}, {Willis}, {Willke}, {Wimmer},
{Winkler}, {Wipf}, {Wittel}, {Woan}, {Woehler}, {Wofford}, {Wong}, {Worden},
{Wright}, {Wu}, {Wysocki}, {Xiao}, {Yamamoto}, {Yancey}, {Yang}, {Yap},
{Yazback}, {Yu}, {Yu}, {Yvert}, {Zadro{\.Z}ny}, {Zanolin}, {Zelenova},
{Zendri}, {Zevin}, {Zhang}, {Zhang}, {Zhang}, {Zhang}, {Zhao}, {Zhou},
{Zhou}, {Zhu}, {Zhu}, {Zimmerman}, {Zucker}, {Zweizig}, {LIGO Scientific
Collaboration}, and {Virgo Collaboration}]{Abbott17}
{Abbott}, B.P.; {Abbott}, R.; {Abbott}, T.D.; {Acernese}, F.; {Ackley}, K.;
{Adams}, C.; {Adams}, T.; {Addesso}, P.; {Adhikari}, R.X.; {Adya}, V.B.;
et~al.
\newblock {GW170817: Observation of Gravitational Waves from a Binary Neutron
Star Inspiral}.
\newblock {\em {Phys. Rev.~Lett.}  } {\bf 2017}, {\em 119},~161101.
%  \href{http://xxx.lanl.gov/abs/1710.05832}{{\normalfont
%  [arXiv:gr-qc/1710.05832]}}. \\
[\href{http://dx.doi.org/10.1103/PhysRevLett.119.161101}{CrossRef}]

\bibitem[{Chase} \em{et~al.}(2022){Chase}, {O'Connor}, {Fryer}, {Troja},
{Korobkin}, {Wollaeger}, {Ristic}, {Fontes}, {Hungerford}, and
{Herring}]{Chase22}
{Chase}, E.A.; {O'Connor}, B.; {Fryer}, C.L.; {Troja}, E.; {Korobkin}, O.;
{Wollaeger}, R.T.; {Ristic}, M.; {Fontes}, C.J.; {Hungerford}, A.L.;
{Herring}, A.M.
\newblock {Kilonova Detectability with Wide-field Instruments}.
\newblock {\em Astrophys. J. Lett.} {\bf 2022}, {\em 927},~163.
%  \href{http://xxx.lanl.gov/abs/2105.12268}{{\normalfont
%  [arXiv:astro-ph.HE/2105.12268]}}. \\
[\href{http://dx.doi.org/10.3847/1538-4357/ac3d25}{CrossRef}]

\bibitem[{Lazzati} \em{et~al.}(2017){Lazzati}, {L{\'o}pez-C{\'a}mara},
{Cantiello}, {Morsony}, {Perna}, and {Workman}]{Lazzati17}
{Lazzati}, D.; {L{\'o}pez-C{\'a}mara}, D.; {Cantiello}, M.; {Morsony}, B.J.;
{Perna}, R.; {Workman}, J.C.
\newblock {Off-axis Prompt X-Ray Transients from the Cocoon of Short Gamma-Ray
Bursts}.
\newblock {\em Astrophys. J. Lett.} {\bf 2017}, {\em 848},~L6.
%  \href{http://xxx.lanl.gov/abs/1709.01468}{{\normalfont
%  [arXiv:astro-ph.HE/1709.01468]}}. \\
[\href{http://dx.doi.org/10.3847/2041-8213/aa8f3d}{CrossRef}]

\bibitem[{Bromberg} \em{et~al.}(2018){Bromberg}, {Tchekhovskoy}, {Gottlieb},
{Nakar}, and {Piran}]{Bromberg18}
{Bromberg}, O.; {Tchekhovskoy}, A.; {Gottlieb}, O.; {Nakar}, E.; {Piran}, T.
\newblock {The {\ensuremath{\gamma}}-rays that accompanied GW170817 and the
observational signature of a magnetic jet breaking out of NS merger ejecta}.
\newblock {\em Mon. Not. R. Astron. Soc.} {\bf 2018}, {\em 475},~2971--2977.
%  \href{http://xxx.lanl.gov/abs/1710.05897}{{\normalfont
%  [arXiv:astro-ph.HE/1710.05897]}}. \\
[\href{http://dx.doi.org/10.1093/mnras/stx3316}{CrossRef}]

\bibitem[{Troja} \em{et~al.}(2010){Troja}, {Rosswog}, and {Gehrels}]{Troja10}
{Troja}, E.; {Rosswog}, S.; {Gehrels}, N.
\newblock {Precursors of Short Gamma-ray Bursts}.
\newblock {\em Astrophys. J. Lett.} {\bf 2010}, {\em 723},~1711--1717.
%  \href{http://xxx.lanl.gov/abs/1009.1385}{{\normalfont
%  [arXiv:astro-ph.HE/1009.1385]}}. \\
[\href{http://dx.doi.org/10.1088/0004-637X/723/2/1711}{CrossRef}]

\bibitem[{Norris} and {Bonnell}(2006)]{NB06}
{Norris}, J.P.; {Bonnell}, J.T.
\newblock {Short Gamma-Ray Bursts with Extended Emission}.
\newblock {\em Astrophys. J. Lett.} {\bf 2006}, {\em 643},~266--275.
%  \href{http://xxx.lanl.gov/abs/astro-ph/0601190}{{\normalfont
%  [arXiv:astro-ph/astro-ph/0601190]}}. \\
[\href{http://dx.doi.org/10.1086/502796}{CrossRef}]

\bibitem[{Ryan} \em{et~al.}(2015){Ryan}, {van Eerten}, {MacFadyen}, and
{Zhang}]{Ryan15}
{Ryan}, G.; {van Eerten}, H.; {MacFadyen}, A.; {Zhang}, B.B.
\newblock {Gamma-Ray Bursts are Observed Off-axis}.
\newblock {\em Astrophys. J. Lett.} {\bf 2015}, {\em 799},~3.
%  \href{http://xxx.lanl.gov/abs/1405.5516}{{\normalfont
%  [arXiv:astro-ph.HE/1405.5516]}}. \\
[\href{http://dx.doi.org/10.1088/0004-637X/799/1/3}{CrossRef}]

\bibitem[{Ryan} \em{et~al.}(2020){Ryan}, {van Eerten}, {Piro}, and
{Troja}]{Ryan20}
{Ryan}, G.; {van Eerten}, H.; {Piro}, L.; {Troja}, E.
\newblock {Gamma-Ray Burst Afterglows in the Multimessenger Era: Numerical
Models and Closure Relations}.
\newblock {\em Astrophys. J. Lett.} {\bf 2020}, {\em 896},~166.
%  \href{http://xxx.lanl.gov/abs/1909.11691}{{\normalfont
%  [arXiv:astro-ph.HE/1909.11691]}}. \\
[\href{http://dx.doi.org/10.3847/1538-4357/ab93cf}{CrossRef}]

\bibitem[{Troja} \em{et~al.}(2017){Troja}, {Piro}, {van Eerten}, {Wollaeger},
{Im}, {Fox}, {Butler}, {Cenko}, {Sakamoto}, {Fryer}, {Ricci}, {Lien}, {Ryan},
{Korobkin}, {Lee}, {Burgess}, {Lee}, {Watson}, {Choi}, {Covino}, {D'Avanzo},
{Fontes}, {Gonz{\'a}lez}, {Khandrika}, {Kim}, {Kim}, {Lee}, {Lee}, {Kutyrev},
{Lim}, {S{\'a}nchez-Ram{\'\i}rez}, {Veilleux}, {Wieringa}, and
{Yoon}]{Troja17b}
{Troja}, E.; {Piro}, L.; {van Eerten}, H.; {Wollaeger}, R.T.; {Im}, M.; {Fox},
O.D.; {Butler}, N.R.; {Cenko}, S.B.; {Sakamoto}, T.; {Fryer}, C.L.;  et~al.
\newblock {The X-ray counterpart to the gravitational-wave event GW170817}.
\newblock {\em {Nature} } {\bf 2017}, {\em 551},~71--74.
%  \href{http://xxx.lanl.gov/abs/1710.05433}{{\normalfont
%  [arXiv:astro-ph.HE/1710.05433]}}. \\
[\href{http://dx.doi.org/10.1038/nature24290}{CrossRef}]

\bibitem[{Nativi} \em{et~al.}(2021){Nativi}, {Bulla}, {Rosswog}, {Lundman},
{Kowal}, {Gizzi}, {Lamb}, and {Perego}]{Nativi21}
{Nativi}, L.; {Bulla}, M.; {Rosswog}, S.; {Lundman}, C.; {Kowal}, G.; {Gizzi},
D.; {Lamb}, G.P.; {Perego}, A.
\newblock {Can jets make the radioactively powered emission from neutron star
mergers bluer?}
\newblock {\em Mon. Not. R. Astron. Soc.} {\bf 2021}, {\em 500},~1772--1783.
%  \href{http://xxx.lanl.gov/abs/2010.08989}{{\normalfont
%  [arXiv:astro-ph.HE/2010.08989]}}. \\
[\href{http://dx.doi.org/10.1093/mnras/staa3337}{CrossRef}]

\bibitem[{Banerjee} \em{et~al.}(2022){Banerjee}, {Tanaka}, {Kato}, {Gaigalas},
{Kawaguchi}, and {Domoto}]{Banerjee22}
{Banerjee}, S.; {Tanaka}, M.; {Kato}, D.; {Gaigalas}, G.; {Kawaguchi}, K.;
{Domoto}, N.
\newblock {Opacity of the Highly Ionized Lanthanides and the Effect on the
Early Kilonova}.
\newblock {\em Astrophys. J. Lett.} {\bf 2022}, {\em 934},~117.
%  \href{http://xxx.lanl.gov/abs/2204.06861}{{\normalfont
%  [arXiv:astro-ph.HE/2204.06861]}}. \\
[\href{http://dx.doi.org/10.3847/1538-4357/ac7565}{CrossRef}]

\bibitem[{Even} \em{et~al.}(2020){Even}, {Korobkin}, {Fryer}, {Fontes},
{Wollaeger}, {Hungerford}, {Lippuner}, {Miller}, {Mumpower}, and
{Misch}]{Even20}
{Even}, W.; {Korobkin}, O.; {Fryer}, C.L.; {Fontes}, C.J.; {Wollaeger}, R.T.;
{Hungerford}, A.; {Lippuner}, J.; {Miller}, J.; {Mumpower}, M.R.; {Misch},
G.W.
\newblock {Composition Effects on Kilonova Spectra and Light Curves. I}.
\newblock {\em Astrophys. J. Lett.} {\bf 2020}, {\em 899},~24.
%  \href{http://xxx.lanl.gov/abs/1904.13298}{{\normalfont
%  [arXiv:astro-ph.HE/1904.13298]}}. \\
[\href{http://dx.doi.org/10.3847/1538-4357/ab70b9}{CrossRef}]

\bibitem[{Wollaeger} \em{et~al.}(2019){Wollaeger}, {Fryer}, {Fontes},
{Lippuner}, {Vestrand}, {Mumpower}, {Korobkin}, {Hungerford}, and
{Even}]{Wollager19}
{Wollaeger}, R.T.; {Fryer}, C.L.; {Fontes}, C.J.; {Lippuner}, J.; {Vestrand},
W.T.; {Mumpower}, M.R.; {Korobkin}, O.; {Hungerford}, A.L.; {Even}, W.P.
\newblock {Impact of Pulsar and Fallback Sources on Multifrequency Kilonova
Models}.
\newblock {\em Astrophys. J. Lett.} {\bf 2019}, {\em 880},~22.
%  \href{http://xxx.lanl.gov/abs/1904.05934}{{\normalfont
%  [arXiv:astro-ph.HE/1904.05934]}}. \\
[\href{http://dx.doi.org/10.3847/1538-4357/ab25f5}{CrossRef}]

\bibitem[{Waxman} \em{et~al.}(2018){Waxman}, {Ofek}, {Kushnir}, and
{Gal-Yam}]{Waxman18}
{Waxman}, E.; {Ofek}, E.O.; {Kushnir}, D.; {Gal-Yam}, A.
\newblock {Constraints on the ejecta of the GW170817 neutron star merger from
its electromagnetic emission}.
\newblock {\em Mon. Not. R. Astron. Soc.} {\bf 2018}, {\em 481},~3423--3441.
%  \href{http://xxx.lanl.gov/abs/1711.09638}{{\normalfont
%  [arXiv:astro-ph.HE/1711.09638]}}. \\
[\href{http://dx.doi.org/10.1093/mnras/sty2441}{CrossRef}]

\bibitem[{Piro} and {Kollmeier}(2018)]{Piro18}
{Piro}, A.L.; {Kollmeier}, J.A.
\newblock {Evidence for Cocoon Emission from the Early Light Curve of SSS17a}.
\newblock {\em Astrophys. J. Lett.} {\bf 2018}, {\em 855},~103.
%  \href{http://xxx.lanl.gov/abs/1710.05822}{{\normalfont
%  [arXiv:astro-ph.HE/1710.05822]}}. \\
[\href{http://dx.doi.org/10.3847/1538-4357/aaaab3}{CrossRef}]

\bibitem[{Arcavi}(2018)]{Arcavi18}
{Arcavi}, I.
\newblock {The First Hours of the GW170817 Kilonova and the Importance of Early
Optical and Ultraviolet Observations for Constraining Emission Models}.
\newblock {\em Astrophys. J. Lett.} {\bf 2018}, {\em 855},~L23.
%  \href{http://xxx.lanl.gov/abs/1802.02164}{{\normalfont
%  [arXiv:astro-ph.HE/1802.02164]}}. \\
[\href{http://dx.doi.org/10.3847/2041-8213/aab267}{CrossRef}]

\bibitem[{Evans} \em{et~al.}(2012){Evans}, {Fridriksson}, {Gehrels}, {Homan},
{Osborne}, {Siegel}, {Beardmore}, {Handbauer}, {Gelbord}, {Kennea}, {Smith},
{Zhu}, {LIGO Scientific Collaboration}, and {Virgo Collaboration}]{Evans12}
{Evans}, P.A.; {Fridriksson}, J.K.; {Gehrels}, N.; {Homan}, J.; {Osborne},
J.P.; {Siegel}, M.; {Beardmore}, A.; {Handbauer}, P.; {Gelbord}, J.;
{Kennea}, J.A.;  et~al.
\newblock {Swift Follow-up Observations of Candidate Gravitational-wave
Transient Events}.
\newblock {\em {Astrophys. J. Suppl.}} {\bf 2012}, {\em 203},~28.
%  \href{http://xxx.lanl.gov/abs/1205.1124}{{\normalfont
%  [arXiv:astro-ph.HE/1205.1124]}}. \\
[\href{http://dx.doi.org/10.1088/0067-0049/203/2/28}{CrossRef}]

\bibitem[{LIGO Scientific Collaboration} and {Virgo
Collaboration}(2016)]{150914}
{LIGO Scientific Collaboration}.; {Virgo Collaboration}.
\newblock {Observation of Gravitational Waves from a Binary Black Hole Merger}.
\newblock {\em {Phys. Rev.~Lett.}  } {\bf 2016}, {\em 116},~061102.
%  \href{http://xxx.lanl.gov/abs/1602.03837}{{\normalfont
%  [arXiv:gr-qc/1602.03837]}}. \\
[\href{http://dx.doi.org/10.1103/PhysRevLett.116.061102}{CrossRef}] [\href{http://www.ncbi.nlm.nih.gov/pubmed/26918975}{PubMed}]

\bibitem[{Abbott} \em{et~al.}(2017){Abbott}, {Abbott}, {Abbott}, {Acernese},
{Ackley}, {Adams}, {Adams}, {Addesso}, {Adhikari}, {Adya}, {Affeldt},
{Afrough}, {Agarwal}, {Agathos}, {Agatsuma}, {Aggarwal}, {Aguiar}, {Aiello},
{Ain}, {Ajith}, {Allen}, {Allen}, {Allocca}, {Altin}, {Amato}, {Ananyeva},
{Anderson}, {Anderson}, {Angelova}, {Antier}, {Appert}, {Arai}, {Araya},
{Areeda}, {Arnaud}, {Arun}, {Ascenzi}, {Ashton}, {Ast}, {Aston}, {Astone},
{Atallah}, {Aufmuth}, {Aulbert}, {AultONeal}, {Austin}, {Avila-Alvarez},
{Babak}, {Bacon}, {Bader}, {Bae}, {Baker}, {Baldaccini}, {Ballardin},
{Ballmer}, {Banagiri}, {Barayoga}, {Barclay}, {Barish}, {Barker}, {Barkett},
{Barone}, {Barr}, {Barsotti}, {Barsuglia}, {Barta}, {Barthelmy}, {Bartlett},
{Bartos}, {Bassiri}, {Basti}, {Batch}, {Bawaj}, {Bayley}, {Bazzan},
{B{\'e}csy}, {Beer}, {Bejger}, {Belahcene}, {Bell}, {Berger}, {Bergmann},
{Bero}, {Berry}, {Bersanetti}, {Bertolini}, {Betzwieser}, {Bhagwat},
{Bhandare}, {Bilenko}, {Billingsley}, {Billman}, {Birch}, {Birney},
{Birnholtz}, {Biscans}, {Biscoveanu}, {Bisht}, {Bitossi}, {Biwer},
{Bizouard}, {Blackburn}, {Blackman}, {Blair}, {Blair}, {Blair}, {Bloemen},
{Bock}, {Bode}, {Boer}, {Bogaert}, {Bohe}, {Bondu}, {Bonilla}, {Bonnand},
{Boom}, {Bork}, {Boschi}, {Bose}, {Bossie}, {Bouffanais}, {Bozzi},
{Bradaschia}, {Brady}, {Branchesi}, {Brau}, {Briant}, {Brillet}, {Brinkmann},
{Brisson}, {Brockill}, {Broida}, {Brooks}, {Brown}, {Brown}, {Brunett},
{Buchanan}, {Buikema}, {Bulik}, {Bulten}, {Buonanno}, {Buskulic}, {Buy},
{Byer}, {Cabero}, {Cadonati}, {Cagnoli}, {Cahillane}, {Calder{\'o}n
Bustillo}, {Callister}, {Calloni}, {Camp}, {Canepa}, {Canizares}, {Cannon},
{Cao}, {Cao}, {Capano}, {Capocasa}, {Carbognani}, {Caride}, {Carney},
{Casanueva Diaz}, {Casentini}, {Caudill}, {Cavagli{\`a}}, {Cavalier},
{Cavalieri}, {Cella}, {Cepeda}, {Cerd{\'a}-Dur{\'a}n}, {Cerretani},
{Cesarini}, {Chamberlin}, {Chan}, {Chao}, {Charlton}, {Chase},
{Chassande-Mottin}, {Chatterjee}, {Chatziioannou}, {Cheeseboro}, {Chen},
{Chen}, {Chen}, {Cheng}, {Chia}, {Chincarini}, {Chiummo}, {Chmiel}, {Cho},
{Cho}, {Chow}, {Christensen}, {Chu}, {Chua}, {Chua}, {Chung}, {Chung},
{Ciani}, {Ciolfi}, {Cirelli}, {Cirone}, {Clara}, {Clark}, {Clearwater},
{Cleva}, {Cocchieri}, {Coccia}, {Cohadon}, {Cohen}, {Colla}, {Collette},
{Cominsky}, {Constancio}, {Conti}, {Cooper}, {Corban}, {Corbitt},
{Cordero-Carri{\'o}n}, {Corley}, {Cornish}, {Corsi}, {Cortese}, {Costa},
{Coughlin}, {Coughlin}, {Coulon}, {Countryman}, {Couvares}, {Covas}, {Cowan},
{Coward}, {Cowart}, {Coyne}, {Coyne}, {Creighton}, {Creighton}, {Cripe},
{Crowder}, {Cullen}, {Cumming}, {Cunningham}, {Cuoco}, {Dal Canton},
{D{\'a}lya}, {Danilishin}, {D'Antonio}, {Danzmann}, {Dasgupta}, {Da Silva
Costa}, {Dattilo}, {Dave}, {Davier}, {Davis}, {Daw}, {Day}, {De}, {DeBra},
{Degallaix}, {De Laurentis}, {Del{\'e}glise}, {Del Pozzo}, {Demos}, {Denker},
{Dent}, {De Pietri}, {Dergachev}, {De Rosa}, {DeRosa}, {De Rossi}, {DeSalvo},
{de Varona}, {Devenson}, {Dhurandhar}, {D{\'\i}az}, {Di Fiore}, {Di
Giovanni}, {Di Girolamo}, {Di Lieto}, {Di Pace}, {Di Palma}, {Di Renzo},
{Doctor}, {Dolique}, {Donovan}, {Dooley}, {Doravari}, {Dorrington},
{Douglas}, {Dovale {\'A}lvarez}, {Downes}, {Drago}, {Dreissigacker},
{Driggers}, {Du}, {Ducrot}, {Dupej}, {Dwyer}, {Edo}, {Edwards}, {Effler},
{Ehrens}, {Eichholz}, {Eikenberry}, {Eisenstein}, {Essick}, {Estevez},
{Etienne}, {Etzel}, {Evans}, {Evans}, {Factourovich}, {Fafone}, {Fair},
{Fairhurst}, {Fan}, {Farinon}, {Farr}, {Farr}, {Fauchon-Jones}, {Favata},
{Fays}, {Fee}, {Fehrmann}, {Feicht}, {Fejer}, {Fernandez-Galiana},
{Ferrante}, {Ferreira}, {Ferrini}, {Fidecaro}, {Finstad}, {Fiori},
{Fiorucci}, {Fishbach}, {Fisher}, {Fitz-Axen}, {Flaminio}, {Fletcher},
{Fong}, {Font}, {Forsyth}, {Forsyth}, {Fournier}, {Frasca}, {Frasconi},
{Frei}, {Freise}, {Frey}, {Frey}, {Fries}, {Fritschel}, {Frolov}, {Fulda},
{Fyffe}, {Gabbard}, {Gadre}, {Gaebel}, {Gair}, {Gammaitoni}, {Ganija},
{Gaonkar}, {Garcia-Quiros}, {Garufi}, {Gateley}, {Gaudio}, {Gaur},
{Gayathri}, {Gehrels}, {Gemme}, {Genin}, {Gennai}, {George}, {George},
{Gergely}, {Germain}, {Ghonge}, {Ghosh}, {Ghosh}, {Ghosh}, {Giaime},
{Giardina}, {Giazotto}, {Gill}, {Glover}, {Goetz}, {Goetz}, {Gomes},
{Goncharov}, {Gonz{\'a}lez}, {Gonzalez Castro}, {Gopakumar}, {Gorodetsky},
{Gossan}, {Gosselin}, {Gouaty}, {Grado}, {Graef}, {Granata}, {Grant}, {Gras},
{Gray}, {Greco}, {Green}, {Gretarsson}, {Griswold}, {Groot}, {Grote},
{Grunewald}, {Gruning}, {Guidi}, {Guo}, {Gupta}, {Gupta}, {Gushwa},
{Gustafson}, {Gustafson}, {Halim}, {Hall}, {Hall}, {Hamilton}, {Hammond},
{Haney}, {Hanke}, {Hanks}, {Hanna}, {Hannam}, {Hannuksela}, {Hanson},
{Hardwick}, {Harms}, {Harry}, {Harry}, {Hart}, {Haster}, {Haughian}, {Healy},
{Heidmann}, {Heintze}, {Heitmann}, {Hello}, {Hemming}, {Hendry}, {Heng},
{Hennig}, {Heptonstall}, {Heurs}, {Hild}, {Hinderer}, {Hoak}, {Hofman},
{Holt}, {Holz}, {Hopkins}, {Horst}, {Hough}, {Houston}, {Howell}, {Hreibi},
{Hu}, {Huerta}, {Huet}, {Hughey}, {Husa}, {Huttner}, {Huynh-Dinh}, {Indik},
{Inta}, {Intini}, {Isa}, {Isac}, {Isi}, {Iyer}, {Izumi}, {Jacqmin}, {Jani},
{Jaranowski}, {Jawahar}, {Jim{\'e}nez-Forteza}, {Johnson}, {Jones}, {Jones},
{Jonker}, {Ju}, {Junker}, {Kalaghatgi}, {Kalogera}, {Kamai}, {Kandhasamy},
{Kang}, {Kanner}, {Kapadia}, {Karki}, {Karvinen}, {Kasprzack}, {Katolik},
{Katsavounidis}, {Katzman}, {Kaufer}, {Kawabe}, {K{\'e}f{\'e}lian}, {Keitel},
{Kemball}, {Kennedy}, {Kent}, {Key}, {Khalili}, {Khan}, {Khan}, {Khan},
{Khazanov}, {Kijbunchoo}, {Kim}, {Kim}, {Kim}, {Kim}, {Kim}, {Kim},
{Kimbrell}, {King}, {King}, {Kinley-Hanlon}, {Kirchhoff}, {Kissel},
{Kleybolte}, {Klimenko}, {Knowles}, {Koch}, {Koehlenbeck}, {Koley},
{Kondrashov}, {Kontos}, {Korobko}, {Korth}, {Kowalska}, {Kozak},
{Kr{\"a}mer}, {Kringel}, {Krishnan}, {Kr{\'o}lak}, {Kuehn}, {Kumar}, {Kumar},
{Kumar}, {Kuo}, {Kutynia}, {Kwang}, {Lackey}, {Lai}, {Landry}, {Lang},
{Lange}, {Lantz}, {Lanza}, {Larson}, {Lartaux-Vollard}, {Lasky}, {Laxen},
{Lazzarini}, {Lazzaro}, {Leaci}, {Leavey}, {Lee}, {Lee}, {Lee}, {Lee}, {Lee},
{Lehmann}, {Lenon}, {Leonardi}, {Leroy}, {Letendre}, {Levin}, {Li}, {Linker},
{Littenberg}, {Liu}, {Lo}, {Lockerbie}, {London}, {Lord}, {Lorenzini},
{Loriette}, {Lormand}, {Losurdo}, {Lough}, {Lousto}, {Lovelace}, {L{\"u}ck},
{Lumaca}, {Lundgren}, {Lynch}, {Ma}, {Macas}, {Macfoy}, {Machenschalk},
{MacInnis}, {Macleod}, {Maga{\~n}a Hernandez}, {Maga{\~n}a-Sandoval},
{Maga{\~n}a Zertuche}, {Magee}, {Majorana}, {Maksimovic}, {Man}, {Mandic},
{Mangano}, {Mansell}, {Manske}, {Mantovani}, {Marchesoni}, {Marion},
{M{\'a}rka}, {M{\'a}rka}, {Markakis}, {Markosyan}, {Markowitz}, {Maros},
{Marquina}, {Marsh}, {Martelli}, {Martellini}, {Martin}, {Martin},
{Martynov}, {Mason}, {Massera}, {Masserot}, {Massinger}, {Masso-Reid},
{Mastrogiovanni}, {Matas}, {Matichard}, {Matone}, {Mavalvala}, {Mazumder},
{McCarthy}, {McClelland}, {McCormick}, {McCuller}, {McGuire}, {McIntyre},
{McIver}, {McManus}, {McNeill}, {McRae}, {McWilliams}, {Meacher}, {Meadors},
{Mehmet}, {Meidam}, {Mejuto-Villa}, {Melatos}, {Mendell}, {Mercer}, {Merilh},
{Merzougui}, {Meshkov}, {Messenger}, {Messick}, {Metzdorff}, {Meyers},
{Miao}, {Michel}, {Middleton}, {Mikhailov}, {Milano}, {Miller}, {Miller},
{Miller}, {Millhouse}, {Milovich-Goff}, {Minazzoli}, {Minenkov}, {Ming},
{Mishra}, {Mitra}, {Mitrofanov}, {Mitselmakher}, {Mittleman}, {Moffa},
{Moggi}, {Mogushi}, {Mohan}, {Mohapatra}, {Montani}, {Moore}, {Moraru},
{Moreno}, {Morriss}, {Mours}, {Mow-Lowry}, {Mueller}, {Muir}, {Mukherjee},
{Mukherjee}, {Mukherjee}, {Mukund}, {Mullavey}, {Munch}, {Mu{\~n}iz},
{Muratore}, {Murray}, {Napier}, {Nardecchia}, {Naticchioni}, {Nayak},
{Neilson}, {Nelemans}, {Nelson}, {Nery}, {Neunzert}, {Nevin}, {Newport},
{Newton}, {Ng}, {Nguyen}, {Nguyen}, {Nichols}, {Nielsen}, {Nissanke}, {Nitz},
{Noack}, {Nocera}, {Nolting}, {North}, {Nuttall}, {Oberling}, {O'Dea},
{Ogin}, {Oh}, {Oh}, {Ohme}, {Okada}, {Oliver}, {Oppermann}, {Oram},
{O'Reilly}, {Ormiston}, {Ortega}, {O'Shaughnessy}, {Ossokine}, {Ottaway},
{Overmier}, {Owen}, {Pace}, {Page}, {Page}, {Pai}, {Pai}, {Palamos},
{Palashov}, {Palomba}, {Pal-Singh}, {Pan}, {Pan}, {Pang}, {Pang}, {Pankow},
{Pannarale}, {Pant}, {Paoletti}, {Paoli}, {Papa}, {Parida}, {Parker},
{Pascucci}, {Pasqualetti}, {Passaquieti}, {Passuello}, {Patil}, {Patricelli},
{Pearlstone}, {Pedraza}, {Pedurand}, {Pekowsky}, {Pele}, {Penn}, {Perez},
{Perreca}, {Perri}, {Pfeiffer}, {Phelps}, {Piccinni}, {Pichot},
{Piergiovanni}, {Pierro}, {Pillant}, {Pinard}, {Pinto}, {Pirello}, {Pitkin},
{Poe}, {Poggiani}, {Popolizio}, {Porter}, {Post}, {Powell}, {Prasad},
{Pratt}, {Pratten}, {Predoi}, {Prestegard}, {Price}, {Prijatelj}, {Principe},
{Privitera}, {Prodi}, {Prokhorov}, {Puncken}, {Punturo}, {Puppo},
{P{\"u}rrer}, {Qi}, {Quetschke}, {Quintero}, {Quitzow-James}, {Raab},
{Rabeling}, {Radkins}, {Raffai}, {Raja}, {Rajan}, {Rajbhandari}, {Rakhmanov},
{Ramirez}, {Ramos-Buades}, {Rapagnani}, {Raymond}, {Razzano}, {Read},
{Regimbau}, {Rei}, {Reid}, {Reitze}, {Ren}, {Reyes}, {Ricci}, {Ricker},
{Rieger}, {Riles}, {Rizzo}, {Robertson}, {Robie}, {Robinet}, {Rocchi},
{Rolland}, {Rollins}, {Roma}, {Romano}, {Romel}, {Romie}, {Rosi{\'n}ska},
{Ross}, {Rowan}, {R{\"u}diger}, {Ruggi}, {Rutins}, {Ryan}, {Sachdev},
{Sadecki}, {Sadeghian}, {Sakellariadou}, {Salconi}, {Saleem}, {Salemi},
{Samajdar}, {Sammut}, {Sampson}, {Sanchez}, {Sanchez}, {Sanchis-Gual},
{Sandberg}, {Sanders}, {Sassolas}, {Sathyaprakash}, {Saulson}, {Sauter},
{Savage}, {Sawadsky}, {Schale}, {Scheel}, {Scheuer}, {Schmidt}, {Schmidt},
{Schnabel}, {Schofield}, {Sch{\"o}nbeck}, {Schreiber}, {Schuette}, {Schulte},
{Schutz}, {Schwalbe}, {Scott}, {Scott}, {Seidel}, {Sellers}, {Sengupta},
{Sentenac}, {Sequino}, {Sergeev}, {Shaddock}, {Shaffer}, {Shah}, {Shahriar},
{Shaner}, {Shao}, {Shapiro}, {Shawhan}, {Sheperd}, {Shoemaker}, {Shoemaker},
{Siellez}, {Siemens}, {Sieniawska}, {Sigg}, {Silva}, {Singer}, {Singh},
{Singhal}, {Sintes}, {Slagmolen}, {Smith}, {Smith}, {Smith}, {Somala}, {Son},
{Sonnenberg}, {Sorazu}, {Sorrentino}, {Souradeep}, {Spencer}, {Srivastava},
{Staats}, {Staley}, {Steinke}, {Steinlechner}, {Steinlechner}, {Steinmeyer},
{Stevenson}, {Stone}, {Stops}, {Strain}, {Stratta}, {Strigin}, {Strunk},
{Sturani}, {Stuver}, {Summerscales}, {Sun}, {Sunil}, {Suresh}, {Sutton},
{Swinkels}, {Szczepa{\'n}czyk}, {Tacca}, {Tait}, {Talbot}, {Talukder},
{Tanner}, {T{\'a}pai}, {Taracchini}, {Tasson}, {Taylor}, {Taylor}, {Tewari},
{Theeg}, {Thies}, {Thomas}, {Thomas}, {Thomas}, {Thorne}, {Thorne}, {Thrane},
{Tiwari}, {Tiwari}, {Tokmakov}, {Toland}, {Tonelli}, {Tornasi},
{Torres-Forn{\'e}}, {Torrie}, {T{\"o}yr{\"a}}, {Travasso}, {Traylor},
{Trinastic}, {Tringali}, {Trozzo}, {Tsang}, {Tse}, {Tso}, {Tsukada}, {Tsuna},
{Tuyenbayev}, {Ueno}, {Ugolini}, {Unnikrishnan}, {Urban}, {Usman},
{Vahlbruch}, {Vajente}, {Valdes}, {van Bakel}, {van Beuzekom}, {van den
Brand}, {Van Den Broeck}, {Vander-Hyde}, {van der Schaaf}, {van Heijningen},
{van Veggel}, {Vardaro}, {Varma}, {Vass}, {Vas{\'u}th}, {Vecchio},
{Vedovato}, {Veitch}, {Veitch}, {Venkateswara}, {Venugopalan}, {Verkindt},
{Vetrano}, {Vicer{\'e}}, {Viets}, {Vinciguerra}, {Vine}, {Vinet}, {Vitale},
{Vo}, {Vocca}, {Vorvick}, {Vyatchanin}, {Wade}, {Wade}, {Wade}, {Walet},
{Walker}, {Wallace}, {Walsh}, {Wang}, {Wang}, {Wang}, {Wang}, {Wang}, {Ward},
{Warner}, {Was}, {Watchi}, {Weaver}, {Wei}, {Weinert}, {Weinstein}, {Weiss},
{Wen}, {Wessel}, {Wessels}, {Westerweck}, {Westphal}, {Wette}, {Whelan},
{Whitcomb}, {Whiting}, {Whittle}, {Wilken}, {Williams}, {Williams},
{Williamson}, {Willis}, {Willke}, {Wimmer}, {Winkler}, {Wipf}, {Wittel},
{Woan}, {Woehler}, {Wofford}, {Wong}, {Worden}, {Wright}, {Wu}, {Wysocki},
{Xiao}, {Yamamoto}, {Yancey}, {Yang}, {Yap}, {Yazback}, {Yu}, {Yu}, {Yvert},
{Zadro{\.z}ny}, {Zanolin}, {Zelenova}, {Zendri}, {Zevin}, {Zhang}, {Zhang},
{Zhang}, {Zhang}, {Zhao}, {Zhou}, {Zhou}, {Zhu}, {Zhu}, {Zimmerman},
{Zucker}, {Zweizig}, {LIGO Scientific Collaboration}, {Virgo Collaboration},
{Wilson-Hodge}, {Bissaldi}, {Blackburn}, {Briggs}, {Burns}, {Cleveland},
{Connaughton}, {Gibby}, {Giles}, {Goldstein}, {Hamburg}, {Jenke}, {Hui},
{Kippen}, {Kocevski}, {McBreen}, {Meegan}, {Paciesas}, {Poolakkil}, {Preece},
{Racusin}, {Roberts}, {Stanbro}, {Veres}, {von Kienlin}, {GBM}, {Savchenko},
{Ferrigno}, {Kuulkers}, {Bazzano}, {Bozzo}, {Brandt}, {Chenevez},
{Courvoisier}, {Diehl}, {Domingo}, {Hanlon}, {Jourdain}, {Laurent}, {Lebrun},
{Lutovinov}, {Martin-Carrillo}, {Mereghetti}, {Natalucci}, {Rodi}, {Roques},
{Sunyaev}, {Ubertini}, {INTEGRAL}, {Aartsen}, {Ackermann}, {Adams},
{Aguilar}, {Ahlers}, {Ahrens}, {Samarai}, {Altmann}, {Andeen}, {Anderson},
{Ansseau}, {Anton}, {Arg{\"u}elles}, {Auffenberg}, {Axani}, {Bagherpour},
{Bai}, {Barron}, {Barwick}, {Baum}, {Bay}, {Beatty}, {Becker Tjus},
{Bernardini}, {Besson}, {Binder}, {Bindig}, {Blaufuss}, {Blot}, {Bohm},
{B{\"o}rner}, {Bos}, {Bose}, {B{\"o}ser}, {Botner}, {Bourbeau}, {Bourbeau},
{Bradascio}, {Braun}, {Brayeur}, {Brenzke}, {Bretz}, {Bron},
{Brostean-Kaiser}, {Burgman}, {Carver}, {Casey}, {Casier}, {Cheung},
{Chirkin}, {Christov}, {Clark}, {Classen}, {Coenders}, {Collin}, {Conrad},
{Cowen}, {Cross}, {Day}, {de Andr{\'e}}, {De Clercq}, {DeLaunay},
{Dembinski}, {De Ridder}, {Desiati}, {de Vries}, {de Wasseige}, {de With},
{DeYoung}, {D{\'\i}az-V{\'e}lez}, {di Lorenzo}, {Dujmovic}, {Dumm},
{Dunkman}, {Dvorak}, {Eberhardt}, {Ehrhardt}, {Eichmann}, {Eller}, {Evenson},
{Fahey}, {Fazely}, {Felde}, {Filimonov}, {Finley}, {Flis}, {Franckowiak},
{Friedman}, {Fuchs}, {Gaisser}, {Gallagher}, {Gerhardt}, {Ghorbani}, {Giang},
{Glauch}, {Gl{\"u}senkamp}, {Goldschmidt}, {Gonzalez}, {Grant}, {Griffith},
{Haack}, {Hallgren}, {Halzen}, {Hanson}, {Hebecker}, {Heereman}, {Helbing},
{Hellauer}, {Hickford}, {Hignight}, {Hill}, {Hoffman}, {Hoffmann},
{Hokanson-Fasig}, {Hoshina}, {Huang}, {Huber}, {Hultqvist}, {H{\"u}nnefeld},
{In}, {Ishihara}, {Jacobi}, {Japaridze}, {Jeong}, {Jero}, {Jones},
{Kalaczynski}, {Kang}, {Kappes}, {Karg}, {Karle}, {Kauer}, {Keivani},
{Kelley}, {Kheirandish}, {Kim}, {Kim}, {Kintscher}, {Kiryluk}, {Kittler},
{Klein}, {Kohnen}, {Koirala}, {Kolanoski}, {K{\"o}pke}, {Kopper}, {Kopper},
{Koschinsky}, {Koskinen}, {Kowalski}, {Krings}, {Kroll}, {Kr{\"u}ckl},
{Kunnen}, {Kunwar}, {Kurahashi}, {Kuwabara}, {Kyriacou}, {Labare},
{Lanfranchi}, {Larson}, {Lauber}, {Lesiak-Bzdak}, {Leuermann}, {Liu}, {Lu},
{L{\"u}nemann}, {Luszczak}, {Madsen}, {Maggi}, {Mahn}, {Mancina}, {Maruyama},
{Mase}, {Maunu}, {McNally}, {Meagher}, {Medici}, {Meier}, {Menne}, {Merino},
{Meures}, {Miarecki}, {Micallef}, {Moment{\'e}}, {Montaruli}, {Moore},
{Moulai}, {Nahnhauer}, {Nakarmi}, {Naumann}, {Neer}, {Niederhausen},
{Nowicki}, {Nygren}, {Obertacke Pollmann}, {Olivas}, {O'Murchadha},
{Palczewski}, {Pandya}, {Pankova}, {Peiffer}, {Pepper}, {P{\'e}rez de los
Heros}, {Pieloth}, {Pinat}, {Price}, {Przybylski}, {Raab}, {R{\"a}del},
{Rameez}, {Rawlins}, {Rea}, {Reimann}, {Relethford}, {Relich}, {Resconi},
{Rhode}, {Richman}, {Robertson}, {Rongen}, {Rott}, {Ruhe}, {Ryckbosch},
{Rysewyk}, {S{\"a}lzer}, {Sanchez Herrera}, {Sandrock}, {Sandroos},
{Santander}, {Sarkar}, {Sarkar}, {Satalecka}, {Schlunder}, {Schmidt},
{Schneider}, {Schoenen}, {Sch{\"o}neberg}, {Schumacher}, {Seckel},
{Seunarine}, {Soedingrekso}, {Soldin}, {Song}, {Spiczak}, {Spiering},
{Stachurska}, {Stamatikos}, {Stanev}, {Stasik}, {Stettner}, {Steuer},
{Stezelberger}, {Stokstad}, {St{\"o}ssl}, {Strotjohann}, {Stuttard},
{Sullivan}, {Sutherland}, {Taboada}, {Tatar}, {Tenholt}, {Ter-Antonyan},
{Terliuk}, {Te{\v{s}}i{\'c}}, {Tilav}, {Toale}, {Tobin}, {Toscano}, {Tosi},
{Tselengidou}, {Tung}, {Turcati}, {Turley}, {Ty}, {Unger}, {Usner},
{Vandenbroucke}, {Van Driessche}, {van Eijndhoven}, {Vanheule}, {van Santen},
{Vehring}, {Vogel}, {Vraeghe}, {Walck}, {Wallace}, {Wallraff}, {Wandler},
{Wandkowsky}, {Waza}, {Weaver}, {Weiss}, {Wendt}, {Werthebach}, {Whelan},
{Wiebe}, {Wiebusch}, {Wille}, {Williams}, {Wills}, {Wolf}, {Wood}, {Woolsey},
{Woschnagg}, {Xu}, {Xu}, {Xu}, {Yanez}, {Yodh}, {Yoshida}, {Yuan}, {Zoll},
{IceCube Collaboration}, {Balasubramanian}, {Mate}, {Bhalerao},
{Bhattacharya}, {Vibhute}, {Dewangan}, {Rao}, {Vadawale}, {AstroSat Cadmium
Zinc Telluride Imager Team}, {Svinkin}, {Hurley}, {Aptekar}, {Frederiks},
{Golenetskii}, {Kozlova}, {Lysenko}, {Oleynik}, {Tsvetkova}, {Ulanov},
{Cline}, {IPN Collaboration}, {Li}, {Xiong}, {Zhang}, {Lu}, {Song}, {Cao},
{Chang}, {Chen}, {Chen}, {Chen}, {Chen}, {Chen}, {Chen}, {Cui}, {Cui},
{Deng}, {Dong}, {Du}, {Fu}, {Gao}, {Gao}, {Gao}, {Ge}, {Gu}, {Guan}, {Guo},
{Han}, {Hu}, {Huang}, {Huo}, {Jia}, {Jiang}, {Jiang}, {Jin}, {Jin}, {Li},
{Li}, {Li}, {Li}, {Li}, {Li}, {Li}, {Li}, {Li}, {Li}, {Li}, {Liang}, {Liao},
{Liu}, {Liu}, {Liu}, {Liu}, {Liu}, {Liu}, {Liu}, {Lu}, {Lu}, {Luo}, {Ma},
{Meng}, {Nang}, {Nie}, {Ou}, {Qu}, {Sai}, {Sun}, {Tan}, {Tao}, {Tao}, {Tuo},
{Wang}, {Wang}, {Wang}, {Wang}, {Wang}, {Wen}, {Wu}, {Wu}, {Xiao}, {Xu},
{Xu}, {Yan}, {Yang}, {Yang}, {Yang}, {Zhang}, {Zhang}, {Zhang}, {Zhang},
{Zhang}, {Zhang}, {Zhang}, {Zhang}, {Zhang}, {Zhang}, {Zhang}, {Zhang},
{Zhang}, {Zhang}, {Zhang}, {Zhang}, {Zhang}, {Zhang}, {Zhao}, {Zhao}, {Zhao},
{Zheng}, {Zhu}, {Zhu}, {Zou}, {Insight-HXMT Collaboration}, {Albert},
{Andr{\'e}}, {Anghinolfi}, {Ardid}, {Aubert}, {Aublin}, {Avgitas}, {Baret},
{Barrios-Mart{\'\i}}, {Basa}, {Belhorma}, {Bertin}, {Biagi}, {Bormuth},
{Bourret}, {Bouwhuis}, {Br{\^a}nza{\c{s}}}, {Bruijn}, {Brunner}, {Busto},
{Capone}, {Caramete}, {Carr}, {Celli}, {Cherkaoui El Moursli}, {Chiarusi},
{Circella}, {Coelho}, {Coleiro}, {Coniglione}, {Costantini}, {Coyle},
{Creusot}, {D{\'\i}az}, {Deschamps}, {De Bonis}, {Distefano}, {Di Palma},
{Domi}, {Donzaud}, {Dornic}, {Drouhin}, {Eberl}, {El Bojaddaini}, {El
Khayati}, {Els{\"a}sser}, {Enzenh{\"o}fer}, {Ettahiri}, {Fassi}, {Felis},
{Fusco}, {Gay}, {Giordano}, {Glotin}, {Gr{\'e}goire}, {Ruiz}, {Graf},
{Hallmann}, {van Haren}, {Heijboer}, {Hello}, {Hern{\'a}ndez-Rey},
{H{\"o}ssl}, {Hofest{\"a}dt}, {Hugon}, {Illuminati}, {James}, {de Jong},
{Jongen}, {Kadler}, {Kalekin}, {Katz}, {Kiessling}, {Kouchner}, {Kreter},
{Kreykenbohm}, {Kulikovskiy}, {Lachaud}, {Lahmann}, {Lef{\`e}vre}, {Leonora},
{Lotze}, {Loucatos}, {Marcelin}, {Margiotta}, {Marinelli},
{Mart{\'\i}nez-Mora}, {Mele}, {Melis}, {Michael}, {Migliozzi}, {Moussa},
{Navas}, {Nezri}, {Organokov}, {P{\u{a}}v{\u{a}}la{\c{s}}}, {Pellegrino},
{Perrina}, {Piattelli}, {Popa}, {Pradier}, {Quinn}, {Racca}, {Riccobene},
{S{\'a}nchez-Losa}, {Salda{\~n}a}, {Salvadori}, {Samtleben}, {Sanguineti},
{Sapienza}, {Sieger}, {Spurio}, {Stolarczyk}, {Taiuti}, {Tayalati},
{Trovato}, {Turpin}, {T{\"o}nnis}, {Vallage}, {Van Elewyck}, {Versari},
{Vivolo}, {Vizzoca}, {Wilms}, {Zornoza}, {Z{\'u}{\~n}iga}, {ANTARES
Collaboration}, {Beardmore}, {Breeveld}, {Burrows}, {Cenko}, {Cusumano},
{D'A{\`\i}}, {de Pasquale}, {Emery}, {Evans}, {Giommi}, {Gronwall}, {Kennea},
{Krimm}, {Kuin}, {Lien}, {Marshall}, {Melandri}, {Nousek}, {Oates},
{Osborne}, {Pagani}, {Page}, {Palmer}, {Perri}, {Siegel}, {Sbarufatti},
{Tagliaferri}, {Tohuvavohu}, {Swift Collaboration}, {Tavani}, {Verrecchia},
{Bulgarelli}, {Evangelista}, {Pacciani}, {Feroci}, {Pittori}, {Giuliani},
{Del Monte}, {Donnarumma}, {Argan}, {Trois}, {Ursi}, {Cardillo}, {Piano},
{Longo}, {Lucarelli}, {Munar-Adrover}, {Fuschino}, {Labanti}, {Marisaldi},
{Minervini}, {Fioretti}, {Parmiggiani}, {Gianotti}, {Trifoglio}, {Di Persio},
{Antonelli}, {Barbiellini}, {Caraveo}, {Cattaneo}, {Costa}, {Colafrancesco},
{D'Amico}, {Ferrari}, {Morselli}, {Paoletti}, {Picozza}, {Pilia}, {Rappoldi},
{Soffitta}, {Vercellone}, {AGILE Team}, {Foley}, {Coulter}, {Kilpatrick},
{Drout}, {Piro}, {Shappee}, {Siebert}, {Simon}, {Ulloa}, {Kasen}, {Madore},
{Murguia-Berthier}, {Pan}, {Prochaska}, {Ramirez-Ruiz}, {Rest},
{Rojas-Bravo}, {1M2H Team}, {Berger}, {Soares-Santos}, {Annis}, {Alexander},
{Allam}, {Balbinot}, {Blanchard}, {Brout}, {Butler}, {Chornock}, {Cook},
{Cowperthwaite}, {Diehl}, {Drlica-Wagner}, {Drout}, {Durret}, {Eftekhari},
{Finley}, {Fong}, {Frieman}, {Fryer}, {Garc{\'\i}a-Bellido}, {Gruendl},
{Hartley}, {Herner}, {Kessler}, {Lin}, {Lopes}, {Louren{\c{c}}o}, {Margutti},
{Marshall}, {Matheson}, {Medina}, {Metzger}, {Mu{\~n}oz}, {Muir}, {Nicholl},
{Nugent}, {Palmese}, {Paz-Chinch{\'o}n}, {Quataert}, {Sako}, {Sauseda},
{Schlegel}, {Scolnic}, {Secco}, {Smith}, {Sobreira}, {Villar}, {Vivas},
{Wester}, {Williams}, {Yanny}, {Zenteno}, {Zhang}, {Abbott}, {Banerji},
{Bechtol}, {Benoit-L{\'e}vy}, {Bertin}, {Brooks}, {Buckley-Geer}, {Burke},
{Capozzi}, {Carnero Rosell}, {Carrasco Kind}, {Castander}, {Crocce}, {Cunha},
{D'Andrea}, {da Costa}, {Davis}, {DePoy}, {Desai}, {Dietrich}, {Eifler},
{Fernandez}, {Flaugher}, {Fosalba}, {Gaztanaga}, {Gerdes}, {Giannantonio},
{Goldstein}, {Gruen}, {Gschwend}, {Gutierrez}, {Honscheid}, {James},
{Jeltema}, {Johnson}, {Johnson}, {Kent}, {Krause}, {Kron}, {Kuehn}, {Lahav},
{Lima}, {Maia}, {March}, {Martini}, {McMahon}, {Menanteau}, {Miller},
{Miquel}, {Mohr}, {Nichol}, {Ogando}, {Plazas}, {Romer}, {Roodman}, {Rykoff},
{Sanchez}, {Scarpine}, {Schindler}, {Schubnell}, {Sevilla-Noarbe}, {Sheldon},
{Smith}, {Smith}, {Stebbins}, {Suchyta}, {Swanson}, {Tarle}, {Thomas},
{Troxel}, {Tucker}, {Vikram}, {Walker}, {Wechsler}, {Weller}, {Carlin},
{Gill}, {Li}, {Marriner}, {Neilsen}, {Dark Energy Camera GW-EM
Collaboration}, {DES Collaboration}, {Haislip}, {Kouprianov}, {Reichart},
{Sand}, {Tartaglia}, {Valenti}, {Yang}, {DLT40 Collaboration}, {Benetti},
{Brocato}, {Campana}, {Cappellaro}, {Covino}, {D'Avanzo}, {D'Elia}, {Getman},
{Ghirlanda}, {Ghisellini}, {Limatola}, {Nicastro}, {Palazzi}, {Pian},
{Piranomonte}, {Possenti}, {Rossi}, {Salafia}, {Tomasella}, {Amati},
{Antonelli}, {Bernardini}, {Bufano}, {Capaccioli}, {Casella}, {Dadina}, {De
Cesare}, {Di Paola}, {Giuffrida}, {Giunta}, {Israel}, {Lisi}, {Maiorano},
{Mapelli}, {Masetti}, {Pescalli}, {Pulone}, {Salvaterra}, {Schipani},
{Spera}, {Stamerra}, {Stella}, {Testa}, {Turatto}, {Vergani}, {Aresu},
{Bachetti}, {Buffa}, {Burgay}, {Buttu}, {Caria}, {Carretti}, {Casasola},
{Castangia}, {Carboni}, {Casu}, {Concu}, {Corongiu}, {Deiana}, {Egron},
{Fara}, {Gaudiomonte}, {Gusai}, {Ladu}, {Loru}, {Leurini}, {Marongiu},
{Melis}, {Melis}, {Migoni}, {Milia}, {Navarrini}, {Orlati}, {Ortu}, {Palmas},
{Pellizzoni}, {Perrodin}, {Pisanu}, {Poppi}, {Righini}, {Saba}, {Serra},
{Serrau}, {Stagni}, {Surcis}, {Vacca}, {Vargiu}, {Hunt}, {Jin}, {Klose},
{Kouveliotou}, {Mazzali}, {M{\o}ller}, {Nava}, {Piran}, {Selsing}, {Vergani},
{Wiersema}, {Toma}, {Higgins}, {Mundell}, {di Serego Alighieri}, {G{\'o}tz},
{Gao}, {Gomboc}, {Kaper}, {Kobayashi}, {Kopac}, {Mao}, {Starling}, {Steele},
{van der Horst}, {GRAWITA: GRAvitational Wave Inaf TeAm}, {Acero}, {Atwood},
{Baldini}, {Barbiellini}, {Bastieri}, {Berenji}, {Bellazzini}, {Bissaldi},
{Blandford}, {Bloom}, {Bonino}, {Bottacini}, {Bregeon}, {Buehler}, {Buson},
{Cameron}, {Caputo}, {Caraveo}, {Cavazzuti}, {Chekhtman}, {Cheung}, {Chiang},
{Ciprini}, {Cohen-Tanugi}, {Cominsky}, {Costantin}, {Cuoco}, {D'Ammando}, {de
Palma}, {Digel}, {Di Lalla}, {Di Mauro}, {Di Venere}, {Dubois}, {Fegan},
{Focke}, {Franckowiak}, {Fukazawa}, {Funk}, {Fusco}, {Gargano}, {Gasparrini},
{Giglietto}, {Giordano}, {Giroletti}, {Glanzman}, {Green}, {Grondin},
{Guillemot}, {Guiriec}, {Harding}, {Horan}, {J{\'o}hannesson}, {Kamae},
{Kensei}, {Kuss}, {La Mura}, {Latronico}, {Lemoine-Goumard}, {Longo},
{Loparco}, {Lovellette}, {Lubrano}, {Magill}, {Maldera}, {Manfreda},
{Mazziotta}, {McEnery}, {Meyer}, {Michelson}, {Mirabal}, {Monzani},
{Moretti}, {Morselli}, {Moskalenko}, {Negro}, {Nuss}, {Ojha}, {Omodei},
{Orienti}, {Orlando}, {Palatiello}, {Paliya}, {Paneque}, {Pesce-Rollins},
{Piron}, {Porter}, {Principe}, {Rain{\`o}}, {Rando}, {Razzano}, {Razzaque},
{Reimer}, {Reimer}, {Reposeur}, {Rochester}, {Saz Parkinson}, {Sgr{\`o}},
{Siskind}, {Spada}, {Spandre}, {Suson}, {Takahashi}, {Tanaka}, {Thayer},
{Thayer}, {Thompson}, {Tibaldo}, {Torres}, {Torresi}, {Troja}, {Venters},
{Vianello}, {Zaharijas}, {Fermi Large Area Telescope Collaboration},
{Allison}, {Bannister}, {Dobie}, {Kaplan}, {Lenc}, {Lynch}, {Murphy},
{Sadler}, {Australia Telescope Compact Array}, {Hotan}, {James}, {Oslowski},
{Raja}, {Shannon}, {Whiting}, {Australian SKA Pathfinder}, {Arcavi},
{Howell}, {McCully}, {Hosseinzadeh}, {Hiramatsu}, {Poznanski}, {Barnes},
{Zaltzman}, {Vasylyev}, {Maoz}, {Las Cumbres Observatory Group}, {Cooke},
{Bailes}, {Wolf}, {Deller}, {Lidman}, {Wang}, {Gendre}, {Andreoni}, {Ackley},
{Pritchard}, {Bessell}, {Chang}, {M{\"o}ller}, {Onken}, {Scalzo},
{Ridden-Harper}, {Sharp}, {Tucker}, {Farrell}, {Elmer}, {Johnston},
{Venkatraman Krishnan}, {Keane}, {Green}, {Jameson}, {Hu}, {Ma}, {Sun}, {Wu},
{Wang}, {Shang}, {Hu}, {Ashley}, {Yuan}, {Li}, {Tao}, {Zhu}, {Zhang},
{Suntzeff}, {Zhou}, {Yang}, {Orange}, {Morris}, {Cucchiara}, {Giblin},
{Klotz}, {Staff}, {Thierry}, {Schmidt}, {OzGrav}, {(Deeper}, {Wider},
{program}, {AST3}, {CAASTRO Collaborations}, {Tanvir}, {Levan}, {Cano}, {de
Ugarte-Postigo}, {Gonz{\'a}lez-Fern{\'a}ndez}, {Greiner}, {Hjorth}, {Irwin},
{Kr{\"u}hler}, {Mandel}, {Milvang-Jensen}, {O'Brien}, {Rol}, {Rosetti},
{Rosswog}, {Rowlinson}, {Steeghs}, {Th{\"o}ne}, {Ulaczyk}, {Watson}, {Bruun},
{Cutter}, {Figuera Jaimes}, {Fujii}, {Fruchter}, {Gompertz}, {Jakobsson},
{Hodosan}, {J{\`e}rgensen}, {Kangas}, {Kann}, {Rabus}, {Schr{\o}der},
{Stanway}, {Wijers}, {VINROUGE Collaboration}, {Lipunov}, {Gorbovskoy},
{Kornilov}, {Tyurina}, {Balanutsa}, {Kuznetsov}, {Vlasenko}, {Podesta},
{Lopez}, {Podesta}, {Levato}, {Saffe}, {Mallamaci}, {Budnev}, {Gress},
{Kuvshinov}, {Gorbunov}, {Vladimirov}, {Zimnukhov}, {Gabovich}, {Yurkov},
{Sergienko}, {Rebolo}, {Serra-Ricart}, {Tlatov}, {Ishmuhametova}, {MASTER
Collaboration}, {Abe}, {Aoki}, {Aoki}, {Asakura}, {Baar}, {Barway}, {Bond},
{Doi}, {Finet}, {Fujiyoshi}, {Furusawa}, {Honda}, {Itoh}, {Kanda},
{Kawabata}, {Kawabata}, {Kim}, {Koshida}, {Kuroda}, {Lee}, {Liu},
{Matsubayashi}, {Miyazaki}, {Morihana}, {Morokuma}, {Motohara}, {Murata},
{Nagai}, {Nagashima}, {Nagayama}, {Nakaoka}, {Nakata}, {Ohsawa}, {Ohshima},
{Ohta}, {Okita}, {Saito}, {Saito}, {Sako}, {Sekiguchi}, {Sumi}, {Tajitsu},
{Takahashi}, {Takayama}, {Tamura}, {Tanaka}, {Tanaka}, {Terai}, {Tominaga},
{Tristram}, {Uemura}, {Utsumi}, {Yamaguchi}, {Yasuda}, {Yoshida}, {Zenko},
{J-GEM}, {Adams}, {Anupama}, {Bally}, {Barway}, {Bellm}, {Blagorodnova},
{Cannella}, {Chandra}, {Chatterjee}, {Clarke}, {Cobb}, {Cook}, {Copperwheat},
{De}, {Emery}, {Feindt}, {Foster}, {Fox}, {Frail}, {Fremling}, {Frohmaier},
{Garcia}, {Ghosh}, {Giacintucci}, {Goobar}, {Gottlieb}, {Grefenstette},
{Hallinan}, {Harrison}, {Heida}, {Helou}, {Ho}, {Horesh}, {Hotokezaka}, {Ip},
{Itoh}, {Jacobs}, {Jencson}, {Kasen}, {Kasliwal}, {Kassim}, {Kim}, {Kiran},
{Kuin}, {Kulkarni}, {Kupfer}, {Lau}, {Madsen}, {Mazzali}, {Miller},
{Miyasaka}, {Mooley}, {Myers}, {Nakar}, {Ngeow}, {Nugent}, {Ofek},
{Palliyaguru}, {Pavana}, {Perley}, {Peters}, {Pike}, {Piran}, {Qi}, {Quimby},
{Rana}, {Rosswog}, {Rusu}, {Sadler}, {Van Sistine}, {Sollerman}, {Xu}, {Yan},
{Yatsu}, {Yu}, {Zhang}, {Zhao}, {GROWTH}, {JAGWAR}, {Caltech-NRAO},
{TTU-NRAO}, {NuSTAR Collaborations}, {Chambers}, {Huber}, {Schultz},
{Bulger}, {Flewelling}, {Magnier}, {Lowe}, {Wainscoat}, {Waters}, {Willman},
{Pan-STARRS}, {Ebisawa}, {Hanyu}, {Harita}, {Hashimoto}, {Hidaka}, {Hori},
{Ishikawa}, {Isobe}, {Iwakiri}, {Kawai}, {Kawai}, {Kawamuro}, {Kawase},
{Kitaoka}, {Makishima}, {Matsuoka}, {Mihara}, {Morita}, {Morita}, {Nakahira},
{Nakajima}, {Nakamura}, {Negoro}, {Oda}, {Sakamaki}, {Sasaki}, {Serino},
{Shidatsu}, {Shimomukai}, {Sugawara}, {Sugita}, {Sugizaki}, {Tachibana},
{Takao}, {Tanimoto}, {Tomida}, {Tsuboi}, {Tsunemi}, {Ueda}, {Ueno}, {Yamada},
{Yamaoka}, {Yamauchi}, {Yatabe}, {Yoneyama}, {Yoshii}, {MAXI Team}, {Coward},
{Crisp}, {Macpherson}, {Andreoni}, {Laugier}, {Noysena}, {Klotz}, {Gendre},
{Thierry}, {Turpin}, {Consortium}, {Im}, {Choi}, {Kim}, {Yoon}, {Lim}, {Lee},
{Lee}, {Kim}, {Ko}, {Joe}, {Kwon}, {Kim}, {Lim}, {Choi}, {KU Collaboration},
{Fynbo}, {Malesani}, {Xu}, {Optical Telescope}, {Smartt}, {Jerkstrand},
{Kankare}, {Sim}, {Fraser}, {Inserra}, {Maguire}, {Leloudas}, {Magee},
{Shingles}, {Smith}, {Young}, {Kotak}, {Gal-Yam}, {Lyman}, {Homan},
{Agliozzo}, {Anderson}, {Angus}, {Ashall}, {Barbarino}, {Bauer}, {Berton},
{Botticella}, {Bulla}, {Cannizzaro}, {Cartier}, {Cikota}, {Clark}, {De Cia},
{Della Valle}, {Dennefeld}, {Dessart}, {Dimitriadis}, {Elias-Rosa}, {Firth},
{Fl{\"o}rs}, {Frohmaier}, {Galbany}, {Gonz{\'a}lez-Gait{\'a}n}, {Gromadzki},
{Guti{\'e}rrez}, {Hamanowicz}, {Harmanen}, {Heintz}, {Hernandez}, {Hodgkin},
{Hook}, {Izzo}, {James}, {Jonker}, {Kerzendorf}, {Kostrzewa-Rutkowska},
{Kromer}, {Kuncarayakti}, {Lawrence}, {Manulis}, {Mattila}, {McBrien},
{M{\"u}ller}, {Nordin}, {O'Neill}, {Onori}, {Palmerio}, {Pastorello},
{Patat}, {Pignata}, {Podsiadlowski}, {Razza}, {Reynolds}, {Roy}, {Ruiter},
{Rybicki}, {Salmon}, {Pumo}, {Prentice}, {Seitenzahl}, {Smith}, {Sollerman},
{Sullivan}, {Szegedi}, {Taddia}, {Taubenberger}, {Terreran}, {Van Soelen},
{Vos}, {Walton}, {Wright}, {Wyrzykowski}, {Yaron}, {pre=''(''>ePESSTO},
{Chen}, {Kr{\"u}hler}, {Schady}, {Wiseman}, {Greiner}, {Rau}, {Schweyer},
{Klose}, {Nicuesa Guelbenzu}, {GROND}, {Palliyaguru}, {Tech University},
{Shara}, {Williams}, {Vaisanen}, {Potter}, {Romero Colmenero}, {Crawford},
{Buckley}, {Mao}, {SALT Group}, {D{\'\i}az}, {Macri}, {Garc{\'\i}a Lambas},
{Mendes de Oliveira}, {Nilo Castell{\'o}n}, {Ribeiro}, {S{\'a}nchez},
{Schoenell}, {Abramo}, {Akras}, {Alcaniz}, {Artola}, {Beroiz}, {Bonoli},
{Cabral}, {Camuccio}, {Chavushyan}, {Coelho}, {Colazo}, {Costa-Duarte},
{Cuevas Larenas}, {Dom{\'\i}nguez Romero}, {Dultzin}, {Fern{\'a}ndez},
{Garc{\'\i}a}, {Girardini}, {Gon{\c{c}}alves}, {Gon{\c{c}}alves}, {Gurovich},
{Jim{\'e}nez-Teja}, {Kanaan}, {Lares}, {Lopes de Oliveira}, {L{\'o}pez-Cruz},
{Melia}, {Molino}, {Padilla}, {Pe{\~n}uela}, {Placco}, {Qui{\~n}ones},
{Ram{\'\i}rez Rivera}, {Renzi}, {Riguccini}, {R{\'\i}os-L{\'o}pez},
{Rodriguez}, {Sampedro}, {Schneiter}, {Sodr{\'e}}, {Starck}, {Torres-Flores},
{Tornatore}, {Zadro{\.z}ny}, {Castillo}, {TOROS: Transient Robotic
Observatory of South Collaboration}, {Castro-Tirado}, {Tello}, {Hu}, {Zhang},
{Cunniffe}, {Castell{\'o}n}, {Hiriart}, {Caballero-Garc{\'\i}a},
{Jel{\'\i}nek}, {Kub{\'a}nek}, {P{\'e}rez del Pulgar}, {Park}, {Jeong},
{Castro Cer{\'o}n}, {Pandey}, {Yock}, {Querel}, {Fan}, {Wang}, {BOOTES
Collaboration}, {Beardsley}, {Brown}, {Crosse}, {Emrich}, {Franzen},
{Gaensler}, {Horsley}, {Johnston-Hollitt}, {Kenney}, {Morales}, {Pallot},
{Sokolowski}, {Steele}, {Tingay}, {Trott}, {Walker}, {Wayth}, {Williams},
{Wu}, {Murchison Widefield Array}, {Yoshida}, {Sakamoto}, {Kawakubo},
{Yamaoka}, {Takahashi}, {Asaoka}, {Ozawa}, {Torii}, {Shimizu}, {Tamura},
{Ishizaki}, {Cherry}, {Ricciarini}, {Penacchioni}, {Marrocchesi}, {CALET
Collaboration}, {Pozanenko}, {Volnova}, {Mazaeva}, {Minaev}, {Krugov},
{Kusakin}, {Reva}, {Moskvitin}, {Rumyantsev}, {Inasaridze}, {Klunko},
{Tungalag}, {Schmalz}, {Burhonov}, {IKI-GW Follow-up Collaboration},
{Abdalla}, {Abramowski}, {Aharonian}, {Ait Benkhali}, {Ang{\"u}ner},
{Arakawa}, {Arrieta}, {Aubert}, {Backes}, {Balzer}, {Barnard}, {Becherini},
{Becker Tjus}, {Berge}, {Bernhard}, {Bernl{\"o}hr}, {Blackwell},
{B{\"o}ttcher}, {Boisson}, {Bolmont}, {Bonnefoy}, {Bordas}, {Bregeon},
{Brun}, {Brun}, {Bryan}, {B{\"u}chele}, {Bulik}, {Capasso}, {Caroff},
{Carosi}, {Casanova}, {Cerruti}, {Chakraborty}, {Chaves}, {Chen},
{Chevalier}, {Colafrancesco}, {Condon}, {Conrad}, {Davids}, {Decock}, {Deil},
{Devin}, {deWilt}, {Dirson}, {Djannati-Ata{\"\i}}, {Donath}, {O'C. Drury},
{Dutson}, {Dyks}, {Edwards}, {Egberts}, {Emery}, {Ernenwein}, {Eschbach},
{Farnier}, {Fegan}, {Fernandes}, {Fiasson}, {Fontaine}, {Funk},
{F{\"u}ssling}, {Gabici}, {Gallant}, {Garrigoux}, {Gat{\'e}}, {Giavitto},
{Giebels}, {Glawion}, {Glicenstein}, {Gottschall}, {Grondin}, {Hahn},
{Haupt}, {Hawkes}, {Heinzelmann}, {Henri}, {Hermann}, {Hinton}, {Hofmann},
{Hoischen}, {Holch}, {Holler}, {Horns}, {Ivascenko}, {Iwasaki},
{Jacholkowska}, {Jamrozy}, {Jankowsky}, {Jankowsky}, {Jingo}, {Jouvin},
{Jung-Richardt}, {Kastendieck}, {Katarzy{\'n}ski}, {Katsuragawa},
{Kerszberg}, {Khangulyan}, {Kh{\'e}lifi}, {King}, {Klepser}, {Klochkov},
{Klu{\'z}niak}, {Komin}, {Kosack}, {Krakau}, {Kraus}, {Kr{\"u}ger}, {Laffon},
{Lamanna}, {Lau}, {Lees}, {Lefaucheur}, {Lemi{\`e}re}, {Lemoine-Goumard},
{Lenain}, {Leser}, {Lohse}, {Lorentz}, {Liu}, {Lypova}, {Malyshev},
{Marandon}, {Marcowith}, {Mariaud}, {Marx}, {Maurin}, {Maxted}, {Mayer},
{Meintjes}, {Meyer}, {Mitchell}, {Moderski}, {Mohamed}, {Mohrmann},
{Mor{\r{a}}}, {Moulin}, {Murach}, {Nakashima}, {de Naurois}, {Ndiyavala},
{Niederwanger}, {Niemiec}, {Oakes}, {O'Brien}, {Odaka}, {Ohm}, {Ostrowski},
{Oya}, {Padovani}, {Panter}, {Parsons}, {Pekeur}, {Pelletier}, {Perennes},
{Petrucci}, {Peyaud}, {Piel}, {Pita}, {Poireau}, {Poon}, {Prokhorov},
{Prokoph}, {P{\"u}hlhofer}, {Punch}, {Quirrenbach}, {Raab}, {Rauth},
{Reimer}, {Reimer}, {Renaud}, {de los Reyes}, {Rieger}, {Rinchiuso},
{Romoli}, {Rowell}, {Rudak}, {Rulten}, {Sahakian}, {Saito}, {Sanchez},
{Santangelo}, {Sasaki}, {Schlickeiser}, {Sch{\"u}ssler}, {Schulz},
{Schwanke}, {Schwemmer}, {Seglar-Arroyo}, {Settimo}, {Seyffert}, {Shafi},
{Shilon}, {Shiningayamwe}, {Simoni}, {Sol}, {Spanier}, {Spir-Jacob},
{Stawarz}, {Steenkamp}, {Stegmann}, {Steppa}, {Sushch}, {Takahashi},
{Tavernet}, {Tavernier}, {Taylor}, {Terrier}, {Tibaldo}, {Tiziani},
{Tluczykont}, {Trichard}, {Tsirou}, {Tsuji}, {Tuffs}, {Uchiyama}, {van der
Walt}, {van Eldik}, {van Rensburg}, {van Soelen}, {Vasileiadis}, {Veh},
{Venter}, {Viana}, {Vincent}, {Vink}, {Voisin}, {V{\"o}lk}, {Vuillaume},
{Wadiasingh}, {Wagner}, {Wagner}, {Wagner}, {White}, {Wierzcholska},
{Willmann}, {W{\"o}rnlein}, {Wouters}, {Yang}, {Zaborov}, {Zacharias},
{Zanin}, {Zdziarski}, {Zech}, {Zefi}, {Ziegler}, {Zorn}, {{\.Z}ywucka},
{H.~E.~S.~S. Collaboration}, {Fender}, {Broderick}, {Rowlinson}, {Wijers},
{Stewart}, {ter Veen}, {Shulevski}, {LOFAR Collaboration}, {Kavic},
{Simonetti}, {League}, {Tsai}, {Obenberger}, {Nathaniel}, {Taylor}, {Dowell},
{Liebling}, {Estes}, {Lippert}, {Sharma}, {Vincent}, {Farella}, {Wavelength
Array}, {Abeysekara}, {Albert}, {Alfaro}, {Alvarez}, {Arceo},
{Arteaga-Vel{\'a}zquez}, {Avila Rojas}, {Ayala Solares}, {Barber}, {Becerra
Gonzalez}, {Becerril}, {Belmont-Moreno}, {BenZvi}, {Berley}, {Bernal},
{Braun}, {Brisbois}, {Caballero-Mora}, {Capistr{\'a}n}, {Carrami{\~n}ana},
{Casanova}, {Castillo}, {Cotti}, {Cotzomi}, {Couti{\~n}o de Le{\'o}n}, {De
Le{\'o}n}, {De la Fuente}, {Diaz Hernandez}, {Dichiara}, {Dingus},
{DuVernois}, {D{\'\i}az-V{\'e}lez}, {Ellsworth}, {Engel},
{Enr{\'\i}quez-Rivera}, {Fiorino}, {Fleischhack}, {Fraija},
{Garc{\'\i}a-Gonz{\'a}lez}, {Garfias}, {Gerhardt}, {Gonz{\~o}lez Mu{\~n}oz},
{Gonz{\'a}lez}, {Goodman}, {Hampel-Arias}, {Harding}, {Hernandez},
{Hernandez-Almada}, {Hona}, {H{\"u}ntemeyer}, {Iriarte}, {Jardin-Blicq},
{Joshi}, {Kaufmann}, {Kieda}, {Lara}, {Lauer}, {Lennarz}, {Le{\'o}n Vargas},
{Linnemann}, {Longinotti}, {Raya}, {Luna-Garc{\'\i}a}, {L{\'o}pez-Coto},
{Malone}, {Marinelli}, {Martinez}, {Martinez-Castellanos},
{Mart{\'\i}nez-Castro}, {Mart{\'\i}nez-Huerta}, {Matthews},
{Miranda-Romagnoli}, {Moreno}, {Mostaf{\'a}}, {Nellen}, {Newbold}, {Nisa},
{Noriega-Papaqui}, {Pelayo}, {Pretz}, {P{\'e}rez-P{\'e}rez}, {Ren}, {Rho},
{Rivi{\`e}re}, {Rosa-Gonz{\'a}lez}, {Rosenberg}, {Ruiz-Velasco}, {Salazar},
{Salesa Greus}, {Sandoval}, {Schneider}, {Schoorlemmer}, {Sinnis}, {Smith},
{Springer}, {Surajbali}, {Tibolla}, {Tollefson}, {Torres}, {Ukwatta},
{Weisgarber}, {Westerhoff}, {Wisher}, {Wood}, {Yapici}, {Yodh}, {Younk},
{Zhou}, {{\'A}lvarez}, {HAWC Collaboration}, {Aab}, {Abreu}, {Aglietta},
{Albuquerque}, {Albury}, {Allekotte}, {Almela}, {Alvarez Castillo},
{Alvarez-Mu{\~n}iz}, {Anastasi}, {Anchordoqui}, {Andrada}, {Andringa},
{Aramo}, {Arsene}, {Asorey}, {Assis}, {Avila}, {Badescu}, {Balaceanu},
{Barbato}, {Barreira Luz}, {Becker}, {Bellido}, {Berat}, {Bertaina},
{Bertou}, {Biermann}, {Biteau}, {Blaess}, {Blanco}, {Blazek}, {Bleve},
{Boh{\'a}{\v{c}}ov{\'a}}, {Bonifazi}, {Borodai}, {Botti}, {Brack}, {Brancus},
{Bretz}, {Bridgeman}, {Briechle}, {Buchholz}, {Bueno}, {Buitink}, {Buscemi},
{Caballero-Mora}, {Caccianiga}, {Cancio}, {Canfora}, {Caruso}, {Castellina},
{Catalani}, {Cataldi}, {Cazon}, {Chavez}, {Chinellato}, {Chudoba}, {Clay},
{Cobos Cerutti}, {Colalillo}, {Coleman}, {Collica}, {Coluccia},
{Concei{\c{c}}{\~a}o}, {Consolati}, {Contreras}, {Cooper}, {Coutu},
{Covault}, {Cronin}, {D'Amico}, {Daniel}, {Dasso}, {Daumiller}, {Dawson},
{Day}, {de Almeida}, {de Jong}, {De Mauro}, {de Mello Neto}, {De Mitri}, {de
Oliveira}, {de Souza}, {Debatin}, {Deligny}, {D{\'\i}az Castro}, {Diogo},
{Dobrigkeit}, {D'Olivo}, {Dorosti}, {Dos Anjos}, {Dova}, {Dundovic}, {Ebr},
{Engel}, {Erdmann}, {Erfani}, {Escobar}, {Espadanal}, {Etchegoyen}, {Falcke},
{Farmer}, {Farrar}, {Fauth}, {Fazzini}, {Feldbusch}, {Fenu}, {Fick},
{Figueira}, {Filip{\v{c}}i{\v{c}}}, {Freire}, {Fujii}, {Fuster},
{Ga{\"\i}or}, {Garc{\'\i}a}, {Gat{\'e}}, {Gemmeke}, {Gherghel-Lascu}, {Ghia},
{Giaccari}, {Giammarchi}, {Giller}, {G{\l}as}, {Glaser}, {Golup}, {G{\'o}mez
Berisso}, {G{\'o}mez Vitale}, {Gonz{\'a}lez}, {Gorgi}, {Gottowik}, {Grillo},
{Grubb}, {Guarino}, {Guedes}, {Halliday}, {Hampel}, {Hansen}, {Harari},
{Harrison}, {Harvey}, {Haungs}, {Hebbeker}, {Heck}, {Heimann}, {Herve},
{Hill}, {Hojvat}, {Holt}, {Homola}, {H{\"o}randel}, {Horvath},
{Hrabovsk{\'y}}, {Huege}, {Hulsman}, {Insolia}, {Isar}, {Jandt}, {Johnsen},
{Josebachuili}, {Jurysek}, {K{\"a}{\"a}p{\"a}}, {Kampert}, {Keilhauer},
{Kemmerich}, {Kemp}, {Kieckhafer}, {Klages}, {Kleifges}, {Kleinfeller},
{Krause}, {Krohm}, {Kuempel}, {Kukec Mezek}, {Kunka}, {Kuotb Awad}, {Lago},
{LaHurd}, {Lang}, {Lauscher}, {Legumina}, {Leigui de Oliveira},
{Letessier-Selvon}, {Lhenry-Yvon}, {Link}, {Lo Presti}, {Lopes}, {L{\'o}pez},
{L{\'o}pez Casado}, {Lorek}, {Luce}, {Lucero}, {Malacari}, {Mallamaci},
{Mandat}, {Mantsch}, {Mariazzi}, {Maris}, {Marsella}, {Martello}, {Martinez},
{Mart{\'\i}nez Bravo}, {Mas{\'\i}as Meza}, {Mathes}, {Mathys}, {Matthews},
{Matthiae}, {Mayotte}, {Mazur}, {Medina}, {Medina-Tanco}, {Melo},
{Menshikov}, {Merenda}, {Michal}, {Micheletti}, {Middendorf}, {Miramonti},
{Mitrica}, {Mockler}, {Mollerach}, {Montanet}, {Morello}, {Morlino},
{M{\"u}ller}, {M{\"u}ller}, {Muller}, {M{\"u}ller}, {Mussa}, {Naranjo},
{Nguyen}, {Niculescu-Oglinzanu}, {Niechciol}, {Niemietz}, {Niggemann},
{Nitz}, {Nosek}, {Novotny}, {No{\v{z}}ka}, {N{\'u}{\~n}ez}, {Oikonomou},
{Olinto}, {Palatka}, {Pallotta}, {Papenbreer}, {Parente}, {Parra}, {Paul},
{Pech}, {Pedreira}, {P{\c{e}}kala}, {Pe{\~n}a-Rodriguez}, {Pereira},
{Perlin}, {Perrone}, {Peters}, {Petrera}, {Phuntsok}, {Pierog}, {Pimenta},
{Pirronello}, {Platino}, {Plum}, {Poh}, {Porowski}, {Prado}, {Privitera},
{Prouza}, {Quel}, {Querchfeld}, {Quinn}, {Ramos-Pollan}, {Rautenberg},
{Ravignani}, {Ridky}, {Riehn}, {Risse}, {Ristori}, {Rizi}, {Rodrigues de
Carvalho}, {Rodriguez Fernandez}, {Rodriguez Rojo}, {Roncoroni}, {Roth},
{Roulet}, {Rovero}, {Ruehl}, {Saffi}, {Saftoiu}, {Salamida}, {Salazar},
{Saleh}, {Salina}, {S{\'a}nchez}, {Sanchez-Lucas}, {Santos}, {Santos},
{Sarazin}, {Sarmento}, {Sarmiento-Cano}, {Sato}, {Schauer}, {Scherini},
{Schieler}, {Schimp}, {Schmidt}, {Scholten}, {Schov{\'a}nek}, {Schr{\"o}der},
{Schr{\"o}der}, {Schulz}, {Schumacher}, {Sciutto}, {Segreto}, {Shadkam},
{Shellard}, {Sigl}, {Silli}, {{\v{S}}m{\'\i}da}, {Snow}, {Sommers},
{Sonntag}, {Soriano}, {Squartini}, {Stanca}, {Stani{\v{c}}}, {Stasielak},
{Stassi}, {Stolpovskiy}, {Strafella}, {Streich}, {Suarez},
{Suarez-Dur{\'a}n}, {Sudholz}, {Suomij{\"a}rvi}, {Supanitsky},
{{\v{S}}up{\'\i}k}, {Swain}, {Szadkowski}, {Taboada}, {Taborda},
{Timmermans}, {Todero Peixoto}, {Tomankova}, {Tom{\'e}}, {Torralba Elipe},
{Travnicek}, {Trini}, {Tueros}, {Ulrich}, {Unger}, {Urban}, {Vald{\'e}s
Galicia}, {Vali{\~n}o}, {Valore}, {van Aar}, {van Bodegom}, {van den Berg},
{van Vliet}, {Varela}, {Vargas C{\'a}rdenas}, {V{\'a}zquez}, {Veberi{\v{c}}},
{Ventura}, {Vergara Quispe}, {Verzi}, {Vicha}, {Villase{\~n}or}, {Vorobiov},
{Wahlberg}, {Wainberg}, {Walz}, {Watson}, {Weber}, {Weindl}, {Wiede{\'n}ski},
{Wiencke}, {Wilczy{\'n}ski}, {Wirtz}, {Wittkowski}, {Wundheiler}, {Yang},
{Yushkov}, {Zas}, {Zavrtanik}, {Zavrtanik}, {Zepeda}, {Zimmermann},
{Ziolkowski}, {Zong}, {Zuccarello}, {Pierre Auger Collaboration}, {Kim},
{Schulze}, {Bauer}, {Corral-Santana}, {de Gregorio-Monsalvo},
{Gonz{\'a}lez-L{\'o}pez}, {Hartmann}, {Ishwara-Chandra}, {Mart{\'\i}n},
{Mehner}, {Misra}, {Micha{\l}owski}, {Resmi}, {ALMA Collaboration}, {Paragi},
{Agudo}, {An}, {Beswick}, {Casadio}, {Frey}, {Jonker}, {Kettenis}, {Marcote},
{Moldon}, {Szomoru}, {van Langevelde}, {Yang}, {Euro VLBI Team}, {Cwiek},
{Cwiok}, {Czyrkowski}, {Dabrowski}, {Kasprowicz}, {Mankiewicz}, {Nawrocki},
{Opiela}, {Piotrowski}, {Wrochna}, {Zaremba}, {{\.Z}arnecki}, {Pi of Sky
Collaboration}, {Haggard}, {Nynka}, {Ruan}, {Chandra Team at McGill
University}, {Bland}, {Booler}, {Devillepoix}, {de Gois}, {Hancock}, {Howie},
{Paxman}, {Sansom}, {Towner}, {Desert Fireball Network}, {Tonry}, {Coughlin},
{Stubbs}, {Denneau}, {Heinze}, {Stalder}, {Weiland}, {ATLAS}, {Eatough},
{Kramer}, {Kraus}, {Time Resolution Universe Survey}, {Troja}, {Piro},
{Becerra Gonz{\'a}lez}, {Butler}, {Fox}, {Khandrika}, {Kutyrev}, {Lee},
{Ricci}, {Ryan}, {S{\'a}nchez-Ram{\'\i}rez}, {Veilleux}, {Watson},
{Wieringa}, {Burgess}, {van Eerten}, {Fontes}, {Fryer}, {Korobkin},
{Wollaeger}, {RIMAS}, {RATIR}, {Camilo}, {Foley}, {Goedhart}, {Makhathini},
{Oozeer}, {Smirnov}, {Fender}, {Woudt}, and {South
Africa/MeerKAT}]{170817MMA}
{Abbott}, B.P.; {Abbott}, R.; {Abbott}, T.D.; {Acernese}, F.; {Ackley}, K.;
{Adams}, C.; {Adams}, T.; {Addesso}, P.; {Adhikari}, R.X.; {Adya}, V.B.;
et~al.
\newblock {Multi-messenger Observations of a Binary Neutron Star Merger}.
\newblock {\em Astrophys. J. Lett.} {\bf 2017}, {\em 848},~L12.
%  \href{http://xxx.lanl.gov/abs/1710.05833}{{\normalfont
%  [arXiv:astro-ph.HE/1710.05833]}}. \\
[\href{http://dx.doi.org/10.3847/2041-8213/aa91c9}{CrossRef}]

\bibitem[{Gehrels} \em{et~al.}(2016){Gehrels}, {Cannizzo}, {Kanner},
{Kasliwal}, {Nissanke}, and {Singer}]{Gehrels16}
{Gehrels}, N.; {Cannizzo}, J.K.; {Kanner}, J.; {Kasliwal}, M.M.; {Nissanke},
S.; {Singer}, L.P.
\newblock {Galaxy Strategy for LIGO-Virgo Gravitational Wave Counterpart
Searches}.
\newblock {\em Astrophys. J. Lett.} {\bf 2016}, {\em 820},~136.
%  \href{http://xxx.lanl.gov/abs/1508.03608}{{\normalfont
%  [arXiv:astro-ph.HE/1508.03608]}}. \\
[\href{http://dx.doi.org/10.3847/0004-637X/820/2/136}{CrossRef}]

\bibitem[{Coulter} \em{et~al.}(2017){Coulter}, {Foley}, {Kilpatrick}, {Drout},
{Piro}, {Shappee}, {Siebert}, {Simon}, {Ulloa}, {Kasen}, {Madore},
{Murguia-Berthier}, {Pan}, {Prochaska}, {Ramirez-Ruiz}, {Rest}, and
{Rojas-Bravo}]{Coulter17}
{Coulter}, D.A.; {Foley}, R.J.; {Kilpatrick}, C.D.; {Drout}, M.R.; {Piro},
A.L.; {Shappee}, B.J.; {Siebert}, M.R.; {Simon}, J.D.; {Ulloa}, N.; {Kasen},
D.;  et~al.
\newblock {Swope Supernova Survey 2017a (SSS17a), the optical counterpart to a
gravitational wave source}.
\newblock {\em Science} {\bf 2017}, {\em 358},~1556--1558.
%  \href{http://xxx.lanl.gov/abs/1710.05452}{{\normalfont
%  [arXiv:astro-ph.HE/1710.05452]}}. \\
[\href{http://dx.doi.org/10.1126/science.aap9811}{CrossRef}]

\bibitem[{Grossman} \em{et~al.}(2014){Grossman}, {Korobkin}, {Rosswog}, and
{Piran}]{Grossman14}
{Grossman}, D.; {Korobkin}, O.; {Rosswog}, S.; {Piran}, T.
\newblock {The long-term evolution of neutron star merger remnants - II.
Radioactively powered transients}.
\newblock {\em Mon. Not. R. Astron. Soc.} {\bf 2014}, {\em 439},~757--770.
%  \href{http://xxx.lanl.gov/abs/1307.2943}{{\normalfont
%  [arXiv:astro-ph.HE/1307.2943]}}. \\
[\href{http://dx.doi.org/10.1093/mnras/stt2503}{CrossRef}]

\bibitem[{Arnett}(1982)]{Arnett82}
{Arnett}, W.D.
\newblock {Type I supernovae. I - Analytic solutions for the early part of the
light curve}.
\newblock {\em Astrophys. J. Lett.} {\bf 1982}, {\em 253},~785--797. 
[\href{http://dx.doi.org/10.1086/159681}{CrossRef}]

\bibitem[{Goriely} \em{et~al.}(2011){Goriely}, {Bauswein}, and
{Janka}]{Goriely11}
{Goriely}, S.; {Bauswein}, A.; {Janka}, H.T.
\newblock {r-process Nucleosynthesis in Dynamically Ejected Matter of Neutron
Star Mergers}.
\newblock {\em Astrophys. J. Lett.} {\bf 2011}, {\em 738},~L32.
%  \href{http://xxx.lanl.gov/abs/1107.0899}{{\normalfont
%  [arXiv:astro-ph.SR/1107.0899]}}. \\
[\href{http://dx.doi.org/10.1088/2041-8205/738/2/L32}{CrossRef}]

\bibitem[{Radice} \em{et~al.}(2018){Radice}, {Perego}, {Hotokezaka}, {Fromm},
{Bernuzzi}, and {Roberts}]{Radice18}
{Radice}, D.; {Perego}, A.; {Hotokezaka}, K.; {Fromm}, S.A.; {Bernuzzi}, S.;
{Roberts}, L.F.
\newblock {Binary Neutron Star Mergers: Mass Ejection, Electromagnetic
Counterparts, and Nucleosynthesis}.
\newblock {\em Astrophys. J. Lett.} {\bf 2018}, {\em 869},~130.
%  \href{http://xxx.lanl.gov/abs/1809.11161}{{\normalfont
%  [arXiv:astro-ph.HE/1809.11161]}}. \\
[\href{http://dx.doi.org/10.3847/1538-4357/aaf054}{CrossRef}]

\bibitem[{Shibata} and {Hotokezaka}(2019)]{Shibata19}
{Shibata}, M.; {Hotokezaka}, K.
\newblock {Merger and Mass Ejection of Neutron Star Binaries}.
\newblock {\em Annual Review of Nuclear and Particle Science} {\bf 2019}, {\em
69},~41--64.
%  \href{http://xxx.lanl.gov/abs/1908.02350}{{\normalfont
%  [arXiv:astro-ph.HE/1908.02350]}}. \\
[\href{http://dx.doi.org/10.1146/annurev-nucl-101918-023625}{CrossRef}]

\bibitem[{Korobkin} \em{et~al.}(2021){Korobkin}, {Wollaeger}, {Fryer},
{Hungerford}, {Rosswog}, {Fontes}, {Mumpower}, {Chase}, {Even}, {Miller},
{Misch}, and {Lippuner}]{Korobkin21}
{Korobkin}, O.; {Wollaeger}, R.T.; {Fryer}, C.L.; {Hungerford}, A.L.;
{Rosswog}, S.; {Fontes}, C.J.; {Mumpower}, M.R.; {Chase}, E.A.; {Even}, W.P.;
{Miller}, J.;  et~al.
\newblock {Axisymmetric Radiative Transfer Models of Kilonovae}.
\newblock {\em Astrophys. J. Lett.} {\bf 2021}, {\em 910},~116.
%  \href{http://xxx.lanl.gov/abs/2004.00102}{{\normalfont
%  [arXiv:astro-ph.HE/2004.00102]}}. \\
[\href{http://dx.doi.org/10.3847/1538-4357/abe1b5}{CrossRef}]

\bibitem[{Barnes} \em{et~al.}(2021){Barnes}, {Zhu}, {Lund}, {Sprouse}, {Vassh},
{McLaughlin}, {Mumpower}, and {Surman}]{Barnes21}
{Barnes}, J.; {Zhu}, Y.L.; {Lund}, K.A.; {Sprouse}, T.M.; {Vassh}, N.;
{McLaughlin}, G.C.; {Mumpower}, M.R.; {Surman}, R.
\newblock {Kilonovae Across the Nuclear Physics Landscape: The Impact of
Nuclear Physics Uncertainties on r-process-powered Emission}.
\newblock {\em Astrophys. J. Lett.} {\bf 2021}, {\em 918},~44.
%  \href{http://xxx.lanl.gov/abs/2010.11182}{{\normalfont
%  [arXiv:astro-ph.HE/2010.11182]}}. \\
[\href{http://dx.doi.org/10.3847/1538-4357/ac0aec}{CrossRef}]

\bibitem[{Yu} \em{et~al.}(2013){Yu}, {Zhang}, and {Gao}]{Yu13}
{Yu}, Y.W.; {Zhang}, B.; {Gao}, H.
\newblock {Bright ``Merger-nova'' from the Remnant of a Neutron Star Binary
Merger: A Signature of a Newly Born, Massive, Millisecond Magnetar}.
\newblock {\em Astrophys. J. Lett.} {\bf 2013}, {\em 776},~L40.
%  \href{http://xxx.lanl.gov/abs/1308.0876}{{\normalfont
%  [arXiv:astro-ph.SR/1308.0876]}}. \\
[\href{http://dx.doi.org/10.1088/2041-8205/776/2/L40}{CrossRef}]

\bibitem[{Tarumi} \em{et~al.}(2023){Tarumi}, {Hotokezaka}, {Domoto}, and
{Tanaka}]{Tarumi23}
{Tarumi}, Y.; {Hotokezaka}, K.; {Domoto}, N.; {Tanaka}, M.
\newblock {Non-LTE analysis for Helium and Strontium lines in the kilonova
AT2017gfo}.
\newblock {\em arXiv} {\bf 2023}, arXiv:2302.13061.
%  \href{http://xxx.lanl.gov/abs/2302.13061}{{\normalfont
%  [arXiv:astro-ph.HE/2302.13061]}}.
%\newblock {\url{https://doi.org/10.48550/arXiv.2302.13061}}.

\bibitem[{Gillanders} \em{et~al.}(2021){Gillanders}, {McCann}, {Sim}, {Smartt},
and {Ballance}]{Gillanders21}
{Gillanders}, J.H.; {McCann}, M.; {Sim}, S.A.; {Smartt}, S.J.; {Ballance}, C.P.
\newblock {Constraints on the presence of platinum and gold in the spectra of
the kilonova AT2017gfo}.
\newblock {\em Mon. Not. R. Astron. Soc.} {\bf 2021}, {\em 506},~3560--3577.
%  \href{http://xxx.lanl.gov/abs/2101.08271}{{\normalfont
%  [arXiv:astro-ph.HE/2101.08271]}}. \\
[\href{http://dx.doi.org/10.1093/mnras/stab1861}{CrossRef}]

\bibitem[{Watson} \em{et~al.}(2019){Watson}, {Hansen}, {Selsing}, {Koch},
{Malesani}, {Andersen}, {Fynbo}, {Arcones}, {Bauswein}, {Covino}, {Grado},
{Heintz}, {Hunt}, {Kouveliotou}, {Leloudas}, {Levan}, {Mazzali}, and
{Pian}]{Watson19}
{Watson}, D.; {Hansen}, C.J.; {Selsing}, J.; {Koch}, A.; {Malesani}, D.B.;
{Andersen}, A.C.; {Fynbo}, J.P.U.; {Arcones}, A.; {Bauswein}, A.; {Covino},
S.;  et~al.
\newblock {Identification of strontium in the merger of two neutron stars}.
\newblock {\em {Nature} } {\bf 2019}, {\em 574},~497--500.
%  \href{http://xxx.lanl.gov/abs/1910.10510}{{\normalfont
%  [arXiv:astro-ph.HE/1910.10510]}}. \\
[\href{http://dx.doi.org/10.1038/s41586-019-1676-3}{CrossRef}] [\href{http://www.ncbi.nlm.nih.gov/pubmed/31645733}{PubMed}]

\bibitem[{Pian} \em{et~al.}(2017){Pian}, {D'Avanzo}, {Benetti}, {Branchesi},
{Brocato}, {Campana}, {Cappellaro}, {Covino}, {D'Elia}, {Fynbo}, {Getman},
{Ghirlanda}, {Ghisellini}, {Grado}, {Greco}, {Hjorth}, {Kouveliotou},
{Levan}, {Limatola}, {Malesani}, {Mazzali}, {Melandri}, {M{\o}ller},
{Nicastro}, {Palazzi}, {Piranomonte}, {Rossi}, {Salafia}, {Selsing},
{Stratta}, {Tanaka}, {Tanvir}, {Tomasella}, {Watson}, {Yang}, {Amati},
{Antonelli}, {Ascenzi}, {Bernardini}, {Bo{\"e}r}, {Bufano}, {Bulgarelli},
{Capaccioli}, {Casella}, {Castro-Tirado}, {Chassande-Mottin}, {Ciolfi},
{Copperwheat}, {Dadina}, {De Cesare}, {di Paola}, {Fan}, {Gendre},
{Giuffrida}, {Giunta}, {Hunt}, {Israel}, {Jin}, {Kasliwal}, {Klose}, {Lisi},
{Longo}, {Maiorano}, {Mapelli}, {Masetti}, {Nava}, {Patricelli}, {Perley},
{Pescalli}, {Piran}, {Possenti}, {Pulone}, {Razzano}, {Salvaterra},
{Schipani}, {Spera}, {Stamerra}, {Stella}, {Tagliaferri}, {Testa}, {Troja},
{Turatto}, {Vergani}, and {Vergani}]{Pian17}
{Pian}, E.; {D'Avanzo}, P.; {Benetti}, S.; {Branchesi}, M.; {Brocato}, E.;
{Campana}, S.; {Cappellaro}, E.; {Covino}, S.; {D'Elia}, V.; {Fynbo}, J.P.U.;
et~al.
\newblock {Spectroscopic identification of r-process nucleosynthesis in a
double neutron-star merger}.
\newblock {\em {Nature} } {\bf 2017}, {\em 551},~67--70.
%  \href{http://xxx.lanl.gov/abs/1710.05858}{{\normalfont
%  [arXiv:astro-ph.HE/1710.05858]}}. \\
[\href{http://dx.doi.org/10.1038/nature24298}{CrossRef}]

\bibitem[{Kuin} \em{et~al.}(2015){Kuin}, {Breeveld}, and {Page}]{Kuin05}
{Kuin}, P.; {Breeveld}, A.; {Page}, M.
\newblock {The Swift UVOT grism calibration and example spectra}.
\newblock {\em arXiv} {\bf 2015}, arXiv:1502.07204.
%  \href{http://xxx.lanl.gov/abs/1502.07204}{{\normalfont
%  [arXiv:astro-ph.IM/1502.07204]}}.
\newblock {\url{https://doi.org/10.48550/arXiv.1502.07204}}.

\bibitem[{Ben-Ami} \em{et~al.}(2022){Ben-Ami}, {Shvartzvald}, {Waxman},
{Netzer}, {Yaniv}, {Algranatti}, {Gal-Yam}, {Lapid}, {Ofek}, {Topaz},
{Arcavi}, {Asif}, {Azaria}, {Bahalul}, {Barschke}, {Bastian-Querner},
{Berge}, {Berlea}, {Buehler}, {Dittmar}, {Gelman}, {Giavitto}, {Guttman},
{Haces Crespo}, {Heilbrunn}, {Kachergincky}, {Kaipachery}, {Kowalski},
{Kulkarni}, {Kumar}, {K{\"u}sters}, {Liran}, {Miron-Salomon}, {Mor}, {Nir},
{Nitzan}, {Philipp}, {Porelli}, {Sagiv}, {Schliwinski}, {Sprecher}, {De
Simone}, {Stern}, {Stone}, {Trakhtenbrot}, {Vasilev}, {Watson}, and
{Zappon}]{benami2022}
{Ben-Ami}, S.; {Shvartzvald}, Y.; {Waxman}, E.; {Netzer}, U.; {Yaniv}, Y.;
{Algranatti}, V.M.; {Gal-Yam}, A.; {Lapid}, O.; {Ofek}, E.; {Topaz}, J.;
et~al.
\newblock {The scientific payload of the Ultraviolet Transient Astronomy
Satellite (ULTRASAT)}.
\newblock In Proceedings of the Space Telescopes and Instrumentation 2022:
Ultraviolet to Gamma Ray, Montreal, QC, Canada, 17--22 July 2022; {den Herder}, J.W.A., {Nikzad}, S., {Nakazawa}, K.,
Eds.; Society of Photo-Optical Instrumentation
Engineers (SPIE) Conference Series; Volume 12181, p. 1218105.
%  \href{http://xxx.lanl.gov/abs/2208.00159}{{\normalfont
%  [arXiv:astro-ph.IM/2208.00159]}}. \\
[\href{http://dx.doi.org/10.1117/12.2629850}{CrossRef}]

\bibitem[{Dorsman} \em{et~al.}(2023){Dorsman}, {Raaijmakers}, {Cenko},
{Nissanke}, {Singer}, {Kasliwal}, {Piro}, {Bellm}, {Hartmann}, {Hotokezaka},
and {Luko{\v{s}}i{\={u}}t{\.{e}}}]{dorsman23}
{Dorsman}, B.; {Raaijmakers}, G.; {Cenko}, S.B.; {Nissanke}, S.; {Singer},
L.P.; {Kasliwal}, M.M.; {Piro}, A.L.; {Bellm}, E.C.; {Hartmann}, D.H.;
{Hotokezaka}, K.;  et~al.
\newblock {Prospects of Gravitational-wave Follow-up through a Wide-field
Ultraviolet Satellite: A Dorado Case Study}.
\newblock {\em Astrophys. J. Lett.} {\bf 2023}, {\em 944},~126.
%  \href{http://xxx.lanl.gov/abs/2206.09696}{{\normalfont
%  [arXiv:astro-ph.HE/2206.09696]}}. \\
[\href{http://dx.doi.org/10.3847/1538-4357/acaa9e}{CrossRef}]

\bibitem[{Lien} \em{et~al.}(2016){Lien}, {Sakamoto}, {Barthelmy},
{Baumgartner}, {Cannizzo}, {Chen}, {Collins}, {Cummings}, {Gehrels}, {Krimm},
{Markwardt}, {Palmer}, {Stamatikos}, {Troja}, and {Ukwatta}]{Lien16}
{Lien}, A.; {Sakamoto}, T.; {Barthelmy}, S.D.; {Baumgartner}, W.H.; {Cannizzo},
J.K.; {Chen}, K.; {Collins}, N.R.; {Cummings}, J.R.; {Gehrels}, N.; {Krimm},
H.A.;  et~al.
\newblock {The Third Swift Burst Alert Telescope Gamma-Ray Burst Catalog}.
\newblock {\em Astrophys. J. Lett.} {\bf 2016}, {\em 829},~7.
%  \href{http://xxx.lanl.gov/abs/1606.01956}{{\normalfont
%  [arXiv:astro-ph.HE/1606.01956]}}. \\
[\href{http://dx.doi.org/10.3847/0004-637X/829/1/7}{CrossRef}]

\bibitem[{Sakamoto} \em{et~al.}(2008){Sakamoto}, {Barthelmy}, {Barbier},
{Cummings}, {Fenimore}, {Gehrels}, {Hullinger}, {Krimm}, {Markwardt},
{Palmer}, {Parsons}, {Sato}, {Stamatikos}, {Tueller}, {Ukwatta}, and
{Zhang}]{Sakamoto08}
{Sakamoto}, T.; {Barthelmy}, S.D.; {Barbier}, L.; {Cummings}, J.R.; {Fenimore},
E.E.; {Gehrels}, N.; {Hullinger}, D.; {Krimm}, H.A.; {Markwardt}, C.B.;
{Palmer}, D.M.;  et~al.
\newblock {The First Swift BAT Gamma-Ray Burst Catalog}.
\newblock {\em {Astrophys. J. Suppl.}} {\bf 2008}, {\em 175},~179--190.
%  \href{http://xxx.lanl.gov/abs/0707.4626}{{\normalfont
%  [arXiv:astro-ph/0707.4626]}}. \\
[\href{http://dx.doi.org/10.1086/523646}{CrossRef}]

\bibitem[{Stratta} \em{et~al.}(2007){Stratta}, {D'Avanzo}, {Piranomonte},
{Cutini}, {Preger}, {Perri}, {Conciatore}, {Covino}, {Stella}, {Guetta},
{Marshall}, {Holland}, {Stamatikos}, {Guidorzi}, {Mangano}, {Antonelli},
{Burrows}, {Campana}, {Capalbi}, {Chincarini}, {Cusumano}, {D'Elia}, {Evans},
{Fiore}, {Fugazza}, {Giommi}, {Osborne}, {La Parola}, {Mineo}, {Moretti},
{Page}, {Romano}, and {Tagliaferri}]{Stratta07}
{Stratta}, G.; {D'Avanzo}, P.; {Piranomonte}, S.; {Cutini}, S.; {Preger}, B.;
{Perri}, M.; {Conciatore}, M.L.; {Covino}, S.; {Stella}, L.; {Guetta}, D.;
et~al.
\newblock {A study of the prompt and afterglow emission of the short GRB
061201}.
\newblock {\em Astron. Astrophys.} {\bf 2007}, {\em 474},~827--835.
  [\href{http://dx.doi.org/10.1051/0004-6361:20078006}{CrossRef}]

\bibitem[{Antonelli} \em{et~al.}(2009){Antonelli}, {D'Avanzo}, {Perna},
{Amati}, {Covino}, {Cutini}, {D'Elia}, {Gallozzi}, {Grazian}, {Palazzi},
{Piranomonte}, {Rossi}, {Spiro}, {Stella}, {Testa}, {Chincarini}, {di Paola},
{Fiore}, {Fugazza}, {Giallongo}, {Maiorano}, {Masetti}, {Pedichini},
{Salvaterra}, {Tagliaferri}, and {Vergani}]{Antonelli09}
{Antonelli}, L.A.; {D'Avanzo}, P.; {Perna}, R.; {Amati}, L.; {Covino}, S.;
{Cutini}, S.; {D'Elia}, V.; {Gallozzi}, S.; {Grazian}, A.; {Palazzi}, E.;
et~al.
\newblock {GRB 090426: The farthest short gamma-ray burst?}
\newblock {\em Astron. Astrophys.} {\bf 2009}, {\em 507},~L45--L48.
%  \href{http://xxx.lanl.gov/abs/0911.0046}{{\normalfont
%  [arXiv:astro-ph.HE/0911.0046]}}. \\
[\href{http://dx.doi.org/10.1051/0004-6361/200913062}{CrossRef}]

\bibitem[{O'Connor} \em{et~al.}(2022){O'Connor}, {Troja}, {Dichiara},
{Beniamini}, {Cenko}, {Kouveliotou}, {Gonz{\'a}lez}, {Durbak}, {Gatkine},
{Kutyrev}, {Sakamoto}, {S{\'a}nchez-Ram{\'\i}rez}, and {Veilleux}]{Oconnor22}
{O'Connor}, B.; {Troja}, E.; {Dichiara}, S.; {Beniamini}, P.; {Cenko}, S.B.;
{Kouveliotou}, C.; {Gonz{\'a}lez}, J.B.; {Durbak}, J.; {Gatkine}, P.;
{Kutyrev}, A.;  et~al.
\newblock {A deep survey of short GRB host galaxies over z 0-2: implications
for offsets, redshifts, and environments}.
\newblock {\em Mon. Not. R. Astron. Soc.} {\bf 2022}, {\em 515},~4890--4928.
%  \href{http://xxx.lanl.gov/abs/2204.09059}{{\normalfont
%  [arXiv:astro-ph.HE/2204.09059]}}. \\
[\href{http://dx.doi.org/10.1093/mnras/stac1982}{CrossRef}]

\bibitem[{Troja} \em{et~al.}(2019){Troja}, {van Eerten}, {Ryan}, {Ricci},
{Burgess}, {Wieringa}, {Piro}, {Cenko}, and {Sakamoto}]{Troja19a}
{Troja}, E.; {van Eerten}, H.; {Ryan}, G.; {Ricci}, R.; {Burgess}, J.M.;
{Wieringa}, M.H.; {Piro}, L.; {Cenko}, S.B.; {Sakamoto}, T.
\newblock {A year in the life of GW 170817: The rise and fall of a structured
jet from a binary neutron star merger}.
\newblock {\em Mon. Not. R. Astron. Soc.} {\bf 2019}, {\em 489},~1919--1926.
%  \href{http://xxx.lanl.gov/abs/1808.06617}{{\normalfont
%  [arXiv:astro-ph.HE/1808.06617]}}. \\
[\href{http://dx.doi.org/10.1093/mnras/stz2248}{CrossRef}]

\bibitem[{Burns} \em{et~al.}(2018){Burns}, {Veres}, {Connaughton}, {Racusin},
{Briggs}, {Christensen}, {Goldstein}, {Hamburg}, {Kocevski}, {McEnery},
{Bissaldi}, {Dal Canton}, {Cleveland}, {Gibby}, {Hui}, {von Kienlin},
{Mailyan}, {Paciesas}, {Roberts}, {Siellez}, {Stanbro}, and
{Wilson-Hodge}]{Burns18}
{Burns}, E.; {Veres}, P.; {Connaughton}, V.; {Racusin}, J.; {Briggs}, M.S.;
{Christensen}, N.; {Goldstein}, A.; {Hamburg}, R.; {Kocevski}, D.; {McEnery},
J.;  et~al.
\newblock {Fermi GBM Observations of GRB 150101B: A Second Nearby Event with a
Short Hard Spike and a Soft Tail}.
\newblock {\em Astrophys. J. Lett.} {\bf 2018}, {\em 863},~L34.
%  \href{http://xxx.lanl.gov/abs/1807.02866}{{\normalfont
%  [arXiv:astro-ph.HE/1807.02866]}}. \\
[\href{http://dx.doi.org/10.3847/2041-8213/aad813}{CrossRef}]

\bibitem[{Mooley} \em{et~al.}(2022){Mooley}, {Anderson}, and {Lu}]{Mooley22}
{Mooley}, K.P.; {Anderson}, J.; {Lu}, W.
\newblock {Optical superluminal motion measurement in the neutron-star merger
GW170817}.
\newblock {\em {Nature} } {\bf 2022}, {\em 610},~273--276.
%  \href{http://xxx.lanl.gov/abs/2210.06568}{{\normalfont
%  [arXiv:astro-ph.HE/2210.06568]}}. \\
[\href{http://dx.doi.org/10.1038/s41586-022-05145-7}{CrossRef}]

\bibitem[{Ricci} \em{et~al.}(2021){Ricci}, {Troja}, {Bruni}, {Matsumoto},
{Piro}, {O'Connor}, {Piran}, {Navaieelavasani}, {Corsi}, {Giacomazzo}, and
{Wieringa}]{Ricci21}
{Ricci}, R.; {Troja}, E.; {Bruni}, G.; {Matsumoto}, T.; {Piro}, L.; {O'Connor},
B.; {Piran}, T.; {Navaieelavasani}, N.; {Corsi}, A.; {Giacomazzo}, B.;
et~al.
\newblock {Searching for the radio remnants of short-duration gamma-ray
bursts}.
\newblock {\em Mon. Not. R. Astron. Soc.} {\bf 2021}, {\em 500},~1708--1720.
%  \href{http://xxx.lanl.gov/abs/2008.03659}{{\normalfont
%  [arXiv:astro-ph.HE/2008.03659]}}. \\
[\href{http://dx.doi.org/10.1093/mnras/staa3241}{CrossRef}]

\bibitem[{Dichiara} \em{et~al.}(2020){Dichiara}, {Troja}, {O'Connor},
{Marshall}, {Beniamini}, {Cannizzo}, {Lien}, and {Sakamoto}]{Dichiara20}
{Dichiara}, S.; {Troja}, E.; {O'Connor}, B.; {Marshall}, F.E.; {Beniamini}, P.;
{Cannizzo}, J.K.; {Lien}, A.Y.; {Sakamoto}, T.
\newblock {Short gamma-ray bursts within 200 Mpc}.
\newblock {\em Mon. Not. R. Astron. Soc.} {\bf 2020}, {\em 492},~5011--5022.
%  \href{http://xxx.lanl.gov/abs/1912.08698}{{\normalfont
%  [arXiv:astro-ph.HE/1912.08698]}}. \\
[\href{http://dx.doi.org/10.1093/mnras/staa124}{CrossRef}]

\bibitem[{Bartos} \em{et~al.}(2019){Bartos}, {Lee}, {Corsi}, {M{\'a}rka}, and
{M{\'a}rka}]{Bartos19}
{Bartos}, I.; {Lee}, K.H.; {Corsi}, A.; {M{\'a}rka}, Z.; {M{\'a}rka}, S.
\newblock {Radio forensics could unmask nearby off-axis gamma-ray bursts}.
\newblock {\em Mon. Not. R. Astron. Soc.} {\bf 2019}, {\em 485},~4150--4159.
%  \href{http://xxx.lanl.gov/abs/1811.11260}{{\normalfont
%  [arXiv:astro-ph.HE/1811.11260]}}. \\
[\href{http://dx.doi.org/10.1093/mnras/stz719}{CrossRef}]

\bibitem[{D{\'a}lya} \em{et~al.}(2018){D{\'a}lya}, {Galg{\'o}czi}, {Dobos},
{Frei}, {Heng}, {Macas}, {Messenger}, {Raffai}, and {de Souza}]{Glade18}
{D{\'a}lya}, G.; {Galg{\'o}czi}, G.; {Dobos}, L.; {Frei}, Z.; {Heng}, I.S.;
{Macas}, R.; {Messenger}, C.; {Raffai}, P.; {de Souza}, R.S.
\newblock {GLADE: A galaxy catalogue for multimessenger searches in the
advanced gravitational-wave detector era}.
\newblock {\em Mon. Not. R. Astron. Soc.} {\bf 2018}, {\em 479},~2374--2381.
%  \href{http://xxx.lanl.gov/abs/1804.05709}{{\normalfont
%  [arXiv:astro-ph.HE/1804.05709]}}. \\
[\href{http://dx.doi.org/10.1093/mnras/sty1703}{CrossRef}]

\bibitem[{Evans} \em{et~al.}(2019){Evans}, {Tohuvavohu}, and {Neil Gehrels
Swift Observatory Team}]{GCN190610}
{Evans}, P.A.; {Tohuvavohu}, A.; {Neil Gehrels Swift Observatory Team}.
\newblock {GRB 190610A: Swift detection of a short burst}.
\newblock {\em GRB Coord. Netw.} {\bf 2019}, {\em 24775},~1.

\bibitem[{LIGO Scientific Collaboration} and {Virgo
Collaboration}(2021)]{LVC21}
{LIGO Scientific Collaboration}; {Virgo Collaboration}.
\newblock {Search for Gravitational Waves Associated with Gamma-Ray Bursts
Detected by Fermi and Swift during the LIGO-Virgo Run O3a}.
\newblock {\em Astrophys. J. Lett.} {\bf 2021}, {\em 915},~86.
%  \href{http://xxx.lanl.gov/abs/2010.14550}{{\normalfont
%  [arXiv:astro-ph.HE/2010.14550]}}. \\
[\href{http://dx.doi.org/10.3847/1538-4357/abee15}{CrossRef}]

\bibitem[{Ghosh} \em{et~al.}(2022){Ghosh}, {Vaishnava}, {Resmi}, {Misra},
{Arun}, {Omar}, and {Chakradhari}]{Ghosh22}
{Ghosh}, A.; {Vaishnava}, C.S.; {Resmi}, L.; {Misra}, K.; {Arun}, K.G.; {Omar},
A.; {Chakradhari}, N.K.
\newblock {Search for merger ejecta emission from late time radio observations
of short GRBs using GMRT}.
\newblock {\em arXiv } {\bf 2022}, arXiv:2207.10001.
%  \href{http://xxx.lanl.gov/abs/2207.10001}{{\normalfont
%  [arXiv:astro-ph.HE/2207.10001]}}.
%\newblock {\url{https://doi.org/10.48550/arXiv.2207.10001}}.

\bibitem[{Bruni} \em{et~al.}(2021){Bruni}, {O'Connor}, {Matsumoto}, {Troja},
{Piran}, {Piro}, and {Ricci}]{Bruni21}
{Bruni}, G.; {O'Connor}, B.; {Matsumoto}, T.; {Troja}, E.; {Piran}, T.; {Piro},
L.; {Ricci}, R.
\newblock {Late-time radio observations of the short GRB 200522A: Constraints
on the magnetar model}.
\newblock {\em Mon. Not. R. Astron. Soc.} {\bf 2021}, {\em 505},~L41--L45.
%  \href{http://xxx.lanl.gov/abs/2105.01440}{{\normalfont
%  [arXiv:astro-ph.HE/2105.01440]}}. \\
[\href{http://dx.doi.org/10.1093/mnrasl/slab046}{CrossRef}]

\bibitem[{Fong} \em{et~al.}(2016){Fong}, {Metzger}, {Berger}, and
{{\"O}zel}]{Fong16}
{Fong}, W.; {Metzger}, B.D.; {Berger}, E.; {{\"O}zel}, F.
\newblock {Radio Constraints on Long-lived Magnetar Remnants in Short Gamma-Ray
Bursts}.
\newblock {\em Astrophys. J. Lett.} {\bf 2016}, {\em 831},~141.
%  \href{http://xxx.lanl.gov/abs/1607.00416}{{\normalfont
%  [arXiv:astro-ph.HE/1607.00416]}}. \\
[\href{http://dx.doi.org/10.3847/0004-637X/831/2/141}{CrossRef}]

\bibitem[{Nakar} and {Piran}(2011)]{NP11}
{Nakar}, E.; {Piran}, T.
\newblock {Detectable radio flares following gravitational waves from mergers
of binary neutron stars}.
\newblock {\em {Nature} } {\bf 2011}, {\em 478},~82--84.
%  \href{http://xxx.lanl.gov/abs/1102.1020}{{\normalfont
%  [arXiv:astro-ph.HE/1102.1020]}}. \\
[\href{http://dx.doi.org/10.1038/nature10365}{CrossRef}]

\bibitem[{Wei} \em{et~al.}(2016){Wei}, {Cordier}, {Antier}, {Antilogus},
{Atteia}, {Bajat}, {Basa}, {Beckmann}, {Bernardini}, {Boissier}, {Bouchet},
{Burwitz}, {Claret}, {Dai}, {Daigne}, {Deng}, {Dornic}, {Feng}, {Foglizzo},
{Gao}, {Gehrels}, {Godet}, {Goldwurm}, {Gonzalez}, {Gosset}, {G{\"o}tz},
{Gouiffes}, {Grise}, {Gros}, {Guilet}, {Han}, {Huang}, {Huang}, {Jouret},
{Klotz}, {La Marle}, {Lachaud}, {Le Floch}, {Lee}, {Leroy}, {Li}, {Li}, {Li},
{Liang}, {Lyu}, {Mercier}, {Migliori}, {Mochkovitch}, {O'Brien}, {Osborne},
{Paul}, {Perinati}, {Petitjean}, {Piron}, {Qiu}, {Rau}, {Rodriguez},
{Schanne}, {Tanvir}, {Vangioni}, {Vergani}, {Wang}, {Wang}, {Wang}, {Wang},
{Watson}, {Webb}, {Wei}, {Willingale}, {Wu}, {Wu}, {Xin}, {Xu}, {Yu}, {Yu},
{Yu}, {Zhang}, {Zhang}, {Zhang}, and {Zhou}]{Wei16}
{Wei}, J.; {Cordier}, B.; {Antier}, S.; {Antilogus}, P.; {Atteia}, J.L.;
{Bajat}, A.; {Basa}, S.; {Beckmann}, V.; {Bernardini}, M.G.; {Boissier}, S.;
et~al.
\newblock {The Deep and Transient Universe in the SVOM Era: New Challenges and
Opportunities---Scientific prospects of the SVOM mission}.
\newblock {\em arXiv} {\bf 2016}, arXiv:1610.06892.
%  \href{http://xxx.lanl.gov/abs/1610.06892}{{\normalfont
%  [arXiv:astro-ph.IM/1610.06892]}}.
%\newblock {\url{https://doi.org/10.48550/arXiv.1610.06892}}.

\bibitem[{Yuan} \em{et~al.}(2015){Yuan}, {Zhang}, {Feng}, {Zhang}, {Ling},
{Zhao}, {Deng}, {Qiu}, {Osborne}, {O'Brien}, {Willingale}, {Lapington},
{Fraser}, and {the Einstein Probe team}]{Weimin15}
{Yuan}, W.; {Zhang}, C.; {Feng}, H.; {Zhang}, S.N.; {Ling}, Z.X.; {Zhao}, D.;
{Deng}, J.; {Qiu}, Y.; {Osborne}, J.P.; {O'Brien}, P.;  et~al.
\newblock {Einstein Probe---A small mission to monitor and explore the dynamic
X-ray Universe}.
\newblock {\em arXiv} {\bf 2015}, arXiv:1506.07735.
%  \href{http://xxx.lanl.gov/abs/1506.07735}{{\normalfont
%  [arXiv:astro-ph.HE/1506.07735]}}.
%\newblock {\url{https://doi.org/10.48550/arXiv.1506.07735}}.

\bibitem[{Schady} \em{et~al.}(2010){Schady}, {Page}, {Oates}, {Still}, {de
Pasquale}, {Dwelly}, {Kuin}, {Holland}, {Marshall}, and {Roming}]{Schady10}
{Schady}, P.; {Page}, M.J.; {Oates}, S.R.; {Still}, M.; {de Pasquale}, M.;
{Dwelly}, T.; {Kuin}, N.P.M.; {Holland}, S.T.; {Marshall}, F.E.; {Roming},
P.W.A.
\newblock {Dust and metal column densities in gamma-ray burst host galaxies}.
\newblock {\em Mon. Not. R. Astron. Soc.} {\bf 2010}, {\em 401},~2773--2792.
%  \href{http://xxx.lanl.gov/abs/0910.2590}{{\normalfont
%  [arXiv:astro-ph.CO/0910.2590]}}. \\
[\href{http://dx.doi.org/10.1111/j.1365-2966.2009.15861.x}{CrossRef}]

\bibitem[{Starling} \em{et~al.}(2007){Starling}, {Wijers}, {Wiersema}, {Rol},
{Curran}, {Kouveliotou}, {van der Horst}, and {Heemskerk}]{Starling07}
{Starling}, R.L.C.; {Wijers}, R.A.M.J.; {Wiersema}, K.; {Rol}, E.; {Curran},
P.A.; {Kouveliotou}, C.; {van der Horst}, A.J.; {Heemskerk}, M.H.M.
\newblock {Gamma-Ray Burst Afterglows as Probes of Environment and Blast Wave
Physics. I. Absorption by Host-Galaxy Gas and Dust}.
\newblock {\em Astrophys. J. Lett.} {\bf 2007}, {\em 661},~787--800.
%  \href{http://xxx.lanl.gov/abs/astro-ph/0610899}{{\normalfont
%  [arXiv:astro-ph/astro-ph/0610899]}}. \\
[\href{http://dx.doi.org/10.1086/511953}{CrossRef}]

\bibitem[{Kann} \em{et~al.}(2006){Kann}, {Klose}, and {Zeh}]{Kann06}
{Kann}, D.A.; {Klose}, S.; {Zeh}, A.
\newblock {Signatures of Extragalactic Dust in Pre-Swift GRB Afterglows}.
\newblock {\em Astrophys. J. Lett.} {\bf 2006}, {\em 641},~993--1009.
%  \href{http://xxx.lanl.gov/abs/astro-ph/0512575}{{\normalfont
%  [arXiv:astro-ph/astro-ph/0512575]}}. \\
[\href{http://dx.doi.org/10.1086/500652}{CrossRef}]

\bibitem[{O'Connor} \em{et~al.}(2021){O'Connor}, {Troja}, {Dichiara}, {Chase},
{Ryan}, {Cenko}, {Fryer}, {Ricci}, {Marshall}, {Kouveliotou}, {Wollaeger},
{Fontes}, {Korobkin}, {Gatkine}, {Kutyrev}, {Veilleux}, {Kawai}, and
{Sakamoto}]{Oconnor21}
{O'Connor}, B.; {Troja}, E.; {Dichiara}, S.; {Chase}, E.A.; {Ryan}, G.;
{Cenko}, S.B.; {Fryer}, C.L.; {Ricci}, R.; {Marshall}, F.; {Kouveliotou}, C.;
et~al.
\newblock {A tale of two mergers: Constraints on kilonova detection in two
short GRBs at z {\ensuremath{\sim}} 0.5}.
\newblock {\em Mon. Not. R. Astron. Soc.} {\bf 2021}, {\em 502},~1279--1298.
%  \href{http://xxx.lanl.gov/abs/2012.00026}{{\normalfont
%  [arXiv:astro-ph.HE/2012.00026]}}. \\
[\href{http://dx.doi.org/10.1093/mnras/stab132}{CrossRef}]

\bibitem[{Troja} \em{et~al.}(2016){Troja}, {Sakamoto}, {Cenko}, {Lien},
{Gehrels}, {Castro-Tirado}, {Ricci}, {Capone}, {Toy}, {Kutyrev}, {Kawai},
{Cucchiara}, {Fruchter}, {Gorosabel}, {Jeong}, {Levan}, {Perley},
{Sanchez-Ramirez}, {Tanvir}, and {Veilleux}]{Troja16}
{Troja}, E.; {Sakamoto}, T.; {Cenko}, S.B.; {Lien}, A.; {Gehrels}, N.;
{Castro-Tirado}, A.J.; {Ricci}, R.; {Capone}, J.; {Toy}, V.; {Kutyrev}, A.;
et~al.
\newblock {An Achromatic Break in the Afterglow of the Short GRB 140903A:
Evidence for a Narrow Jet}.
\newblock {\em Astrophys. J. Lett.} {\bf 2016}, {\em 827},~102.
%  \href{http://xxx.lanl.gov/abs/1605.03573}{{\normalfont
%  [arXiv:astro-ph.HE/1605.03573]}}. \\
[\href{http://dx.doi.org/10.3847/0004-637X/827/2/102}{CrossRef}]

\bibitem[{Fong} \em{et~al.}(2014){Fong}, {Berger}, {Metzger}, {Margutti},
{Chornock}, {Migliori}, {Foley}, {Zauderer}, {Lunnan}, {Laskar}, {Desch},
{Meech}, {Sonnett}, {Dickey}, {Hedlund}, and {Harding}]{Fong14}
{Fong}, W.; {Berger}, E.; {Metzger}, B.D.; {Margutti}, R.; {Chornock}, R.;
{Migliori}, G.; {Foley}, R.J.; {Zauderer}, B.A.; {Lunnan}, R.; {Laskar}, T.;
et~al.
\newblock {Short GRB 130603B: Discovery of a Jet Break in the Optical and Radio
Afterglows, and a Mysterious Late-time X-Ray Excess}.
\newblock {\em Astrophys. J. Lett.} {\bf 2014}, {\em 780},~118.
%  \href{http://xxx.lanl.gov/abs/1309.7479}{{\normalfont
%  [arXiv:astro-ph.HE/1309.7479]}}. \\
[\href{http://dx.doi.org/10.1088/0004-637X/780/2/118}{CrossRef}]

\bibitem[{Berger} \em{et~al.}(2009){Berger}, {Cenko}, {Fox}, and
{Cucchiara}]{Berger09}
{Berger}, E.; {Cenko}, S.B.; {Fox}, D.B.; {Cucchiara}, A.
\newblock {Discovery of the Very Red Near-Infrared and Optical Afterglow of the
Short-Duration GRB 070724A}.
\newblock {\em Astrophys. J. Lett.} {\bf 2009}, {\em 704},~877--882.
%  \href{http://xxx.lanl.gov/abs/0908.0940}{{\normalfont
%  [arXiv:astro-ph.HE/0908.0940]}}. \\
[\href{http://dx.doi.org/10.1088/0004-637X/704/1/877}{CrossRef}]

\bibitem[{Soderberg} \em{et~al.}(2006){Soderberg}, {Berger}, {Kasliwal},
{Frail}, {Price}, {Schmidt}, {Kulkarni}, {Fox}, {Cenko}, {Gal-Yam}, {Nakar},
and {Roth}]{Soderberg06}
{Soderberg}, A.M.; {Berger}, E.; {Kasliwal}, M.; {Frail}, D.A.; {Price}, P.A.;
{Schmidt}, B.P.; {Kulkarni}, S.R.; {Fox}, D.B.; {Cenko}, S.B.; {Gal-Yam}, A.;
et~al.
\newblock {The Afterglow, Energetics, and Host Galaxy of the Short-Hard
Gamma-Ray Burst 051221a}.
\newblock {\em Astrophys. J. Lett.} {\bf 2006}, {\em 650},~261--271.
%  \href{http://xxx.lanl.gov/abs/astro-ph/0601455}{{\normalfont
%  [arXiv:astro-ph/astro-ph/0601455]}}. \\
[\href{http://dx.doi.org/10.1086/506429}{CrossRef}]

\bibitem[{Roming} \em{et~al.}(2006){Roming}, {Vanden Berk}, {Pal'shin},
{Pagani}, {Norris}, {Kumar}, {Krimm}, {Holland}, {Gronwall}, {Blustin},
{Zhang}, {Schady}, {Sakamoto}, {Osborne}, {Nousek}, {Marshall},
{M{\'e}sz{\'a}ros}, {Golenetskii}, {Gehrels}, {Frederiks}, {Campana},
{Burrows}, {Boyd}, {Barthelmy}, and {Aptekar}]{Roming06}
{Roming}, P.W.A.; {Vanden Berk}, D.; {Pal'shin}, V.; {Pagani}, C.; {Norris},
J.; {Kumar}, P.; {Krimm}, H.; {Holland}, S.T.; {Gronwall}, C.; {Blustin},
A.J.;  et~al.
\newblock {GRB 060313: A New Paradigm for Short-Hard Bursts?}
\newblock {\em Astrophys. J. Lett.} {\bf 2006}, {\em 651},~985--993.
%  \href{http://xxx.lanl.gov/abs/astro-ph/0605005}{{\normalfont
%  [arXiv:astro-ph/astro-ph/0605005]}}. \\
[\href{http://dx.doi.org/10.1086/508054}{CrossRef}]

\bibitem[{De Pasquale} \em{et~al.}(2010){De Pasquale}, {Schady}, {Kuin},
{Page}, {Curran}, {Zane}, {Oates}, {Holland}, {Breeveld}, {Hoversten},
{Chincarini}, {Grupe}, {Abdo}, {Ackermann}, {Ajello}, {Axelsson}, {Baldini},
{Ballet}, {Barbiellini}, {Baring}, {Bastieri}, {Bechtol}, {Bellazzini},
{Berenji}, {Bissaldi}, {Blandford}, {Bloom}, {Bonamente}, {Borgland},
{Bouvier}, {Bregeon}, {Brez}, {Briggs}, {Brigida}, {Bruel}, {Burnett},
{Buson}, {Caliandro}, {Cameron}, {Caraveo}, {Carrigan}, {Casandjian},
{Cecchi}, {{\c{C}}elik}, {Chekhtman}, {Chiang}, {Ciprini}, {Claus},
{Cohen-Tanugi}, {Connaughton}, {Conrad}, {Dermer}, {de Angelis}, {de Palma},
{Dingus}, {Silva}, {Drell}, {Dubois}, {Dumora}, {Farnier}, {Favuzzi},
{Fegan}, {Fishman}, {Focke}, {Frailis}, {Fukazawa}, {Funk}, {Fusco},
{Gargano}, {Gasparrini}, {Gehrels}, {Germani}, {Giglietto}, {Giordano},
{Glanzman}, {Godfrey}, {Granot}, {Greiner}, {Grenier}, {Grove}, {Guillemot},
{Guiriec}, {Harding}, {Hayashida}, {Hays}, {Horan}, {Hughes}, {Jackson},
{J{\'o}hannesson}, {Johnson}, {Johnson}, {Kamae}, {Katagiri}, {Kataoka},
{Kawai}, {Kerr}, {Kippen}, {Kn{\"o}dlseder}, {Kocevski}, {Kuss}, {Lande},
{Latronico}, {Lemoine-Goumard}, {Longo}, {Loparco}, {Lott}, {Lovellette},
{Lubrano}, {Makeev}, {Mazziotta}, {McEnery}, {McGlynn}, {Meegan},
{M{\'e}sz{\'a}ros}, {Meurer}, {Michelson}, {Mitthumsiri}, {Mizuno}, {Monte},
{Monzani}, {Moretti}, {Morselli}, {Moskalenko}, {Murgia}, {Nolan}, {Norris},
{Nuss}, {Ohno}, {Ohsugi}, {Omodei}, {Orlando}, {Ormes}, {Paciesas},
{Paneque}, {Panetta}, {Parent}, {Pelassa}, {Pepe}, {Pesce-Rollins}, {Piron},
{Porter}, {Preece}, {Rain{\`o}}, {Rando}, {Razzano}, {Reimer}, {Reimer},
{Reposeur}, {Ritz}, {Rochester}, {Rodriguez}, {Roth}, {Ryde}, {Sadrozinski},
{Sander}, {Saz Parkinson}, {Scargle}, {Schalk}, {Sgr{\`o}}, {Siskind},
{Smith}, {Spandre}, {Spinelli}, {Stamatikos}, {Starck}, {Stecker},
{Strickman}, {Suson}, {Tajima}, {Takahashi}, {Tanaka}, {Thayer}, {Thayer},
{Thompson}, {Tibaldo}, {Toma}, {Torres}, {Tosti}, {Tramacere}, {Uchiyama},
{Uehara}, {Usher}, {van der Horst}, {Vasileiou}, {Vilchez}, {Vitale}, {von
Kienlin}, {Waite}, {Wang}, {Winer}, {Wood}, {Wu}, {Yamazaki}, {Ylinen}, and
{Ziegler}]{DePasquale10}
{De Pasquale}, M.; {Schady}, P.; {Kuin}, N.P.M.; {Page}, M.J.; {Curran}, P.A.;
{Zane}, S.; {Oates}, S.R.; {Holland}, S.T.; {Breeveld}, A.A.; {Hoversten},
E.A.;  et~al.
\newblock {Swift and Fermi Observations of the Early Afterglow of the Short
Gamma-Ray Burst 090510}.
\newblock {\em Astrophys. J. Lett.} {\bf 2010}, {\em 709},~L146--L151.
%  \href{http://xxx.lanl.gov/abs/0910.1629}{{\normalfont
%  [arXiv:astro-ph.HE/0910.1629]}}. \\
[\href{http://dx.doi.org/10.1088/2041-8205/709/2/L146}{CrossRef}]

\bibitem[{Galama} and {Wijers}(2001)]{Galama01}
{Galama}, T.J.; {Wijers}, R.A.M.J.
\newblock {High Column Densities and Low Extinctions of Gamma-Ray Bursts:
Evidence for Hypernovae and Dust Destruction}.
\newblock {\em Astrophys. J. Lett.} {\bf 2001}, {\em 549},~L209--L213.
%  \href{http://xxx.lanl.gov/abs/astro-ph/0009367}{{\normalfont
%  [arXiv:astro-ph/astro-ph/0009367]}}. \\
[\href{http://dx.doi.org/10.1086/319162}{CrossRef}]

\bibitem[{Stratta} \em{et~al.}(2004){Stratta}, {Fiore}, {Antonelli}, {Piro},
and {De Pasquale}]{Stratta04}
{Stratta}, G.; {Fiore}, F.; {Antonelli}, L.A.; {Piro}, L.; {De Pasquale}, M.
\newblock {Absorption in Gamma-Ray Burst Afterglows}.
\newblock {\em Astrophys. J. Lett.} {\bf 2004}, {\em 608},~846--864.
%  \href{http://xxx.lanl.gov/abs/astro-ph/0403149}{{\normalfont
%  [arXiv:astro-ph/astro-ph/0403149]}}. \\
[\href{http://dx.doi.org/10.1086/420836}{CrossRef}]

\bibitem[{Nowak} \em{et~al.}(2012){Nowak}, {Neilsen}, {Markoff}, {Baganoff},
{Porquet}, {Grosso}, {Levin}, {Houck}, {Eckart}, {Falcke}, {Ji}, {Miller},
and {Wang}]{Nowak12}
{Nowak}, M.A.; {Neilsen}, J.; {Markoff}, S.B.; {Baganoff}, F.K.; {Porquet}, D.;
{Grosso}, N.; {Levin}, Y.; {Houck}, J.; {Eckart}, A.; {Falcke}, H.;  et~al.
\newblock {Chandra/HETGS Observations of the Brightest Flare Seen from Sgr A*}.
\newblock {\em Astrophys. J. Lett.} {\bf 2012}, {\em 759},~95.
%  \href{http://xxx.lanl.gov/abs/1209.6354}{{\normalfont
%  [arXiv:astro-ph.HE/1209.6354]}}. \\
[\href{http://dx.doi.org/10.1088/0004-637X/759/2/95}{CrossRef}]

\bibitem[{Covino} \em{et~al.}(2013){Covino}, {Melandri}, {Salvaterra},
{Campana}, {Vergani}, {Bernardini}, {D'Avanzo}, {D'Elia}, {Fugazza},
{Ghirlanda}, {Ghisellini}, {Gomboc}, {Jin}, {Kr{\"u}hler}, {Malesani},
{Nava}, {Sbarufatti}, and {Tagliaferri}]{Covino13}
{Covino}, S.; {Melandri}, A.; {Salvaterra}, R.; {Campana}, S.; {Vergani}, S.D.;
{Bernardini}, M.G.; {D'Avanzo}, P.; {D'Elia}, V.; {Fugazza}, D.; {Ghirlanda},
G.;  et~al.
\newblock {Dust extinctions for an unbiased sample of gamma-ray burst
afterglows}.
\newblock {\em Mon. Not. R. Astron. Soc.} {\bf 2013}, {\em 432},~1231--1244.
%  \href{http://xxx.lanl.gov/abs/1303.4743}{{\normalfont
%  [arXiv:astro-ph.HE/1303.4743]}}. \\
[\href{http://dx.doi.org/10.1093/mnras/stt540}{CrossRef}]

\bibitem[{Kumar} and {Zhang}(2015)]{Kumar15}
{Kumar}, P.; {Zhang}, B.
\newblock {The physics of gamma-ray bursts \& relativistic jets}.
\newblock {\em {Phys.~Rep.}} {\bf 2015}, {\em 561},~1--109.
%  \href{http://xxx.lanl.gov/abs/1410.0679}{{\normalfont
%  [arXiv:astro-ph.HE/1410.0679]}}. \\
[\href{http://dx.doi.org/10.1016/j.physrep.2014.09.008}{CrossRef}]

\bibitem[{O'Connor} \em{et~al.}(2020){O'Connor}, {Beniamini}, and
{Kouveliotou}]{Oconnor20}
{O'Connor}, B.; {Beniamini}, P.; {Kouveliotou}, C.
\newblock {Constraints on the circumburst environments of short gamma-ray
bursts}.
\newblock {\em Mon. Not. R. Astron. Soc.} {\bf 2020}, {\em 495},~4782--4799.
%  \href{http://xxx.lanl.gov/abs/2004.00031}{{\normalfont
%  [arXiv:astro-ph.HE/2004.00031]}}. \\
[\href{http://dx.doi.org/10.1093/mnras/staa1433}{CrossRef}]

\bibitem[{Bigiel} and {Blitz}(2012)]{BB12}
{Bigiel}, F.; {Blitz}, L.
\newblock {A Universal Neutral Gas Profile for nearby Disk Galaxies}.
\newblock {\em Astrophys. J. Lett.} {\bf 2012}, {\em 756},~183.
%  \href{http://xxx.lanl.gov/abs/1208.1505}{{\normalfont
%  [arXiv:astro-ph.CO/1208.1505]}}. \\
[\href{http://dx.doi.org/10.1088/0004-637X/756/2/183}{CrossRef}]

\bibitem[{Fong} and {Berger}(2013)]{Fong13}
{Fong}, W.; {Berger}, E.
\newblock {The Locations of Short Gamma-Ray Bursts as Evidence for Compact
Object Binary Progenitors}.
\newblock {\em Astrophys. J. Lett.} {\bf 2013}, {\em 776},~18.
%  \href{http://xxx.lanl.gov/abs/1307.0819}{{\normalfont
%  [arXiv:astro-ph.HE/1307.0819]}}. \\
[\href{http://dx.doi.org/10.1088/0004-637X/776/1/18}{CrossRef}]

\bibitem[{Dichiara} \em{et~al.}(2021){Dichiara}, {Troja}, {Beniamini},
{O'Connor}, {Moss}, {Lien}, {Ricci}, {Amati}, {Ryan}, and
{Sakamoto}]{Dichiara21}
{Dichiara}, S.; {Troja}, E.; {Beniamini}, P.; {O'Connor}, B.; {Moss}, M.;
{Lien}, A.Y.; {Ricci}, R.; {Amati}, L.; {Ryan}, G.; {Sakamoto}, T.
\newblock {Evidence of Extended Emission in GRB 181123B and Other High-redshift
Short GRBs}.
\newblock {\em Astrophys. J. Lett.} {\bf 2021}, {\em 911},~L28.
%  \href{http://xxx.lanl.gov/abs/2103.02558}{{\normalfont
%  [arXiv:astro-ph.HE/2103.02558]}}. \\
[\href{http://dx.doi.org/10.3847/2041-8213/abf562}{CrossRef}]

\bibitem[{Rowlinson} \em{et~al.}(2010){Rowlinson}, {Wiersema}, {Levan},
{Tanvir}, {O'Brien}, {Rol}, {Hjorth}, {Th{\"o}ne}, {de Ugarte Postigo},
{Fynbo}, {Jakobsson}, {Pagani}, and {Stamatikos}]{Rowlinson10}
{Rowlinson}, A.; {Wiersema}, K.; {Levan}, A.J.; {Tanvir}, N.R.; {O'Brien},
P.T.; {Rol}, E.; {Hjorth}, J.; {Th{\"o}ne}, C.C.; {de Ugarte Postigo}, A.;
{Fynbo}, J.P.U.;  et~al.
\newblock {Discovery of the afterglow and host galaxy of the low-redshift short
GRB 080905A}.
\newblock {\em Mon. Not. R. Astron. Soc.} {\bf 2010}, {\em 408},~383--391.
%  \href{http://xxx.lanl.gov/abs/1006.0487}{{\normalfont
%  [arXiv:astro-ph.HE/1006.0487]}}. \\
[\href{http://dx.doi.org/10.1111/j.1365-2966.2010.17115.x}{CrossRef}]

\bibitem[{de Ugarte Postigo} \em{et~al.}(2014){de Ugarte Postigo}, {Th{\"o}ne},
{Rowlinson}, {Garc{\'\i}a-Benito}, {Levan}, {Gorosabel}, {Goldoni},
{Schulze}, {Zafar}, {Wiersema}, {S{\'a}nchez-Ram{\'\i}rez}, {Melandri},
{D'Avanzo}, {Oates}, {D'Elia}, {De Pasquale}, {Kr{\"u}hler}, {van der Horst},
{Xu}, {Watson}, {Piranomonte}, {Vergani}, {Milvang-Jensen}, {Kaper},
{Malesani}, {Fynbo}, {Cano}, {Covino}, {Flores}, {Greiss}, {Hammer},
{Hartoog}, {Hellmich}, {Heuser}, {Hjorth}, {Jakobsson}, {Mottola}, {Sparre},
{Sollerman}, {Tagliaferri}, {Tanvir}, {Vestergaard}, and
{Wijers}]{deugarte14}
{de Ugarte Postigo}, A.; {Th{\"o}ne}, C.C.; {Rowlinson}, A.;
{Garc{\'\i}a-Benito}, R.; {Levan}, A.J.; {Gorosabel}, J.; {Goldoni}, P.;
{Schulze}, S.; {Zafar}, T.; {Wiersema}, K.;  et~al.
\newblock {Spectroscopy of the short-hard GRB 130603B. The host galaxy and
environment of a compact object merger}.
\newblock {\em Astron. Astrophys.} {\bf 2014}, {\em 563},~A62.
%  \href{http://xxx.lanl.gov/abs/1308.2984}{{\normalfont
%  [arXiv:astro-ph.CO/1308.2984]}}. \\
[\href{http://dx.doi.org/10.1051/0004-6361/201322985}{CrossRef}]

\bibitem[{Izzo} \em{et~al.}(2017){Izzo}, {Cano}, {de Ugarte Postigo}, {Kann},
{Thoene}, and {Geier}]{gcn170428A}
{Izzo}, L.; {Cano}, Z.; {de Ugarte Postigo}, A.; {Kann}, D.A.; {Thoene}, C.;
{Geier}, S.
\newblock {GRB 170428A: GTC spectroscopic redshift of candidate host galaxy}.
\newblock {\em GRB Coord. Netw.} {\bf 2017}, {\em 21059},~1.

\bibitem[{Bloom} \em{et~al.}(2007){Bloom}, {Perley}, {Chen}, {Butler},
{Prochaska}, {Kocevski}, {Blake}, {Szentgyorgyi}, {Falco}, and
{Starr}]{Bloom07}
{Bloom}, J.S.; {Perley}, D.A.; {Chen}, H.W.; {Butler}, N.; {Prochaska}, J.X.;
{Kocevski}, D.; {Blake}, C.H.; {Szentgyorgyi}, A.; {Falco}, E.E.; {Starr},
D.L.
\newblock {A Putative Early-Type Host Galaxy for GRB 060502B: Implications for
the Progenitors of Short-Duration Hard-Spectrum Bursts}.
\newblock {\em Astrophys. J. Lett.} {\bf 2007}, {\em 654},~878--884.
%  \href{http://xxx.lanl.gov/abs/astro-ph/0607223}{{\normalfont
%  [arXiv:astro-ph/astro-ph/0607223]}}. \\
[\href{http://dx.doi.org/10.1086/509114}{CrossRef}]

\bibitem[{Nugent} \em{et~al.}(2022){Nugent}, {Fong}, {Dong}, {Leja}, {Berger},
{Zevin}, {Chornock}, {Cobb}, {Kelley}, {Kilpatrick}, {Levan}, {Margutti},
{Paterson}, {Perley}, {Escorial}, {Smith}, and {Tanvir}]{Nugent22}
{Nugent}, A.E.; {Fong}, W.F.; {Dong}, Y.; {Leja}, J.; {Berger}, E.; {Zevin},
M.; {Chornock}, R.; {Cobb}, B.E.; {Kelley}, L.Z.; {Kilpatrick}, C.D.;  et~al.
\newblock {Short GRB Host Galaxies. II. A Legacy Sample of Redshifts, Stellar
Population Properties, and Implications for Their Neutron Star Merger
Origins}.
\newblock {\em Astrophys. J. Lett.} {\bf 2022}, {\em 940},~57.
%  \href{http://xxx.lanl.gov/abs/2206.01764}{{\normalfont
%  [arXiv:astro-ph.GA/2206.01764]}}. \\
[\href{http://dx.doi.org/10.3847/1538-4357/ac91d1}{CrossRef}]

\bibitem[{Berger}(2010)]{Berger10}
{Berger}, E.
\newblock {A Short Gamma-ray Burst ``No-host'' Problem? Investigating Large
Progenitor Offsets for Short GRBs with Optical Afterglows}.
\newblock {\em Astrophys. J. Lett.} {\bf 2010}, {\em 722},~1946--1961.
%  \href{http://xxx.lanl.gov/abs/1007.0003}{{\normalfont
%  [arXiv:astro-ph.HE/1007.0003]}}. \\
[\href{http://dx.doi.org/10.1088/0004-637X/722/2/1946}{CrossRef}]

\bibitem[{Perley} \em{et~al.}(2012){Perley}, {Modjaz}, {Morgan}, {Cenko},
{Bloom}, {Butler}, {Filippenko}, and {Miller}]{Perley12}
{Perley}, D.A.; {Modjaz}, M.; {Morgan}, A.N.; {Cenko}, S.B.; {Bloom}, J.S.;
{Butler}, N.R.; {Filippenko}, A.V.; {Miller}, A.A.
\newblock {The Luminous Infrared Host Galaxy of Short-duration GRB 100206A}.
\newblock {\em Astrophys. J. Lett.} {\bf 2012}, {\em 758},~122.
%  \href{http://xxx.lanl.gov/abs/1112.3963}{{\normalfont
%  [arXiv:astro-ph.HE/1112.3963]}}. \\
[\href{http://dx.doi.org/10.1088/0004-637X/758/2/122}{CrossRef}]

\bibitem[{Bloom} \em{et~al.}(2002){Bloom}, {Kulkarni}, and
{Djorgovski}]{Bloom02}
{Bloom}, J.S.; {Kulkarni}, S.R.; {Djorgovski}, S.G.
\newblock {The Observed Offset Distribution of Gamma-Ray Bursts from Their Host
Galaxies: A Robust Clue to the Nature of the Progenitors}.
\newblock {\em {Astron. J.} } {\bf 2002}, {\em 123},~1111--1148.
%  \href{http://xxx.lanl.gov/abs/astro-ph/0010176}{{\normalfont
%  [arXiv:astro-ph/astro-ph/0010176]}}. \\
[\href{http://dx.doi.org/10.1086/338893}{CrossRef}]

\bibitem[{Burrows} \em{et~al.}(2006){Burrows}, {Grupe}, {Capalbi},
{Panaitescu}, {Patel}, {Kouveliotou}, {Zhang}, {M{\'e}sz{\'a}ros},
{Chincarini}, {Gehrels}, and {Wijers}]{Burrows06}
{Burrows}, D.N.; {Grupe}, D.; {Capalbi}, M.; {Panaitescu}, A.; {Patel}, S.K.;
{Kouveliotou}, C.; {Zhang}, B.; {M{\'e}sz{\'a}ros}, P.; {Chincarini}, G.;
{Gehrels}, N.;  et~al.
\newblock {Jet Breaks in Short Gamma-Ray Bursts. II. The Collimated Afterglow
of GRB 051221A}.
\newblock {\em Astrophys. J. Lett.} {\bf 2006}, {\em 653},~468--473.
%  \href{http://xxx.lanl.gov/abs/astro-ph/0604320}{{\normalfont
%  [arXiv:astro-ph/astro-ph/0604320]}}. \\
[\href{http://dx.doi.org/10.1086/508740}{CrossRef}]

\bibitem[{Schlafly} and {Finkbeiner}(2011)]{Schlafly11}
{Schlafly}, E.F.; {Finkbeiner}, D.P.
\newblock {Measuring Reddening with Sloan Digital Sky Survey Stellar Spectra
and Recalibrating SFD}.
\newblock {\em Astrophys. J. Lett.} {\bf 2011}, {\em 737},~103.
%  \href{http://xxx.lanl.gov/abs/1012.4804}{{\normalfont
%  [arXiv:astro-ph.GA/1012.4804]}}. \\
[\href{http://dx.doi.org/10.1088/0004-637X/737/2/103}{CrossRef}]

\bibitem[{Hogg} \em{et~al.}(2002){Hogg}, {Baldry}, {Blanton}, and
{Eisenstein}]{Hogg02}
{Hogg}, D.W.; {Baldry}, I.K.; {Blanton}, M.R.; {Eisenstein}, D.J.
\newblock {The K correction}.
\newblock {\em arXiv} {\bf 2002}, arXiv:astro-ph/0210394.
%  \href{http://xxx.lanl.gov/abs/astro-ph/0210394}{{\normalfont
%  [arXiv:astro-ph/astro-ph/0210394]}}.
%\newblock {\url{https://doi.org/10.48550/arXiv.astro-ph/0210394}}.

\bibitem[{Pei}(1992)]{Pei92}
{Pei}, Y.C.
\newblock {Interstellar Dust from the Milky Way to the Magellanic Clouds}.
\newblock {\em Astrophys. J. Lett.} {\bf 1992}, {\em 395},~130. 
[\href{http://dx.doi.org/10.1086/171637}{CrossRef}]

\bibitem[{Lipunov} \em{et~al.}(2017){Lipunov}, {Gorbovskoy}, {Kornilov}, {.
Tyurina}, {Balanutsa}, {Kuznetsov}, {Vlasenko}, {Kuvshinov}, {Gorbunov},
{Buckley}, {Krylov}, {Podesta}, {Lopez}, {Podesta}, {Levato}, {Saffe},
{Mallamachi}, {Potter}, {Budnev}, {Gress}, {Ishmuhametova}, {Vladimirov},
{Zimnukhov}, {Yurkov}, {Sergienko}, {Gabovich}, {Rebolo}, {Serra-Ricart},
{Israelyan}, {Chazov}, {Wang}, {Tlatov}, and {Panchenko}]{lipunov17}
{Lipunov}, V.M.; {Gorbovskoy}, E.; {Kornilov}, V.G.; {Tyurina}, N.;
{Balanutsa}, P.; {Kuznetsov}, A.; {Vlasenko}, D.; {Kuvshinov}, D.;
{Gorbunov}, I.; {Buckley}, D.A.H.;  et~al.
\newblock {MASTER Optical Detection of the First LIGO/Virgo Neutron Star Binary
Merger GW170817}.
\newblock {\em Astrophys. J. Lett.} {\bf 2017}, {\em 850},~L1.
%  \href{http://xxx.lanl.gov/abs/1710.05461}{{\normalfont
%  [arXiv:astro-ph.HE/1710.05461]}}. \\
[\href{http://dx.doi.org/10.3847/2041-8213/aa92c0}{CrossRef}]

\bibitem[{Valenti} \em{et~al.}(2017){Valenti}, {Sand}, {Yang}, {Cappellaro},
{Tartaglia}, {Corsi}, {Jha}, {Reichart}, {Haislip}, and
{Kouprianov}]{valenti17}
{Valenti}, S.; {Sand}, D.J.; {Yang}, S.; {Cappellaro}, E.; {Tartaglia}, L.;
{Corsi}, A.; {Jha}, S.W.; {Reichart}, D.E.; {Haislip}, J.; {Kouprianov}, V.
\newblock {The Discovery of the Electromagnetic Counterpart of GW170817:
Kilonova AT 2017gfo/DLT17ck}.
\newblock {\em Astrophys. J. Lett.} {\bf 2017}, {\em 848},~L24.
%  \href{http://xxx.lanl.gov/abs/1710.05854}{{\normalfont
%  [arXiv:astro-ph.HE/1710.05854]}}. \\
[\href{http://dx.doi.org/10.3847/2041-8213/aa8edf}{CrossRef}]

\bibitem[{Soares-Santos} \em{et~al.}(2017){Soares-Santos}, {Holz}, {Annis},
{Chornock}, {Herner}, {Berger}, {Brout}, {Chen}, {Kessler}, {Sako}, {Allam},
{Tucker}, {Butler}, {Palmese}, {Doctor}, {Diehl}, {Frieman}, {Yanny}, {Lin},
{Scolnic}, {Cowperthwaite}, {Neilsen}, {Marriner}, {Kuropatkin}, {Hartley},
{Paz-Chinch{\'o}n}, {Alexander}, {Balbinot}, {Blanchard}, {Brown}, {Carlin},
{Conselice}, {Cook}, {Drlica-Wagner}, {Drout}, {Durret}, {Eftekhari}, {Farr},
{Finley}, {Foley}, {Fong}, {Fryer}, {Garc{\'\i}a-Bellido}, {Gill}, {Gruendl},
{Hanna}, {Kasen}, {Li}, {Lopes}, {Louren{\c{c}}o}, {Margutti}, {Marshall},
{Matheson}, {Medina}, {Metzger}, {Mu{\~n}oz}, {Muir}, {Nicholl}, {Quataert},
{Rest}, {Sauseda}, {Schlegel}, {Secco}, {Sobreira}, {Stebbins}, {Villar},
{Vivas}, {Walker}, {Wester}, {Williams}, {Zenteno}, {Zhang}, {Abbott},
{Abdalla}, {Banerji}, {Bechtol}, {Benoit-L{\'e}vy}, {Bertin}, {Brooks},
{Buckley-Geer}, {Burke}, {Carnero Rosell}, {Carrasco Kind}, {Carretero},
{Castander}, {Crocce}, {Cunha}, {D'Andrea}, {da Costa}, {Davis}, {Desai},
{Dietrich}, {Doel}, {Eifler}, {Fernandez}, {Flaugher}, {Fosalba},
{Gaztanaga}, {Gerdes}, {Giannantonio}, {Goldstein}, {Gruen}, {Gschwend},
{Gutierrez}, {Honscheid}, {Jain}, {James}, {Jeltema}, {Johnson}, {Johnson},
{Kent}, {Krause}, {Kron}, {Kuehn}, {Kuhlmann}, {Lahav}, {Lima}, {Maia},
{March}, {McMahon}, {Menanteau}, {Miquel}, {Mohr}, {Nichol}, {Nord},
{Ogando}, {Petravick}, {Plazas}, {Romer}, {Roodman}, {Rykoff}, {Sanchez},
{Scarpine}, {Schubnell}, {Sevilla-Noarbe}, {Smith}, {Smith}, {Suchyta},
{Swanson}, {Tarle}, {Thomas}, {Thomas}, {Troxel}, {Vikram}, {Wechsler},
{Weller}, {Dark Energy Survey}, and {Dark Energy Camera GW-EM
Collaboration}]{soares17}
{Soares-Santos}, M.; {Holz}, D.E.; {Annis}, J.; {Chornock}, R.; {Herner}, K.;
{Berger}, E.; {Brout}, D.; {Chen}, H.Y.; {Kessler}, R.; {Sako}, M.;  et~al.
\newblock {The Electromagnetic Counterpart of the Binary Neutron Star Merger
LIGO/Virgo GW170817. I. Discovery of the Optical Counterpart Using the Dark
Energy Camera}.
\newblock {\em Astrophys. J. Lett.} {\bf 2017}, {\em 848},~L16.
%  \href{http://xxx.lanl.gov/abs/1710.05459}{{\normalfont
%  [arXiv:astro-ph.HE/1710.05459]}}. \\
[\href{http://dx.doi.org/10.3847/2041-8213/aa9059}{CrossRef}]

\bibitem[{Gompertz} \em{et~al.}(2018){Gompertz}, {Levan}, {Tanvir}, {Hjorth},
{Covino}, {Evans}, {Fruchter}, {Gonz{\'a}lez-Fern{\'a}ndez}, {Jin}, {Lyman},
{Oates}, {O'Brien}, and {Wiersema}]{Gompertz18}
{Gompertz}, B.P.; {Levan}, A.J.; {Tanvir}, N.R.; {Hjorth}, J.; {Covino}, S.;
{Evans}, P.A.; {Fruchter}, A.S.; {Gonz{\'a}lez-Fern{\'a}ndez}, C.; {Jin},
Z.P.; {Lyman}, J.D.;  et~al.
\newblock {The Diversity of Kilonova Emission in Short Gamma-Ray Bursts}.
\newblock {\em Astrophys. J. Lett.} {\bf 2018}, {\em 860},~62.
%  \href{http://xxx.lanl.gov/abs/1710.05442}{{\normalfont
%  [arXiv:astro-ph.HE/1710.05442]}}. \\
[\href{http://dx.doi.org/10.3847/1538-4357/aac206}{CrossRef}]

\bibitem[{Pandey} \em{et~al.}(2019){Pandey}, {Hu}, {Castro-Tirado},
{Pozanenko}, {S{\'a}nchez-Ram{\'\i}rez}, {Gorosabel}, {Guziy}, {Jelinek},
{Tello}, {Jeong}, {Oates}, {Zhang}, {Mazaeva}, {Volnova}, {Minaev}, {van
Eerten}, {Caballero-Garc{\'\i}a}, {P{\'e}rez-Ram{\'\i}rez}, {Bremer},
{Winters}, {Park}, {Guelbenzu}, {Klose}, {Moskvitin}, {Sokolov}, {Sonbas},
{Ayala}, {Cepa}, {Butler}, {Troja}, {Chernenko}, {Molkov}, {Volvach},
{Inasaridze}, {Egamberdiyev}, {Burkhonov}, {Reva}, {Polyakov}, {Matkin},
{Ivanov}, {Molotov}, {Guver}, {Watson}, {Kutyrev}, {Lee}, {Fox},
{Littlejohns}, {Cucchiara}, {Gonzalez}, {Richer}, {Rom{\'a}n-Z{\'u}{\~n}iga},
{Tanvir}, {Bloom}, {Prochaska}, {Gehrels}, {Moseley}, {de Diego},
{Ram{\'\i}rez-Ruiz}, {Klunko}, {Fan}, {Zhao}, {Bai}, {Wang}, {Xin}, {Cui},
{Tungalag}, {Peng}, {Kumar}, {Gupta}, {Aryan}, {Kumar}, {Volvach}, {Lamb},
and {Valeev}]{Pandey19}
{Pandey}, S.B.; {Hu}, Y.; {Castro-Tirado}, A.J.; {Pozanenko}, A.S.;
{S{\'a}nchez-Ram{\'\i}rez}, R.; {Gorosabel}, J.; {Guziy}, S.; {Jelinek}, M.;
{Tello}, J.C.; {Jeong}, S.;  et~al.
\newblock {A multiwavelength analysis of a collection of short-duration GRBs
observed between 2012 and 2015}.
\newblock {\em Mon. Not. R. Astron. Soc.} {\bf 2019}, {\em 485},~5294--5318.
%  \href{http://xxx.lanl.gov/abs/1902.07900}{{\normalfont
%  [arXiv:astro-ph.HE/1902.07900]}}. \\
[\href{http://dx.doi.org/10.1093/mnras/stz530}{CrossRef}]

\bibitem[{Waxman} \em{et~al.}(2022){Waxman}, {Ofek}, and {Kushnir}]{Waxman22}
{Waxman}, E.; {Ofek}, E.O.; {Kushnir}, D.
\newblock {Strong NIR emission following the long duration GRB 211211A: Dust
heating as an alternative to a kilonova}.
\newblock {\em arXiv} {\bf 2022}, arXiv:2206.10710.
%  \href{http://xxx.lanl.gov/abs/2206.10710}{{\normalfont
%  [arXiv:astro-ph.HE/2206.10710]}}.
%\newblock {\url{https://doi.org/10.48550/arXiv.2206.10710}}.

\bibitem[{Lu} \em{et~al.}(2021){Lu}, {McKee}, and {Mooley}]{Lu21}
{Lu}, W.; {McKee}, C.F.; {Mooley}, K.P.
\newblock {Infrared dust echoes from neutron star mergers}.
\newblock {\em Mon. Not. R. Astron. Soc.} {\bf 2021}, {\em 507},~3672--3689.
%  \href{http://xxx.lanl.gov/abs/2108.04243}{{\normalfont
%  [arXiv:astro-ph.HE/2108.04243]}}. \\
[\href{http://dx.doi.org/10.1093/mnras/stab2388}{CrossRef}]

\bibitem[{Kasliwal} \em{et~al.}(2017){Kasliwal}, {Korobkin}, {Lau},
{Wollaeger}, and {Fryer}]{Kasliwal17}
{Kasliwal}, M.M.; {Korobkin}, O.; {Lau}, R.M.; {Wollaeger}, R.; {Fryer}, C.L.
\newblock {Infrared Emission from Kilonovae: The Case of the Nearby Short Hard
Burst GRB 160821B}.
\newblock {\em Astrophys. J. Lett.} {\bf 2017}, {\em 843},~L34.
%  \href{http://xxx.lanl.gov/abs/1706.04647}{{\normalfont
%  [arXiv:astro-ph.HE/1706.04647]}}. \\
[\href{http://dx.doi.org/10.3847/2041-8213/aa799d}{CrossRef}]

\bibitem[{Jin} \em{et~al.}(2018){Jin}, {Li}, {Wang}, {Wang}, {He}, {Yuan},
{Zhang}, {Zou}, {Fan}, and {Wei}]{Jin18}
{Jin}, Z.P.; {Li}, X.; {Wang}, H.; {Wang}, Y.Z.; {He}, H.N.; {Yuan}, Q.;
{Zhang}, F.W.; {Zou}, Y.C.; {Fan}, Y.Z.; {Wei}, D.M.
\newblock {Short GRBs: Opening Angles, Local Neutron Star Merger Rate, and
Off-axis Events for GRB/GW Association}.
\newblock {\em Astrophys. J. Lett.} {\bf 2018}, {\em 857},~128.
%  \href{http://xxx.lanl.gov/abs/1708.07008}{{\normalfont
%  [arXiv:astro-ph.HE/1708.07008]}}. \\
[\href{http://dx.doi.org/10.3847/1538-4357/aab76d}{CrossRef}]

\bibitem[{Rowlinson} \em{et~al.}(2010){Rowlinson}, {O'Brien}, {Tanvir},
{Zhang}, {Evans}, {Lyons}, {Levan}, {Willingale}, {Page}, {Onal}, {Burrows},
{Beardmore}, {Ukwatta}, {Berger}, {Hjorth}, {Fruchter}, {Tunnicliffe}, {Fox},
and {Cucchiara}]{Rowlinson10b}
{Rowlinson}, A.; {O'Brien}, P.T.; {Tanvir}, N.R.; {Zhang}, B.; {Evans}, P.A.;
{Lyons}, N.; {Levan}, A.J.; {Willingale}, R.; {Page}, K.L.; {Onal}, O.;
et~al.
\newblock {The unusual X-ray emission of the short Swift GRB 090515: Evidence
for the formation of a magnetar?}
\newblock {\em Mon. Not. R. Astron. Soc.} {\bf 2010}, {\em 409},~531--540.
%  \href{http://xxx.lanl.gov/abs/1007.2185}{{\normalfont
%  [arXiv:astro-ph.HE/1007.2185]}}. \\
[\href{http://dx.doi.org/10.1111/j.1365-2966.2010.17354.x}{CrossRef}]

\bibitem[{Gehrels} \em{et~al.}(2006){Gehrels}, {Norris}, {Barthelmy}, {Granot},
{Kaneko}, {Kouveliotou}, {Markwardt}, {M{\'e}sz{\'a}ros}, {Nakar}, {Nousek},
{O'Brien}, {Page}, {Palmer}, {Parsons}, {Roming}, {Sakamoto}, {Sarazin},
{Schady}, {Stamatikos}, and {Woosley}]{Gehrels06}
{Gehrels}, N.; {Norris}, J.P.; {Barthelmy}, S.D.; {Granot}, J.; {Kaneko}, Y.;
{Kouveliotou}, C.; {Markwardt}, C.B.; {M{\'e}sz{\'a}ros}, P.; {Nakar}, E.;
{Nousek}, J.A.;  et~al.
\newblock {A new {\ensuremath{\gamma}}-ray burst classification scheme from
GRB060614}.
\newblock {\em {Nature} } {\bf 2006}, {\em 444},~1044--1046.
%  \href{http://xxx.lanl.gov/abs/astro-ph/0610635}{{\normalfont
%  [arXiv:astro-ph/astro-ph/0610635]}}. \\
[\href{http://dx.doi.org/10.1038/nature05376}{CrossRef}] [\href{http://www.ncbi.nlm.nih.gov/pubmed/17183315}{PubMed}]

\bibitem[{Norris}(2002)]{Norris02}
{Norris}, J.P.
\newblock {Implications of the Lag-Luminosity Relationship for Unified
Gamma-Ray Burst Paradigms}.
\newblock {\em Astrophys. J. Lett.} {\bf 2002}, {\em 579},~386--403.
%  \href{http://xxx.lanl.gov/abs/astro-ph/0201503}{{\normalfont
%  [arXiv:astro-ph/astro-ph/0201503]}}. \\
[\href{http://dx.doi.org/10.1086/342747}{CrossRef}]

\bibitem[{Golkhou} and {Butler}(2014)]{GB14}
{Golkhou}, V.Z.; {Butler}, N.R.
\newblock {Uncovering the Intrinsic Variability of Gamma-Ray Bursts}.
\newblock {\em Astrophys. J. Lett.} {\bf 2014}, {\em 787},~90.
%  \href{http://xxx.lanl.gov/abs/1403.4254}{{\normalfont
%  [arXiv:astro-ph.HE/1403.4254]}}. \\
[\href{http://dx.doi.org/10.1088/0004-637X/787/1/90}{CrossRef}]

\bibitem[{Gompertz} \em{et~al.}(2023){Gompertz}, {Ravasio}, {Nicholl}, {Levan},
{Metzger}, {Oates}, {Lamb}, {Fong}, {Malesani}, {Rastinejad}, {Tanvir},
{Evans}, {Jonker}, {Page}, and {Pe'er}]{Gompertz23}
{Gompertz}, B.P.; {Ravasio}, M.E.; {Nicholl}, M.; {Levan}, A.J.; {Metzger},
B.D.; {Oates}, S.R.; {Lamb}, G.P.; {Fong}, W.f.; {Malesani}, D.B.;
{Rastinejad}, J.C.;  et~al.
\newblock {The case for a minute-long merger-driven gamma-ray burst from
fast-cooling synchrotron emission}.
\newblock {\em Nat. Astron.} {\bf 2023}, {\em 7},~67--79.
%  \href{http://xxx.lanl.gov/abs/2205.05008}{{\normalfont
%  [arXiv:astro-ph.HE/2205.05008]}}. \\
[\href{http://dx.doi.org/10.1038/s41550-022-01819-4}{CrossRef}]

\bibitem[{Lazzati} \em{et~al.}(2001){Lazzati}, {Ramirez-Ruiz}, and
{Ghisellini}]{Lazzati01}
{Lazzati}, D.; {Ramirez-Ruiz}, E.; {Ghisellini}, G.
\newblock {Possible detection of hard X-ray afterglows of short gamma -ray
bursts}.
\newblock {\em Astron. Astrophys.} {\bf 2001}, {\em 379},~L39--L43.
%  \href{http://xxx.lanl.gov/abs/astro-ph/0110215}{{\normalfont
%  [arXiv:astro-ph/astro-ph/0110215]}}.
 [\href{http://dx.doi.org/10.1051/0004-6361:20011485}{CrossRef}]

\bibitem[{Della Valle} \em{et~al.}(2006){Della Valle}, {Chincarini}, {Panagia},
{Tagliaferri}, {Malesani}, {Testa}, {Fugazza}, {Campana}, {Covino},
{Mangano}, {Antonelli}, {D'Avanzo}, {Hurley}, {Mirabel}, {Pellizza},
{Piranomonte}, and {Stella}]{DV06}
{Della Valle}, M.; {Chincarini}, G.; {Panagia}, N.; {Tagliaferri}, G.;
{Malesani}, D.; {Testa}, V.; {Fugazza}, D.; {Campana}, S.; {Covino}, S.;
{Mangano}, V.;  et~al.
\newblock {An enigmatic long-lasting {\ensuremath{\gamma}}-ray burst not
accompanied by a bright supernova}.
\newblock {\em {Nature} } {\bf 2006}, {\em 444},~1050--1052.
%  \href{http://xxx.lanl.gov/abs/astro-ph/0608322}{{\normalfont
%  [arXiv:astro-ph/astro-ph/0608322]}}. \\
[\href{http://dx.doi.org/10.1038/nature05374}{CrossRef}]

\bibitem[{Gal-Yam} \em{et~al.}(2006){Gal-Yam}, {Fox}, {Price}, {Ofek}, {Davis},
{Leonard}, {Soderberg}, {Schmidt}, {Lewis}, {Peterson}, {Kulkarni}, {Berger},
{Cenko}, {Sari}, {Sharon}, {Frail}, {Moon}, {Brown}, {Cucchiara}, {Harrison},
{Piran}, {Persson}, {McCarthy}, {Penprase}, {Chevalier}, and
{MacFadyen}]{Galyam06}
{Gal-Yam}, A.; {Fox}, D.B.; {Price}, P.A.; {Ofek}, E.O.; {Davis}, M.R.;
{Leonard}, D.C.; {Soderberg}, A.M.; {Schmidt}, B.P.; {Lewis}, K.M.;
{Peterson}, B.A.;  et~al.
\newblock {A novel explosive process is required for the
{\ensuremath{\gamma}}-ray burst GRB 060614}.
\newblock {\em {Nature} } {\bf 2006}, {\em 444},~1053--1055.
%  \href{http://xxx.lanl.gov/abs/astro-ph/0608257}{{\normalfont
%  [arXiv:astro-ph/astro-ph/0608257]}}. \\
[\href{http://dx.doi.org/10.1038/nature05373}{CrossRef}]

\bibitem[{Gompertz} \em{et~al.}(2020){Gompertz}, {Levan}, and
{Tanvir}]{Gompertz20}
{Gompertz}, B.P.; {Levan}, A.J.; {Tanvir}, N.R.
\newblock {A Search for Neutron Star-Black Hole Binary Mergers in the Short
Gamma-Ray Burst Population}.
\newblock {\em Astrophys. J. Lett.} {\bf 2020}, {\em 895},~58.
%  \href{http://xxx.lanl.gov/abs/2001.08706}{{\normalfont
%  [arXiv:astro-ph.HE/2001.08706]}}. \\
[\href{http://dx.doi.org/10.3847/1538-4357/ab8d24}{CrossRef}]

\bibitem[{Troja} \em{et~al.}(2008){Troja}, {King}, {O'Brien}, {Lyons}, and
{Cusumano}]{Troja08}
{Troja}, E.; {King}, A.R.; {O'Brien}, P.T.; {Lyons}, N.; {Cusumano}, G.
\newblock {Different progenitors of short hard gamma-ray bursts}.
\newblock {\em Mon. Not. R. Astron. Soc.} {\bf 2008}, {\em 385},~L10--L14.
%  \href{http://xxx.lanl.gov/abs/0711.3034}{{\normalfont
%  [arXiv:astro-ph/0711.3034]}}. \\
[\href{http://dx.doi.org/10.1111/j.1745-3933.2007.00421.x}{CrossRef}]

\bibitem[{Gao} \em{et~al.}(2017){Gao}, {Zhang}, {L{\"u}}, and {Li}]{Gao17}
{Gao}, H.; {Zhang}, B.; {L{\"u}}, H.J.; {Li}, Y.
\newblock {Searching for Magnetar-powered Merger-novae from Short GRBS}.
\newblock {\em Astrophys. J. Lett.} {\bf 2017}, {\em 837},~50.
%  \href{http://xxx.lanl.gov/abs/1608.03375}{{\normalfont
%  [arXiv:astro-ph.HE/1608.03375]}}. \\
[\href{http://dx.doi.org/10.3847/1538-4357/aa5be3}{CrossRef}]

\bibitem[{Jin} \em{et~al.}(2021){Jin}, {Zhou}, {Covino}, {Liao}, {Li}, {Lei},
{D'Avanzo}, {Fan}, and {Wei}]{Jin21}
{Jin}, Z.P.; {Zhou}, H.; {Covino}, S.; {Liao}, N.H.; {Li}, X.; {Lei}, L.;
{D'Avanzo}, P.; {Fan}, Y.Z.; {Wei}, D.M.
\newblock {A kilonova from an ultra-quick merger of a neutron star binary}.
\newblock {\em arXiv} {\bf 2021},  arXiv:2109.07694.
%  \href{http://xxx.lanl.gov/abs/2109.07694}{{\normalfont
%  [arXiv:astro-ph.HE/2109.07694]}}.
%\newblock {\url{https://doi.org/10.48550/arXiv.2109.07694}}.

\bibitem[{Villasenor} \em{et~al.}(2005){Villasenor}, {Lamb}, {Ricker},
{Atteia}, {Kawai}, {Butler}, {Nakagawa}, {Jernigan}, {Boer}, {Crew},
{Donaghy}, {Doty}, {Fenimore}, {Galassi}, {Graziani}, {Hurley}, {Levine},
{Martel}, {Matsuoka}, {Olive}, {Prigozhin}, {Sakamoto}, {Shirasaki},
{Suzuki}, {Tamagawa}, {Vanderspek}, {Woosley}, {Yoshida}, {Braga},
{Manchanda}, {Pizzichini}, {Takagishi}, and {Yamauchi}]{Villasenor05}
{Villasenor}, J.S.; {Lamb}, D.Q.; {Ricker}, G.R.; {Atteia}, J.L.; {Kawai}, N.;
{Butler}, N.; {Nakagawa}, Y.; {Jernigan}, J.G.; {Boer}, M.; {Crew}, G.B.;
et~al.
\newblock {Discovery of the short {\ensuremath{\gamma}}-ray burst GRB 050709}.
\newblock {\em {Nature} } {\bf 2005}, {\em 437},~855--858.
%  \href{http://xxx.lanl.gov/abs/astro-ph/0510190}{{\normalfont
%  [arXiv:astro-ph/astro-ph/0510190]}}. \\
[\href{http://dx.doi.org/10.1038/nature04213}{CrossRef}]

\bibitem[{Covino} \em{et~al.}(2006){Covino}, {Malesani}, {Israel}, {D'Avanzo},
{Antonelli}, {Chincarini}, {Fugazza}, {Conciatore}, {Della Valle}, {Fiore},
{Guetta}, {Hurley}, {Lazzati}, {Stella}, {Tagliaferri}, {Vietri}, {Campana},
{Burrows}, {D'Elia}, {Filliatre}, {Gehrels}, {Goldoni}, {Melandri},
{Mereghetti}, {Mirabel}, {Moretti}, {Nousek}, {O'Brien}, {Pellizza}, {Perna},
{Piranomonte}, {Romano}, and {Zerbi}]{Covino06}
{Covino}, S.; {Malesani}, D.; {Israel}, G.L.; {D'Avanzo}, P.; {Antonelli},
L.A.; {Chincarini}, G.; {Fugazza}, D.; {Conciatore}, M.L.; {Della Valle}, M.;
{Fiore}, F.;  et~al.
\newblock {Optical emission from GRB 050709: A short/hard GRB in a star-forming
galaxy}.
\newblock {\em Astron. Astrophys.} {\bf 2006}, {\em 447},~L5--L8.
%  \href{http://xxx.lanl.gov/abs/astro-ph/0509144}{{\normalfont
%  [arXiv:astro-ph/astro-ph/0509144]}}.
[\href{http://dx.doi.org/10.1051/0004-6361:200500228}{CrossRef}]

\bibitem[{D'Avanzo} \em{et~al.}(2009){D'Avanzo}, {Malesani}, {Covino},
{Piranomonte}, {Grazian}, {Fugazza}, {Margutti}, {D'Elia}, {Antonelli},
{Campana}, {Chincarini}, {Della Valle}, {Fiore}, {Goldoni}, {Mao}, {Perna},
{Salvaterra}, {Stella}, {Stratta}, and {Tagliaferri}]{Davanzo09}
{D'Avanzo}, P.; {Malesani}, D.; {Covino}, S.; {Piranomonte}, S.; {Grazian}, A.;
{Fugazza}, D.; {Margutti}, R.; {D'Elia}, V.; {Antonelli}, L.A.; {Campana},
S.;  et~al.
\newblock {The optical afterglows and host galaxies of three short/hard
gamma-ray bursts}.
\newblock {\em Astron. Astrophys.} {\bf 2009}, {\em 498},~711--721.
%  \href{http://xxx.lanl.gov/abs/0901.4038}{{\normalfont
%  [arXiv:astro-ph.CO/0901.4038]}}. \\
[\href{http://dx.doi.org/10.1051/0004-6361/200811294}{CrossRef}]

\bibitem[{Jordana-Mitjans} \em{et~al.}(2022){Jordana-Mitjans}, {Mundell},
{Guidorzi}, {Smith}, {Ram{\'\i}rez-Ruiz}, {Metzger}, {Kobayashi}, {Gomboc},
{Steele}, {Shrestha}, {Marongiu}, {Rossi}, and {Rothberg}]{Jordana22}
{Jordana-Mitjans}, N.; {Mundell}, C.G.; {Guidorzi}, C.; {Smith}, R.J.;
{Ram{\'\i}rez-Ruiz}, E.; {Metzger}, B.D.; {Kobayashi}, S.; {Gomboc}, A.;
{Steele}, I.A.; {Shrestha}, M.;  et~al.
\newblock {A Short Gamma-Ray Burst from a Protomagnetar Remnant}.
\newblock {\em Astrophys. J. Lett.} {\bf 2022}, {\em 939},~106.
%  \href{http://xxx.lanl.gov/abs/2211.05810}{{\normalfont
%  [arXiv:astro-ph.HE/2211.05810]}}. \\
[\href{http://dx.doi.org/10.3847/1538-4357/ac972b}{CrossRef}]



\bibitem[{Becerra} \em{et~al.}(2023){Becerra}, {Troja}, {Watson}, {O'Connor},
{Veres}, {Dichiara}, {Butler}, {Sakamoto}, {Lopez}, {De Colle}, {Aoki},
{Fraija}, {Im}, {Kutyrev}, {Lee}, {Paek}, {Pereyra}, {Ravi}, and
{Urata}]{Becerra23}
{Becerra}, R.L.; {Troja}, E.; {Watson}, A.M.; {O'Connor}, B.; {Veres}, P.;
{Dichiara}, S.; {Butler}, N.R.; {Sakamoto}, T.; {Lopez}, K.O.C.; {De Colle},
F.;  et~al.
\newblock {Deciphering the unusual stellar progenitor of
GRB\raisebox{-0.5ex}\textasciitilde210704A}.
\newblock {\em arXiv} {\bf 2023},  arXiv:2303.06909.
%  \href{http://xxx.lanl.gov/abs/2303.06909}{{\normalfont
%  [arXiv:astro-ph.HE/2303.06909]}}.
%\newblock {\url{https://doi.org/10.48550/arXiv.2303.06909}}.

\bibitem[{L{\"u}} \em{et~al.}(2022){L{\"u}}, {Yuan}, {Yi}, {Wang}, {Hu},
{Yuan}, {Rice}, {Wang}, {Cao}, {Kong}, {Fernandez-Garc{\'\i}a},
{Castro-Tirado}, {Lian}, {Gan}, {Wang}, {Xin}, {Caballero-Garc{\'\i}a},
{Fan}, and {Liang}]{Lu22}
{L{\"u}}, H.J.; {Yuan}, H.Y.; {Yi}, T.F.; {Wang}, X.G.; {Hu}, Y.D.; {Yuan}, Y.;
{Rice}, J.; {Wang}, J.G.; {Cao}, J.X.; {Kong}, D.F.;  et~al.
\newblock {GRB 211227A as a Peculiar Long Gamma-Ray Burst from a Compact Star
Merger}.
\newblock {\em Astrophys. J. Lett.} {\bf 2022}, {\em 931},~L23.
%  \href{http://xxx.lanl.gov/abs/2201.06395}{{\normalfont
%  [arXiv:astro-ph.HE/2201.06395]}}. \\
[\href{http://dx.doi.org/10.3847/2041-8213/ac6e3a}{CrossRef}]

\bibitem[{Micha{\l}owskI} \em{et~al.}(2018){Micha{\l}owskI}, {Xu}, {Stevens},
{Levan}, {Yang}, {Paragi}, {Kamble}, {Tsai}, {Dannerbauer}, {van der Horst},
{Shao}, {Crosby}, {Gentile}, {Stanway}, {Wiersema}, {Fynbo}, {Tanvir},
{Kamphuis}, {Garrett}, and {Bartczak}]{Michalowski18}
{Micha{\l}owskI}, M.J.; {Xu}, D.; {Stevens}, J.; {Levan}, A.; {Yang}, J.;
{Paragi}, Z.; {Kamble}, A.; {Tsai}, A.L.; {Dannerbauer}, H.; {van der Horst},
A.J.;  et~al.
\newblock {The second-closest gamma-ray burst: Sub-luminous GRB 111005A with no
supernova in a super-solar metallicity environment}.
\newblock {\em Astron. Astrophys.} {\bf 2018}, {\em 616},~A169.
%  \href{http://xxx.lanl.gov/abs/1610.06928}{{\normalfont
%  [arXiv:astro-ph.HE/1610.06928]}}. \\
[\href{http://dx.doi.org/10.1051/0004-6361/201629942}{CrossRef}]

\bibitem[{Gillanders} \em{et~al.}(2023){Gillanders}, {O'Connor}, {Dichiara},
and {Troja}]{gcn230307A}
{Gillanders}, J.; {O'Connor}, B.; {Dichiara}, S.; {Troja}, E.
\newblock {GRB 230307A: Continued Gemini-South observations confirm rapid
optical fading}.
\newblock {\em GRB Coord. Netw.} {\bf 2023}, {\em 33485},~1.

\bibitem[{Mangano} \em{et~al.}(2007){Mangano}, {Holland}, {Malesani}, {Troja},
{Chincarini}, {Zhang}, {La Parola}, {Brown}, {Burrows}, {Campana}, {Capalbi},
{Cusumano}, {Della Valle}, {Gehrels}, {Giommi}, {Grupe}, {Guidorzi}, {Mineo},
{Moretti}, {Osborne}, {Pandey}, {Perri}, {Romano}, {Roming}, and
{Tagliaferri}]{Mangano07}
{Mangano}, V.; {Holland}, S.T.; {Malesani}, D.; {Troja}, E.; {Chincarini}, G.;
{Zhang}, B.; {La Parola}, V.; {Brown}, P.J.; {Burrows}, D.N.; {Campana}, S.;
et~al.
\newblock {Swift observations of GRB 060614: An anomalous burst with a well
behaved afterglow}.
\newblock {\em Astron. Astrophys.} {\bf 2007}, {\em 470},~105--118.
%  \href{http://xxx.lanl.gov/abs/0704.2235}{{\normalfont
%  [arXiv:astro-ph/0704.2235]}}.
  [\href{http://dx.doi.org/10.1051/0004-6361:20077232}{CrossRef}]

\bibitem[{Knust} \em{et~al.}(2017){Knust}, {Greiner}, {van Eerten}, {Schady},
{Kann}, {Chen}, {Delvaux}, {Graham}, {Klose}, {Kr{\"u}hler}, {McConnell},
{Nicuesa Guelbenzu}, {Perley}, {Schmidl}, {Schweyer}, {Tanga}, and
{Varela}]{Knust17}
{Knust}, F.; {Greiner}, J.; {van Eerten}, H.J.; {Schady}, P.; {Kann}, D.A.;
{Chen}, T.W.; {Delvaux}, C.; {Graham}, J.F.; {Klose}, S.; {Kr{\"u}hler}, T.;
et~al.
\newblock {Long optical plateau in the afterglow of the short GRB 150424A with
extended emission. Evidence for energy injection by a magnetar?}
\newblock {\em Astron. Astrophys.} {\bf 2017}, {\em 607},~A84.
%  \href{http://xxx.lanl.gov/abs/1707.01329}{{\normalfont
%  [arXiv:astro-ph.HE/1707.01329]}}. \\
[\href{http://dx.doi.org/10.1051/0004-6361/201730578}{CrossRef}]

\bibitem[{Galama} \em{et~al.}(1998){Galama}, {Vreeswijk}, {van Paradijs},
{Kouveliotou}, {Augusteijn}, {B{\"o}hnhardt}, {Brewer}, {Doublier},
{Gonzalez}, {Leibundgut}, {Lidman}, {Hainaut}, {Patat}, {Heise}, {in't Zand},
{Hurley}, {Groot}, {Strom}, {Mazzali}, {Iwamoto}, {Nomoto}, {Umeda},
{Nakamura}, {Young}, {Suzuki}, {Shigeyama}, {Koshut}, {Kippen}, {Robinson},
{de Wildt}, {Wijers}, {Tanvir}, {Greiner}, {Pian}, {Palazzi}, {Frontera},
{Masetti}, {Nicastro}, {Feroci}, {Costa}, {Piro}, {Peterson}, {Tinney},
{Boyle}, {Cannon}, {Stathakis}, {Sadler}, {Begam}, and {Ianna}]{Galama98}
{Galama}, T.J.; {Vreeswijk}, P.M.; {van Paradijs}, J.; {Kouveliotou}, C.;
{Augusteijn}, T.; {B{\"o}hnhardt}, H.; {Brewer}, J.P.; {Doublier}, V.;
{Gonzalez}, J.F.; {Leibundgut}, B.;  et~al.
\newblock {An unusual supernova in the error box of the
{\ensuremath{\gamma}}-ray burst of 25 April 1998}.
\newblock {\em {Nature} } {\bf 1998}, {\em 395},~670--672.
%  \href{http://xxx.lanl.gov/abs/astro-ph/9806175}{{\normalfont
%  [arXiv:astro-ph/astro-ph/9806175]}}. \\
[\href{http://dx.doi.org/10.1038/27150}{CrossRef}]

\bibitem[{Zhang} \em{et~al.}(2022){Zhang}, {Huang}, {Zheng}, {Liu}, and
{Wang}]{Zhang22}
{Zhang}, H.M.; {Huang}, Y.Y.; {Zheng}, J.H.; {Liu}, R.Y.; {Wang}, X.Y.
\newblock {Fermi-LAT Detection of a GeV Afterglow from a Compact Stellar
Merger}.
\newblock {\em Astrophys. J. Lett.} {\bf 2022}, {\em 933},~L22.
%  \href{http://xxx.lanl.gov/abs/2205.09675}{{\normalfont
%  [arXiv:astro-ph.HE/2205.09675]}}. \\
[\href{http://dx.doi.org/10.3847/2041-8213/ac7b23}{CrossRef}]

\bibitem[{Mei} and et~al.(2022)]{Mei22}
Mei, A.; Banerjee, B.; Oganesyan, G.; Salafia, O.S.; Giarratana, S.; Branchesi, M.; D'Avanzo, P.; Campana, S.; Ghirlanda, G.; Ronchini, S.; et al.
\newblock {GeV emission from a compact binary merger}.
\newblock {\em {Nature} } {\bf 2022}, {\em 610},~230--233.

\bibitem[{Tanaka} \em{et~al.}(2014){Tanaka}, {Hotokezaka}, {Kyutoku}, {Wanajo},
{Kiuchi}, {Sekiguchi}, and {Shibata}]{Tanaka14}
{Tanaka}, M.; {Hotokezaka}, K.; {Kyutoku}, K.; {Wanajo}, S.; {Kiuchi}, K.;
{Sekiguchi}, Y.; {Shibata}, M.
\newblock {Radioactively Powered Emission from Black Hole-Neutron Star
Mergers}.
\newblock {\em Astrophys. J. Lett.} {\bf 2014}, {\em 780},~31.
[\href{http://dx.doi.org/10.1088/0004-637X/780/1/31}{CrossRef}]

\bibitem[{Ma} \em{et~al.}(2018){Ma}, {Lei}, {Gao}, {Xie}, {Chen}, {Zhang}, and
{Wang}]{Ma18}
{Ma}, S.B.; {Lei}, W.H.; {Gao}, H.; {Xie}, W.; {Chen}, W.; {Zhang}, B.; {Wang},
D.X.
\newblock {Bright Merger-nova Emission Powered by Magnetic Wind from a Newborn
Black Hole}.
\newblock {\em Astrophys. J. Lett.} {\bf 2018}, {\em 852},~L5.
%  \href{http://xxx.lanl.gov/abs/1710.06318}{{\normalfont
%  [arXiv:astro-ph.HE/1710.06318]}}. \\
[\href{http://dx.doi.org/10.3847/2041-8213/aaa0cd}{CrossRef}]

\bibitem[{Ai} \em{et~al.}(2022){Ai}, {Zhang}, and {Zhu}]{Ai22}
{Ai}, S.; {Zhang}, B.; {Zhu}, Z.
\newblock {Engine-fed kilonovae (mergernovae)---I. Dynamical evolution and
energy injection/heating efficiencies}.
\newblock {\em Mon. Not. R. Astron. Soc.} {\bf 2022}, {\em 516},~2614--2628.
%  \href{http://xxx.lanl.gov/abs/2203.03045}{{\normalfont
%  [arXiv:astro-ph.HE/2203.03045]}}. \\
[\href{http://dx.doi.org/10.1093/mnras/stac2380}{CrossRef}]

\bibitem[{Giacomazzo} and {Perna}(2013)]{Giacomazzo13}
{Giacomazzo}, B.; {Perna}, R.
\newblock {Formation of Stable Magnetars from Binary Neutron Star Mergers}.
\newblock {\em Astrophys. J. Lett.} {\bf 2013}, {\em 771},~L26.
%  \href{http://xxx.lanl.gov/abs/1306.1608}{{\normalfont
%  [arXiv:astro-ph.HE/1306.1608]}}. \\
[\href{http://dx.doi.org/10.1088/2041-8205/771/2/L26}{CrossRef}]

\bibitem[{Fryer} \em{et~al.}(2015){Fryer}, {Belczynski}, {Ramirez-Ruiz},
{Rosswog}, {Shen}, and {Steiner}]{Fryer15}
{Fryer}, C.L.; {Belczynski}, K.; {Ramirez-Ruiz}, E.; {Rosswog}, S.; {Shen}, G.;
{Steiner}, A.W.
\newblock {The Fate of the Compact Remnant in Neutron Star Mergers}.
\newblock {\em Astrophys. J. Lett.} {\bf 2015}, {\em 812},~24.
%  \href{http://xxx.lanl.gov/abs/1504.07605}{{\normalfont
%  [arXiv:astro-ph.HE/1504.07605]}}. \\
[\href{http://dx.doi.org/10.1088/0004-637X/812/1/24}{CrossRef}]

\bibitem[{Hanauske} \em{et~al.}(2017){Hanauske}, {Takami}, {Bovard},
{Rezzolla}, {Font}, {Galeazzi}, and {St{\"o}cker}]{Hanauske17}
{Hanauske}, M.; {Takami}, K.; {Bovard}, L.; {Rezzolla}, L.; {Font}, J.A.;
{Galeazzi}, F.; {St{\"o}cker}, H.
\newblock {Rotational properties of hypermassive neutron stars from binary
mergers}.
\newblock {\em Phys. Rev.~D} {\bf 2017}, {\em 96},~043004.
[\href{http://dx.doi.org/10.1103/PhysRevD.96.043004}{CrossRef}]

\bibitem[{Becerra} \em{et~al.}(2021){Becerra}, {Dichiara}, {Watson}, {Troja},
{Butler}, {Pereyra}, {Moreno M{\'e}ndez}, {De Colle}, {Lee}, {Kutyrev}, and
{L{\'o}pez}]{Becerra21}
{Becerra}, R.L.; {Dichiara}, S.; {Watson}, A.M.; {Troja}, E.; {Butler}, N.R.;
{Pereyra}, M.; {Moreno M{\'e}ndez}, E.; {De Colle}, F.; {Lee}, W.H.;
{Kutyrev}, A.S.;  et~al.
\newblock {DDOTI observations of gravitational-wave sources discovered in O3}.
\newblock {\em Mon. Not. R. Astron. Soc.} {\bf 2021}, {\em 507},~1401--1420.
[\href{http://dx.doi.org/10.1093/mnras/stab2086}{CrossRef}]

\bibitem[{Kasliwal} \em{et~al.}(2020){Kasliwal}, {Anand}, {Ahumada}, {Stein},
{Carracedo}, {Andreoni}, {Coughlin}, {Singer}, {Kool}, {De}, {Kumar},
{AlMualla}, {Yao}, {Bulla}, {Dobie}, {Reusch}, {Perley}, {Cenko}, {Bhalerao},
{Kaplan}, {Sollerman}, {Goobar}, {Copperwheat}, {Bellm}, {Anupama}, {Corsi},
{Nissanke}, {Agudo}, {Bagdasaryan}, {Barway}, {Belicki}, {Bloom}, {Bolin},
{Buckley}, {Burdge}, {Burruss}, {Caballero-Garc{\'\i}a}, {Cannella},
{Castro-Tirado}, {Cook}, {Cooke}, {Cunningham}, {Dahiwale}, {Deshmukh},
{Dichiara}, {Duev}, {Dutta}, {Feeney}, {Franckowiak}, {Frederick},
{Fremling}, {Gal-Yam}, {Gatkine}, {Ghosh}, {Goldstein}, {Golkhou}, {Graham},
{Graham}, {Hankins}, {Helou}, {Hu}, {Ip}, {Jaodand}, {Karambelkar}, {Kong},
{Kowalski}, {Khandagale}, {Kulkarni}, {Kumar}, {Laher}, {Li}, {Mahabal},
{Masci}, {Miller}, {Mogotsi}, {Mohite}, {Mooley}, {Mroz}, {Newman}, {Ngeow},
{Oates}, {Patil}, {Pandey}, {Pavana}, {Pian}, {Riddle},
{S{\'a}nchez-Ram{\'\i}rez}, {Sharma}, {Singh}, {Smith}, {Soumagnac},
{Taggart}, {Tan}, {Tzanidakis}, {Troja}, {Valeev}, {Walters}, {Waratkar},
{Webb}, {Yu}, {Zhang}, {Zhou}, and {Zolkower}]{Kasliwal20}
{Kasliwal}, M.M.; {Anand}, S.; {Ahumada}, T.; {Stein}, R.; {Carracedo}, A.S.;
{Andreoni}, I.; {Coughlin}, M.W.; {Singer}, L.P.; {Kool}, E.C.; {De}, K.;
et~al.
\newblock {Kilonova Luminosity Function Constraints Based on Zwicky Transient
Facility Searches for 13 Neutron Star Merger Triggers during O3}.
\newblock {\em Astrophys. J. Lett.} {\bf 2020}, {\em 905},~145.
[\href{http://dx.doi.org/10.3847/1538-4357/abc335}{CrossRef}]

\bibitem[{Abbott} \em{et~al.}(2018){Abbott}, {Abbott}, {Abbott}, {Abernathy},
{Acernese}, {Ackley}, {Adams}, {Adams}, {Addesso}, {Adhikari}, {Adya},
{Affeldt}, {Agathos}, {Agatsuma}, {Aggarwal}, {Aguiar}, {Aiello}, {Ain},
{Ajith}, {Akutsu}, {Allen}, {Allocca}, {Altin}, {Ananyeva}, {Anderson},
{Anderson}, {Ando}, {Appert}, {Arai}, {Araya}, {Araya}, {Areeda}, {Arnaud},
{Arun}, {Asada}, {Ascenzi}, {Ashton}, {Aso}, {Ast}, {Aston}, {Astone},
{Atsuta}, {Aufmuth}, {Aulbert}, {Avila-Alvarez}, {Awai}, {Babak}, {Bacon},
{Bader}, {Baiotti}, {Baker}, {Baldaccini}, {Ballardin}, {Ballmer},
{Barayoga}, {Barclay}, {Barish}, {Barker}, {Barone}, {Barr}, {Barsotti},
{Barsuglia}, {Barta}, {Bartlett}, {Barton}, {Bartos}, {Bassiri}, {Basti},
{Batch}, {Baune}, {Bavigadda}, {Bazzan}, {B{\'e}csy}, {Beer}, {Bejger},
{Belahcene}, {Belgin}, {Bell}, {Berger}, {Bergmann}, {Berry}, {Bersanetti},
{Bertolini}, {Betzwieser}, {Bhagwat}, {Bhandare}, {Bilenko}, {Billingsley},
{Billman}, {Birch}, {Birney}, {Birnholtz}, {Biscans}, {Bisht}, {Bitossi},
{Biwer}, {Bizouard}, {Blackburn}, {Blackman}, {Blair}, {Blair}, {Blair},
{Bloemen}, {Bock}, {Boer}, {Bogaert}, {Bohe}, {Bondu}, {Bonnand}, {Boom},
{Bork}, {Boschi}, {Bose}, {Bouffanais}, {Bozzi}, {Bradaschia}, {Brady},
{Braginsky}, {Branchesi}, {Brau}, {Briant}, {Brillet}, {Brinkmann},
{Brisson}, {Brockill}, {Broida}, {Brooks}, {Brown}, {Brown}, {Brown},
{Brunett}, {Buchanan}, {Buikema}, {Bulik}, {Bulten}, {Buonanno}, {Buskulic},
{Buy}, {Byer}, {Cabero}, {Cadonati}, {Cagnoli}, {Cahillane}, {Calder{\'o}n
Bustillo}, {Callister}, {Calloni}, {Camp}, {Cannon}, {Cao}, {Cao}, {Capano},
{Capocasa}, {Carbognani}, {Caride}, {Casanueva Diaz}, {Casentini}, {Caudill},
{Cavagli{\`a}}, {Cavalier}, {Cavalieri}, {Cella}, {Cepeda}, {Cerboni
Baiardi}, {Cerretani}, {Cesarini}, {Chamberlin}, {Chan}, {Chao}, {Charlton},
{Chassande-Mottin}, {Cheeseboro}, {Chen}, {Chen}, {Cheng}, {Chincarini},
{Chiummo}, {Chmiel}, {Cho}, {Cho}, {Chow}, {Christensen}, {Chu}, {Chua},
{Chua}, {Chung}, {Ciani}, {Clara}, {Clark}, {Cleva}, {Cocchieri}, {Coccia},
{Cohadon}, {Colla}, {Collette}, {Cominsky}, {Constancio}, {Conti}, {Cooper},
{Corbitt}, {Cornish}, {Corsi}, {Cortese}, {Costa}, {Coughlin}, {Coughlin},
{Coulon}, {Countryman}, {Couvares}, {Covas}, {Cowan}, {Coward}, {Cowart},
{Coyne}, {Coyne}, {Creighton}, {Creighton}, {Cripe}, {Crowder}, {Cullen},
{Cumming}, {Cunningham}, {Cuoco}, {Dal Canton}, {Danilishin}, {D'Antonio},
{Danzmann}, {Dasgupta}, {da Silva Costa}, {Dattilo}, {Dave}, {Davier},
{Davies}, {Davis}, {Daw}, {Day}, {Day}, {de}, {Debra}, {Debreczeni},
{Degallaix}, {de Laurentis}, {Del{\'e}glise}, {Del Pozzo}, {Denker}, {Dent},
{Dergachev}, {De Rosa}, {Derosa}, {Desalvo}, {Devine}, {Dhurandhar},
{D{\'\i}az}, {di Fiore}, {di Giovanni}, {di Girolamo}, {di Lieto}, {di Pace},
{di Palma}, {di Virgilio}, {Doctor}, {Doi}, {Dolique}, {Donovan}, {Dooley},
{Doravari}, {Dorrington}, {Douglas}, {Dovale {\'A}lvarez}, {Downes}, {Drago},
{Drever}, {Driggers}, {Du}, {Ducrot}, {Dwyer}, {Eda}, {Edo}, {Edwards},
{Effler}, {Eggenstein}, {Ehrens}, {Eichholz}, {Eikenberry}, {Eisenstein},
{Essick}, {Etienne}, {Etzel}, {Evans}, {Evans}, {Everett}, {Factourovich},
{Fafone}, {Fair}, {Fairhurst}, {Fan}, {Farinon}, {Farr}, {Farr},
{Fauchon-Jones}, {Favata}, {Fays}, {Fehrmann}, {Fejer}, {Fern{\'a}ndez
Galiana}, {Ferrante}, {Ferreira}, {Ferrini}, {Fidecaro}, {Fiori}, {Fiorucci},
{Fisher}, {Flaminio}, {Fletcher}, {Fong}, {Forsyth}, {Fournier}, {Frasca},
{Frasconi}, {Frei}, {Freise}, {Frey}, {Frey}, {Fries}, {Fritschel}, {Frolov},
{Fujii}, {Fujimoto}, {Fulda}, {Fyffe}, {Gabbard}, {Gadre}, {Gaebel}, {Gair},
{Gammaitoni}, {Gaonkar}, {Garufi}, {Gaur}, {Gayathri}, {Gehrels}, {Gemme},
{Genin}, {Gennai}, {George}, {Gergely}, {Germain}, {Ghonge}, {Ghosh},
{Ghosh}, {Ghosh}, {Giaime}, {Giardina}, {Giazotto}, {Gill}, {Glaefke},
{Goetz}, {Goetz}, {Gondan}, {Gonz{\'a}lez}, {Gonzalez Castro}, {Gopakumar},
{Gorodetsky}, {Gossan}, {Gosselin}, {Gouaty}, {Grado}, {Graef}, {Granata},
{Grant}, {Gras}, {Gray}, {Greco}, {Green}, {Groot}, {Grote}, {Grunewald},
{Guidi}, {Guo}, {Gupta}, {Gupta}, {Gushwa}, {Gustafson}, {Gustafson},
{Hacker}, {Hagiwara}, {Hall}, {Hall}, {Hammond}, {Haney}, {Hanke}, {Hanks},
{Hanna}, {Hannam}, {Hanson}, {Hardwick}, {Harms}, {Harry}, {Harry}, {Hart},
{Hartman}, {Haster}, {Haughian}, {Hayama}, {Healy}, {Heidmann}, {Heintze},
{Heitmann}, {Hello}, {Hemming}, {Hendry}, {Heng}, {Hennig}, {Henry},
{Heptonstall}, {Heurs}, {Hild}, {Hirose}, {Hoak}, {Hofman}, {Holt}, {Holz},
{Hopkins}, {Hough}, {Houston}, {Howell}, {Hu}, {Huerta}, {Huet}, {Hughey},
{Husa}, {Huttner}, {Huynh-Dinh}, {Indik}, {Ingram}, {Inta}, {Ioka}, {Isa},
{Isac}, {Isi}, {Isogai}, {Itoh}, {Iyer}, {Izumi}, {Jacqmin}, {Jani},
{Jaranowski}, {Jawahar}, {Jim{\'e}nez-Forteza}, {Johnson}, {Jones}, {Jones},
{Jonker}, {Ju}, {Junker}, {Kagawa}, {Kajita}, {Kakizaki}, {Kalaghatgi},
{Kalogera}, {Kamiizumi}, {Kanda}, {Kandhasamy}, {Kanemura}, {Kaneyama},
{Kang}, {Kanner}, {Karki}, {Karvinen}, {Kasprzack}, {Kataoka},
{Katsavounidis}, {Katzman}, {Kaufer}, {Kaur}, {Kawabe}, {Kawai}, {Kawamura},
{K{\'e}f{\'e}lian}, {Keitel}, {Kelley}, {Kennedy}, {Key}, {Khalili}, {Khan},
{Khan}, {Khan}, {Khazanov}, {Kijbunchoo}, {Kim}, {Kim}, {Kim}, {Kim}, {Kim},
{Kim}, {Kimbrell}, {Kimura}, {King}, {King}, {Kirchhoff}, {Kissel}, {Klein},
{Kleybolte}, {Klimenko}, {Koch}, {Koehlenbeck}, {Kojima}, {Kokeyama},
{Koley}, {Komori}, {Kondrashov}, {Kontos}, {Korobko}, {Korth}, {Kotake},
{Kowalska}, {Kozak}, {Kr{\"a}mer}, {Kringel}, {Krishnan}, {Kr{\'o}lak},
{Kuehn}, {Kumar}, {Kumar}, {Kumar}, {Kuo}, {Kuroda}, {Kutynia}, {Kuwahara},
{Lackey}, {Landry}, {Lang}, {Lange}, {Lantz}, {Lanza}, {Lartaux-Vollard},
{Lasky}, {Laxen}, {Lazzarini}, {Lazzaro}, {Leaci}, {Leavey}, {Lebigot},
{Lee}, {Lee}, {Lee}, {Lee}, {Lee}, {Lehmann}, {Lenon}, {Leonardi}, {Leong},
{Leroy}, {Letendre}, {Levin}, {Li}, {Libson}, {Littenberg}, {Liu},
{Lockerbie}, {Lombardi}, {London}, {Lord}, {Lorenzini}, {Loriette},
{Lormand}, {Losurdo}, {Lough}, {Lousto}, {Lovelace}, {L{\"u}ck}, {Lundgren},
{Lynch}, {Ma}, {Macfoy}, {Machenschalk}, {Macinnis}, {MacLeod},
{Maga{\~n}a-Sandoval}, {Majorana}, {Maksimovic}, {Malvezzi}, {Man}, {Mandic},
{Mangano}, {Mano}, {Mansell}, {Manske}, {Mantovani}, {Marchesoni}, {Marchio},
{Marion}, {M{\'a}rka}, {M{\'a}rka}, {Markosyan}, {Maros}, {Martelli},
{Martellini}, {Martin}, {Martynov}, {Mason}, {Masserot}, {Massinger},
{Masso-Reid}, {Mastrogiovanni}, {Matichard}, {Matone}, {Matsumoto},
{Matsushima}, {Mavalvala}, {Mazumder}, {McCarthy}, {McClelland}, {McCormick},
{McGrath}, {McGuire}, {McIntyre}, {McIver}, {McManus}, {McRae}, {McWilliams},
{Meacher}, {Meadors}, {Meidam}, {Melatos}, {Mendell}, {Mendoza-Gandara},
{Mercer}, {Merilh}, {Merzougui}, {Meshkov}, {Messenger}, {Messick},
{Metzdorff}, {Meyers}, {Mezzani}, {Miao}, {Michel}, {Michimura}, {Middleton},
{Mikhailov}, {Milano}, {Miller}, {Miller}, {Miller}, {Miller}, {Millhouse},
{Minenkov}, {Ming}, {Mirshekari}, {Mishra}, {Mitrofanov}, {Mitselmakher},
{Mittleman}, {Miyakawa}, {Miyamoto}, {Miyamoto}, {Miyoki}, {Moggi}, {Mohan},
{Mohapatra}, {Montani}, {Moore}, {Moore}, {Moraru}, {Moreno}, {Morii},
{Morisaki}, {Moriwaki}, {Morriss}, {Mours}, {Mow-Lowry}, {Mueller}, {Muir},
{Mukherjee}, {Mukherjee}, {Mukherjee}, {Mukund}, {Mullavey}, {Munch},
{Muniz}, {Murray}, {Mytidis}, {Nagano}, {Nakamura}, {Nakamura}, {Nakano},
{Nakano}, {Nakano}, {Nakao}, {Napier}, {Nardecchia}, {Narikawa},
{Naticchioni}, {Nelemans}, {Nelson}, {Neri}, {Nery}, {Neunzert}, {Newport},
{Newton}, {Nguyen}, {Ni}, {Nielsen}, {Nissanke}, {Nitz}, {Noack}, {Nocera},
{Nolting}, {Normandin}, {Nuttall}, {Oberling}, {Ochsner}, {Oelker}, {Ogin},
{Oh}, {Oh}, {Ohashi}, {Ohishi}, {Ohkawa}, {Ohme}, {Okutomi}, {Oliver}, {Ono},
{Ono}, {Oohara}, {Oppermann}, {Oram}, {O'Reilly}, {O'Shaughnessy}, {Ottaway},
{Overmier}, {Owen}, {Pace}, {Page}, {Pai}, {Pai}, {Palamos}, {Palashov},
{Palomba}, {Pal-Singh}, {Pan}, {Pankow}, {Pannarale}, {Pant}, {Paoletti},
{Paoli}, {Papa}, {Paris}, {Parker}, {Pascucci}, {Pasqualetti}, {Passaquieti},
{Passuello}, {Patricelli}, {Pearlstone}, {Pedraza}, {Pedurand}, {Pekowsky},
{Pele}, {Pe{\~n}a Arellano}, {Penn}, {Perez}, {Perreca}, {Perri}, {Pfeiffer},
{Phelps}, {Piccinni}, {Pichot}, {Piergiovanni}, {Pierro}, {Pillant},
{Pinard}, {Pinto}, {Pitkin}, {Poe}, {Poggiani}, {Popolizio}, {Post},
{Powell}, {Prasad}, {Pratt}, {Predoi}, {Prestegard}, {Prijatelj}, {Principe},
{Privitera}, {Prodi}, {Prokhorov}, {Puncken}, {Punturo}, {Puppo},
{P{\"u}rrer}, {Qi}, {Qin}, {Qiu}, {Quetschke}, {Quintero}, {Quitzow-James},
{Raab}, {Rabeling}, {Radkins}, {Raffai}, {Raja}, {Rajan}, {Rakhmanov},
{Rapagnani}, {Raymond}, {Razzano}, {Re}, {Read}, {Regimbau}, {Rei}, {Reid},
{Reitze}, {Rew}, {Reyes}, {Rhoades}, {Ricci}, {Riles}, {Rizzo}, {Robertson},
{Robie}, {Robinet}, {Rocchi}, {Rolland}, {Rollins}, {Roma}, {Romano},
{Romie}, {Rosi{\'n}ska}, {Rowan}, {R{\"u}diger}, {Ruggi}, {Ryan}, {Sachdev},
{Sadecki}, {Sadeghian}, {Sago}, {Saijo}, {Saito}, {Sakai}, {Sakellariadou},
{Salconi}, {Saleem}, {Salemi}, {Samajdar}, {Sammut}, {Sampson}, {Sanchez},
{Sandberg}, {Sanders}, {Sasaki}, {Sassolas}, {Sathyaprakash}, {Sato}, {Sato},
{Saulson}, {Sauter}, {Savage}, {Sawadsky}, {Schale}, {Scheuer}, {Schmidt},
{Schmidt}, {Schmidt}, {Schnabel}, {Schofield}, {Sch{\"o}nbeck}, {Schreiber},
{Schuette}, {Schutz}, {Schwalbe}, {Scott}, {Scott}, {Sekiguchi}, {Sekiguchi},
{Sellers}, {Sengupta}, {Sentenac}, {Sequino}, {Sergeev}, {Setyawati},
{Shaddock}, {Shaffer}, {Shahriar}, {Shapiro}, {Shawhan}, {Sheperd},
{Shibata}, {Shikano}, {Shimoda}, {Shoda}, {Shoemaker}, {Shoemaker},
{Siellez}, {Siemens}, {Sieniawska}, {Sigg}, {Silva}, {Singer}, {Singer},
{Singh}, {Singh}, {Singhal}, {Sintes}, {Slagmolen}, {Smith}, {Smith},
{Smith}, {Somiya}, {Son}, {Sorazu}, {Sorrentino}, {Souradeep}, {Spencer},
{Srivastava}, {Staley}, {Steinke}, {Steinlechner}, {Steinlechner},
{Steinmeyer}, {Stephens}, {Stevenson}, {Stone}, {Strain}, {Straniero},
{Stratta}, {Strigin}, {Sturani}, {Stuver}, {Sugimoto}, {Summerscales}, {Sun},
{Sunil}, {Sutton}, {Suzuki}, {Swinkels}, {Szczepa{\'n}czyk}, {Tacca},
{Tagoshi}, {Takada}, {Takahashi}, {Takahashi}, {Takamori}, {Talukder},
{Tanaka}, {Tanaka}, {Tanaka}, {Tanner}, {T{\'a}pai}, {Taracchini}, {Tatsumi},
{Taylor}, {Telada}, {Theeg}, {Thomas}, {Thomas}, {Thomas}, {Thorne},
{Thrane}, {Tippens}, {Tiwari}, {Tiwari}, {Tokmakov}, {Toland}, {Tomaru},
{Tomlinson}, {Tonelli}, {Tornasi}, {Torrie}, {T{\"o}yr{\"a}}, {Travasso},
{Traylor}, {Trifir{\`o}}, {Trinastic}, {Tringali}, {Trozzo}, {Tse}, {Tso},
{Tsubono}, {Tsuzuki}, {Turconi}, {Tuyenbayev}, {Uchiyama}, {Uehara}, {Ueki},
{Ueno}, {Ugolini}, {Unnikrishnan}, {Urban}, {Ushiba}, {Usman}, {Vahlbruch},
{Vajente}, {Valdes}, {van Bakel}, {van Beuzekom}, {van den Brand}, {van den
Broeck}, {Vander-Hyde}, {van der Schaaf}, {van Heijningen}, {van Putten},
{van Veggel}, {Vardaro}, {Varma}, {Vass}, {Vas{\'u}th}, {Vecchio},
{Vedovato}, {Veitch}, {Veitch}, {Venkateswara}, {Venugopalan}, {Verkindt},
{Vetrano}, {Vicer{\'e}}, {Viets}, {Vinciguerra}, {Vine}, {Vinet}, {Vitale},
{Vo}, {Vocca}, {Vorvick}, {Voss}, {Vousden}, {Vyatchanin}, {Wade}, {Wade},
{Wade}, {Wakamatsu}, {Walker}, {Wallace}, {Walsh}, {Wang}, {Wang}, {Wang},
{Wang}, {Ward}, {Warner}, {Was}, {Watchi}, {Weaver}, {Wei}, {Weinert},
{Weinstein}, {Weiss}, {Wen}, {We{\ss}els}, {Westphal}, {Wette}, {Whelan},
{Whiting}, {Whittle}, {Williams}, {Williams}, {Williamson}, {Willis},
{Willke}, {Wimmer}, {Winkler}, {Wipf}, {Wittel}, {Woan}, {Woehler}, {Worden},
{Wright}, {Wu}, {Wu}, {Yam}, {Yamamoto}, {Yamamoto}, {Yamamoto}, {Yancey},
{Yano}, {Yap}, {Yokoyama}, {Yokozawa}, {Yoon}, {Yu}, {Yu}, {Yuzurihara},
{Yvert}, {Zadro{\.z}ny}, {Zangrando}, {Zanolin}, {Zeidler}, {Zendri},
{Zevin}, {Zhang}, {Zhang}, {Zhang}, {Zhang}, {Zhao}, {Zhou}, {Zhou}, {Zhu},
{Zhu}, {Zucker}, {Zweizig}, {Kagra Collaboration}, and {VIRGO
Collaboration}]{LRR2018}
{Abbott}, B.P.; {Abbott}, R.; {Abbott}, T.D.; {Abernathy}, M.R.; {Acernese},
F.; {Ackley}, K.; {Adams}, C.; {Adams}, T.; {Addesso}, P.; {Adhikari}, R.X.;
et~al.
\newblock {Prospects for observing and localizing gravitational-wave transients
with Advanced LIGO, Advanced Virgo and KAGRA}.
\newblock {\em Living Rev. Relativ.} {\bf 2018}, {\em 21},~3.
[\href{http://dx.doi.org/10.1007/s41114-018-0012-9}{CrossRef}] [\href{http://www.ncbi.nlm.nih.gov/pubmed/29725242}{PubMed}]

\bibitem[{Petrov} \em{et~al.}(2022){Petrov}, {Singer}, {Coughlin}, {Kumar},
{Almualla}, {Anand}, {Bulla}, {Dietrich}, {Foucart}, and
{Guessoum}]{petrov22}
{Petrov}, P.; {Singer}, L.P.; {Coughlin}, M.W.; {Kumar}, V.; {Almualla}, M.;
{Anand}, S.; {Bulla}, M.; {Dietrich}, T.; {Foucart}, F.; {Guessoum}, N.
\newblock {Data-driven Expectations for Electromagnetic Counterpart Searches
Based on LIGO/Virgo Public Alerts}.
\newblock {\em Astrophys. J. Lett.} {\bf 2022}, {\em 924},~54.
[\href{http://dx.doi.org/10.3847/1538-4357/ac366d}{CrossRef}]

\end{thebibliography}
\end{document}